\documentclass{aa} 
\usepackage{graphicx}
\usepackage{color}
\usepackage{graphicx}
\usepackage{lscape}
\usepackage{natbib}
\usepackage[varg]{txfonts}
\usepackage{amssymb}
\usepackage{stmaryrd}
\usepackage{url}
\usepackage{xspace}
\usepackage{color}
\usepackage{siunitx}
\usepackage{textcomp}
\usepackage[usenames,dvipsnames]{xcolor}
\usepackage{longtable}
\usepackage{multicol}
\bibpunct{(}{)}{;}{a}{}{,}

\newcounter{Rco}
\newcommand{\Ionst}[1]{\setcounter{Rco}{#1}\Roman{Rco}}

\newcommand{\Ionw}[3]{\mbox{#1\,{\scriptsize\Ionst{#2}}~$\lambda\,#3$\,\AA}}

\newcommand{\Ionww}[3]{\mbox{#1\,{\scriptsize\Ionst{#2}}~$\lambda\lambda\,#3$\,\AA}}

\newcommand{\se}[1]{\mbox{Sect.\,\ref{#1}}}
\newcommand{\logg}{\mbox{$\log g$}\xspace}
\newcommand{\loggw}[1]{\mbox{$\log g\hspace{-0.5mm} =\hspace{-0.5mm}  #1$}}

\newcommand{\fg}[1]{\mbox{Fig.\,\ref{#1}}}

\newcommand{\Teff}{\mbox{$T_\mathrm{eff}$}\xspace}
\newcommand{\Teffw}[1]{\mbox{$\Teff\hspace{-0.5mm} =\hspace{-0.5mm} #1 \,\mathrm{kK}$}}

\newcommand{\Lsol}{$\mathrm{L}_\odot$}
\newcommand{\Msol}{$\mathrm{M}_\odot$}
\newcommand{\Rsol}{$\mathrm{R}_\odot$}

\begin{document} 

\title{The bright blue side of the night sky: \\Spectroscopic survey of bright and hot (pre-) white dwarfs}
\titlerunning{The bright blue side of the night sky}

\author{Nicole Reindl \inst{1}
    \and Ramazan Islami \inst{1}
    \and Klaus Werner \inst{2}
    \and S. O. Kepler \inst{3}
    \and Max Pritzkuleit \inst{1}
    \and Harry Dawson \inst{1}
    \and Matti Dorsch \inst{1}
    \and Alina Istrate\inst{4}
    \and Ingrid Pelisoli \inst{5} 
    \and Stephan Geier \inst{1}
    \and Murat Uzundag \inst{6, 7, 8}
    \and Judith Provencal \inst{9, 10}
    \and Stephen Justham \inst{11, 12, 13, 14}
    }
    
\offprints{Nicole\,Reindl\\ \email{nreindl885@gmail.com}}

\institute{Institut f\"ur Physik und Astronomie, Universit\"at Potsdam, Haus 28, Karl-Liebknecht-Str. 24/25, D-14476 Potsdam-Golm, Germany
\and Institut f\"ur Astronomie und Astrophysik, Kepler Center for Astro and Particle Physics, Eberhard Karls Universit\"at, Sand 1, 72076 T\"ubingen, Germany
\and Instituto de F\'{i}sica, Universidade Federal do Rio Grande do Sul, 91501-970, Porto-Alegre, RS, Brazil
\and Department of Astrophysics/IMAPP, Radboud University, P O Box 9010, NL-6500 GL Nijmegen, The Netherlands
\and Department of Physics, University of Warwick, Coventry, UK
\and Instituto de F\'isica y Astronom\'ia, Universidad de Valpara\'iso, Gran Breta\~na 1111, Playa Ancha, Valpara\'iso 2360102, Chile
\and European Southern Observatory, Alonso de Cordova 3107, Santiago, Chile
\and Institute of Astronomy, Faculty of Physics, Astronomy and Informatics, Nicolaus Copernicus University in Toru\'n, Grudzi\k{a}dzka 5, PL-87-100 Toru\'n, Poland
\and Department of Physics and Astronomy Newark, University of Delaware, DE 19716, USA 
\and Delaware Asteroseismic Research Center, Mt. Cuba Observatory, Greenville, DE 19807, USA
\and Anton Pannekoek Institute for Astronomy and GRAPPA, University of Amsterdam, NL-1090 GE Amsterdam, The Netherlands
\and School of Astronomy \& Space Science, University of the Chinese Academy of Sciences, Beijing 100012, China
\and National Astronomical Observatories, Chinese Academy of Sciences, Beijing 100012, China
\and Max-Planck-Institut f\"ur Astrophysik, Karl-Schwarzschild-Stra{\ss}e 1, 85748 Garching, Germany}

\date{Received \ Accepted}

\abstract
{We report on the spectroscopic confirmation of 68 new bright ($G=13.5-17.2$\,mag) and blue (pre-)white dwarfs (WDs). This finding has allowed us to almost double the number of the hottest (\Teff$\geq60$\,kK) known WDs brighter than $G=16$\,mag. 
We increased the number of known ultra-high excitation (UHE) WDs by 20\%, found one unambiguous close binary system consisting of one DA WD with an irradiated low-mass companion, one DAO, and one DOA WD that are likely in their transformation phase of becoming pure DA WDs, one rare, naked O(H) star, two DA and two DAO WDs with \Teff\ possibly in excess of 100\,kK, three new DOZ WDs, and three of our targets are central stars of (possible) planetary nebulae.\\
Using non-local thermodynamic equilibrium models, we derived the atmospheric parameters of these stars and by fitting their spectral energy distribution we derived their radii, luminosities, and gravity masses. In addition, we derived their masses in the Kiel and Hertzsprung–Russell diagram (HRD). We find that Kiel, HRD, and gravity mass agree only in half of the cases. This is not unexpected and we attribute this to the neglect of metal opacities, possibly stratified atmospheres, as well as possible uncertainties of the parallax zero point determination.\\
Furthermore, we carried out a search for photometric variability in our targets using archival data, finding that 26\% of our targets are variable. This includes 15 new variable stars, with only one of them being clearly an irradiation effect system. Strikingly, the majority of the variable stars exhibit non-sinusoidal light-curve shapes, which are unlikely explained in terms of close binary systems. We propose that a significant fraction of all (not just UHE) WDs develop spots when entering the WD cooling phase. We suggest that this could be related to the on-set of weak magnetic fields and possibly diffusion.} 


\maketitle

\section{Introduction}
\label{sec:intro}
Very hot (pre-) white dwarfs (WDs) represent the beginning of the end for the vast majority of all stars. They cover a huge but sparsely populated region in the Hertzsprung-Russell diagram (HRD) and, thus, they are an important link in stellar evolution between the (post-) asymptotic giant branch (AGB) stars, the (post-) red giant branch (RGB) stars, the (post-) extreme horizontal branch (EHB) stars, and the bulk of the cooler WDs. Because the atmospheres of hot pre-WDs (and possibly the hottest WDs) are not yet affected by gravitational settling of heavier elements \citep{Werner+2020}, they are particularly valuable objects in the work to uncover and quantify non-canonical evolutionary pathways, such as the occurrence of (very) late thermal pulses or various double WD merger scenarios \citep{Sion+2000, WernerHerwig2006, Justham+2011, Zhang+2012a, Zhang+2012b, Reindl+2014b, Reindl+2014c, Werner+2022a, Werner+2022b}.\\
Besides their role as key objects in the reconstruction of the various evolutionary histories of intermediate mass stars, very hot pre-WDs also serve as powerful tools to address a multitude of \mbox{(astro-)physical} questions. They are considered the most reliable and internally consistent flux calibrators \citep{Bohlin+2020} and they serve as laboratories to derive atomic data for highly ionized species of trans-iron elements (e.g., \citealt{Rauch2007, Rauch+2015a, Rauch+2015b}). In addition, newly formed WDs enable the  study of the age structure and star formation history of the Galactic halo \citep{Kalirai2012, Kilic+2019, Fantin+2021}, investigations of the properties of weakly interacting particles via the shape of the WD luminosity function \citep{Isern2008, MillerBertolami2014b, MillerBertolami2014a, Isern+2022}, and they potentially allow us to directly observe variations in fundamental constants at locations of high gravitational potential \citep{Berengut+2013, Bainbridge+2017, Hu+2020}.\\
The metamorphosis of an AGB or RGB star into a WD is still far from understood. Besides different evolutionary channels, we still do not know whether really every star will go through a planetary nebula (PN) phase, whether PNe are mainly the outcome of a binary phenomenon \citep{Moe+2006, Moe+2012, DeMarco+2009, Jones2019, Boffin+Jones+2019}, whether the occurrence of PNe is restricted to post-AGB stars only, or whether PNe could also be observed around post-RGB stars \citep{Hall+2013, Hillwig+2017, Jones+2020, Jones+2022}.\\
The transformation of some ($\approx2/3$) of the He-rich WDs into H-rich at \Teff$\approx 70-30$\,kK is reasonably well understood \citep{Bedard+2020, Bedard+2023}. However, it is not yet clear why for \Teff$>100$\,kK, there are $\approx5\times$ as many H-deficient WDs than H-rich WDs \citep{Fleming+1986, Krzesinski2009, Werner+2019}, while the opposite seems to be the case for pre-WDs \citep{Weidmann+2020} and cooler WDs \citep{Bedard+2020}.\\
The occurrence and evolution of metal abundances in hot WDs is also not yet understood. While \cite{Bedrad+2022} could aptly reproduce  the evolution of the carbon abundances of PG1159 stars into DOZ WDs, their models could not account for the coolest ($\approx45-50$\,kK) DOZ stars. Abundances for elements heavier than C are typically derived from ultraviolet (UV) spectra. Some of the hottest WDs still seem to display their original abundance pattern, which is attributed to the presence of a weak residual stellar wind (e.g., \citealt{Werner+2018a, Werner+2020}). For slightly more evolved WDs, diffusion seems to become apparent (e.g., \citealt{Loebling+2020, Werner+2018b}), meaning light metals (e.g., C, N, and O) have subsolar abundances, while (trans-)iron group elements appear to be enhanced, similar to what is seen in intermediate He-rich hot subdwarfs (e.g. \citealt{Latour+2018, Dorsch+2019}) or (magnetic) chemically peculiar stars \citep{Michaud1970}. It has also been argued that heavy elements are accreted from external sources rather than being intrinsic to the star \citep{Barstow+2014, Schreiber+2019}.\\
Perhaps the strangest phenomenon associated with hot WDs is the sudden appearance of as-yet unidentified absorption lines in the optical spectra of these stars. These absorption lines were tentatively identified as Rydberg lines of ultra-high excited (UHE) metals in ionization stages {\sc v - x}, indicating line formation in a dense environment with temperatures near $10^6$\,K \citep{Werner+1995}. Based on the discovery of photometric and spectroscopic variable UHE WDs, \cite{Reindl+2019} suggested that the UHE 
lines could be created in a wind-fed and shock-heated magnetosphere.
Furthermore, UHE WDs were established a new class of variable stars \citep{Reindl+2021}. Due to the lack of increasing photometric amplitudes towards longer wavelengths, as well as the non-detection of spectral features of a hypothetical secondary, \cite{Reindl+2021} suggested that spots on the surfaces of these stars and/or geometrical effects of circumstellar material might be responsible.\\
Interestingly, in recent years there have been increasing indications that also ordinary (non-UHE) hot WDs could be infested with spots. 
\cite{Hermes+2017} reported about a few hot (\Teff$>30$\,kK) magnetic WDs found to be variable in the \emph{K2} mission. Their light curves are either asymmetrical or, as in the case of the 60\,kK DAH WD, there are two uneven maxima observed. Furthermore, \cite{Werner+2019} found that the extremely hot (105\,kK) DA WD \object{PG\,0948+534} is photometrically variable with a period of 3.45\,d and an asymmetrical light curve shape, which could be explained by spots. Studying the TESS light curves of central stars of PNe, 
\cite{Aller+2020} reported that the light curve of the DO WD \object{PG\,1034+001} shows two uneven maxima, which could also be explained with spots.\\ 

A major obstacle to making progress with the above-mentioned problems is the small number of known hot (\mbox{pre-)}WDs. In particular bright objects that allow for in-depth investigations, such as detailed metal abundance measurements, search for magnetic fields, or pulsations, are very rare. Thanks to the \emph{Gaia} space mission \citep{Gaia+2016, Gaia+2021, Gaia2022} and catalogs compiled for hot subdwarfs (i.e., hot pre-WDs) by \cite{Geier+2019, Culpan+2022} and catalogs compiled for WDs by \cite{GentileFusillo+2019, GentileFusillo+2021}, it has now become possible to search for such objects in a targeted manner.\\
Motivated to increase especially the number of bright (pre-)WDs on record, we carried out spectroscopic surveys using various telescopes, which are described in \se{sec:obs}. In \se{sect:class}, we describe the spectral classifications of our targets as well as their positions in the \emph{Gaia} color-magnitude-diagram (CMD). The spectral analysis and the spectral energy distribution (SED) fitting of our stars is explained in \se{sec:spec} and \se{sec:sed}, respectively. In \se{sec:masses} we derive the gravity, Kiel, and HRD masses of our stars. In \se{sec:lc}, we investigate archival light curves and search for period signals. Notes on individual objects are provided in \se{sect:notes}. Finally, we provide a summary and a discussion in \se{sect:summary}.

\section{Observations} 
\label{sec:obs}
We obtained the spectra of hot (pre-)WD candidates reported by 
\cite{Geier+2019, Culpan+2022, GentileFusillo+2019}, and \cite{GentileFusillo+2021} using various observing programs and telescopes, as listed below. 
In total, we took spectra of 71 individual stars, of which 68 are 
new discoveries. About two third of the stars have \emph{Gaia} 
G-band magnitudes between $15-16$\,mag, twelve have $14<G/\mathrm{mag}<15$, 
six have $13.4<G/\mathrm{mag}<14$, and eight are fainter than $G=16$\,mag.
An overview of all the relevant observations is given in Table~\ref{table:A1}.

\subsection{Isaac Newton Telescope}
We carried out a survey targeting bright and blue WD candidates 
from the \emph{Gaia} DR2 and eDR3 (ProgID: ING.NL.21A.003, PI: Istrate) at 
the 2.54~m Isaac Newton Telescope (INT) in 2021 on February 15, 16, 17, and 
June $10-13, 2021$. We used the long-slit Intermediate Dispersion Spectrograph 
(IDS) with the grating R400V which provides a 
resolution of $R=1450$. This survey obtained spectra of 44 (pre-) 
WDs.\\
In addition, spectra of another ten WDs were taken in December 
2019 in course of a survey targeting hot subluminous stars 
within 500\,pc (ProgID: ING.NL.19B.005; PI: Istrate/Justham). 
This survey also employed the INT using the same instrument 
setup as mentioned above. 
Using the Image Reduction and Analysis Facility (IRAF) bias and 
flat field corrections were applied to the data. The wavelength 
calibration was performed with Cu-Ne-Ar calibration lamp spectra. 
In addition we performed flux calibrations of the instrument 
response function taking atmospheric extinction into account.

\subsection{Southern Astrophysical Research telescope}
One spectrum was taken with the Goodman High-Throughput Spectrograph 
\citep[GHTS,][]{clemens2004} with a SYZY 400 grating ($R=1000$) at the 
Southern Astrophysical Research (SOAR) 4.1-m telescope on Cerro Pach\'on. 
We reduced the spectroscopic data using the instrument 
pipeline\footnote{\url{https://github.com/soar-telescope/goodman_pipeline}} 
including overscan, trim, slit trim, bias, and flat corrections. 
We employed a method developed by \citet{wojtek2004},
included in the pipeline, to identify and remove cosmic rays.
After that we carried out the wavelength calibration using 
the He-Ar-Ne comparison lamp exposure that was taken at the same 
telescope position as our target.
A sixth order Legendre function is used to 
calibrate the pixel-wavelength correspondence using an atlas of 
known He-Ar-Ne lines.

\subsection{Large Binocular Telescope}
We carried out another survey dedicated to the discovery of more UHE WDs 
as a bad weather filler program (ProgIDs: RDS-2021B-010, RDS-2022A-007, PI: Reindl) 
at the twin 8.4m Large Binocular Telescope (LBT) using the Multi-Object 
Double Spectrographs (MODS, \citealt{Pogge+2010}). Our targets were required to 
have $BP-RP<-0.3$\,mag, and absolute \emph{Gaia} magnitude between 
6\,mag$<M_\mathrm{G}<9$\,mag, meaning that they should lie 
approximately in the color magnitude diagram (CMD) region of known UHE WDs as reported by \cite{Reindl+2021}. 
Furthermore, most of our LBT targets are short periodic photometric variables,
which we previously found in the Zwicky Transient Facility (ZTF) survey DR5. In total, the 
spectra of ten WDs were obtained.\\
MODS provides two-channel grating spectroscopy by using a dichroic that splits 
the light at $\approx5650$\,\AA\ into separately optimized red and blue channels. 
The spectra cover the wavelength region $3330 - 5800$\,\AA\ and 
$5500 - 10\,000$\,\AA\ with a resolving power of $R\approx 1850$ and $2300$, 
respectively. We reduced the spectra using the 
modsccdred\footnote{\url{https://github.com/rwpogge/modsCCDRed}} \textsc{PYTHON} 
package \citep{Pogge2019} for basic 2d CCD reductions, and the 
modsidl\footnote{\url{https://github.com/rwpogge/modsIDL}} pipeline 
\citep{Croxall+2019} to extract 1d spectra and apply wavelength and flux
calibrations.

\subsection{Very Large Telescope}
Twelve further hot (pre-)WD candidates were observed within the
Hot Faint Underluminous Sky Survey (HOTFUSS, ProgIDs: 0106.D-0259(A), 0110.D-4098(A); 
PI: Geier), which was carried out as bad weather filler program with the X-shooter 
instrument at ESO’s Very Large Telescope (VLT). 
In addition, one hot pre-WD was observed with VLT/X-shooter in course of a program targeting
hyper-runaway, intermediate-mass stripped helium stars (ProgID: 0109.D-0235(A);  
PI: Pritzkuleit). The spectra cover the wavelength range $3000-10\,000\,\AA$ and have a 
resolving power of $R\approx 10\,000$. We downloaded the extracted, wavelength- and 
flux-calibrated 1D spectra from the ESO science archive.

\begin{figure}
\centering
\includegraphics[width=\hsize]{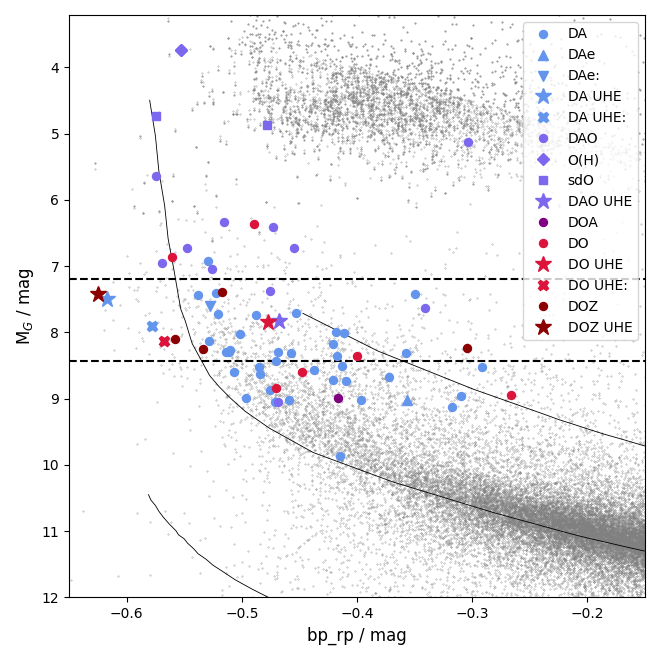}
\caption{Locations of our targets in the \emph{Gaia} CMD. Hot subdwarf and WD candidates from \cite{Culpan+2022} and \cite{GentileFusillo+2021}, respectively, that have parallaxes better than 10\% are shown in gray. The dashed lines indicate the absolute \emph{Gaia} magnitude region in which the UHE phenomenon occurs as uncovered by \cite{Reindl+2021}. The thin solid lines correspond to Montreal WD cooling tracks for WD masses of 0.2 (top right), 0.6 (middle), and 1.3\,\Msol\ (bottom left) .}
\label{fig:cmd}
\end{figure}

\begin{figure*}[ht]
 \centering  
 \includegraphics[width=\textwidth]{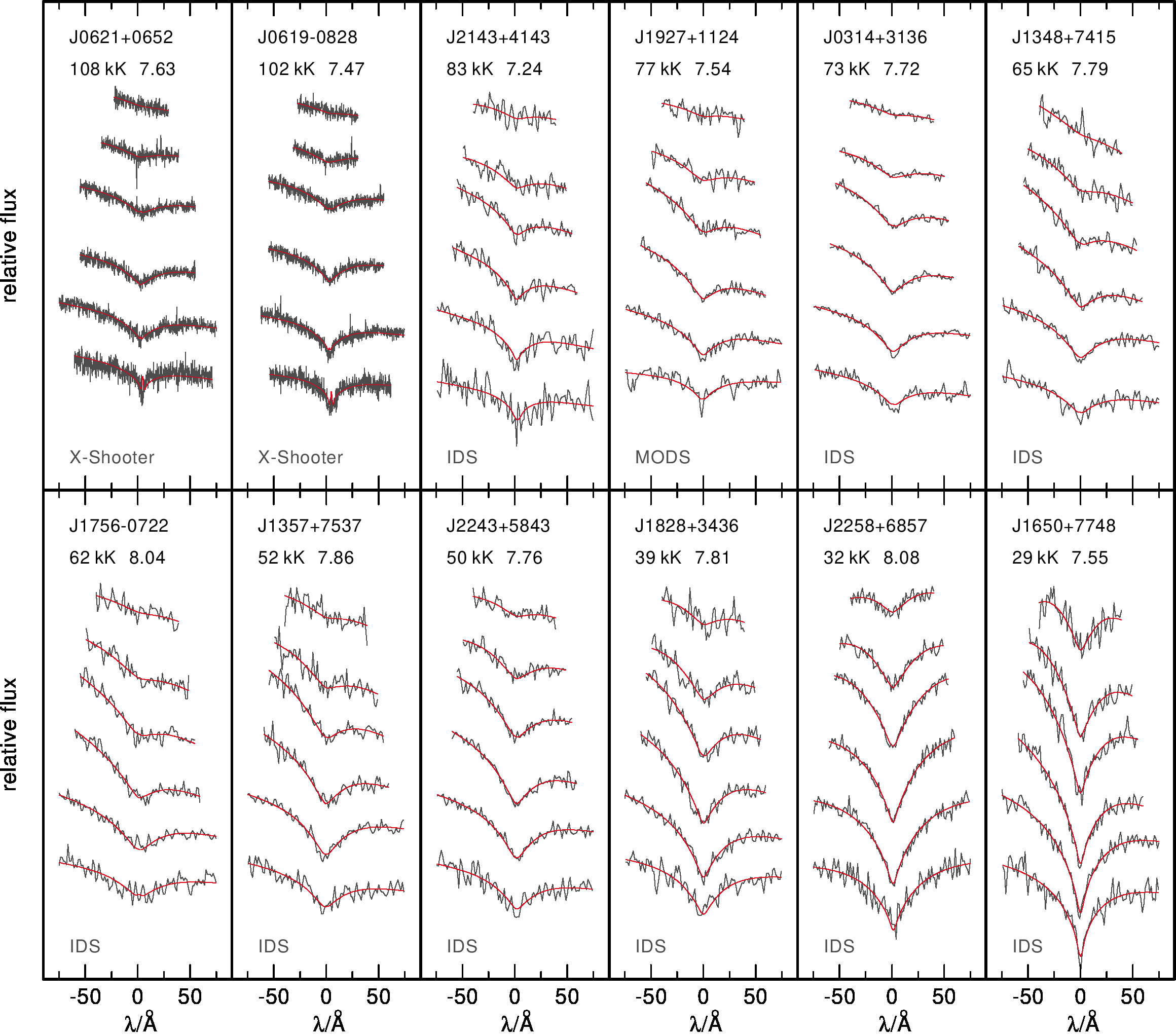}
  \caption{Fits to the Balmer lines (H\,$\zeta$ to H\,$\alpha$, from top to bottom) of some of our DA WDs, overplotted with the best fit TMAP models (red). The names of the stars, the derived effective temperatures and surface gravities as well as the spectrograph used for the observations are indicated.}
\label{fig:da}
\end{figure*}

\section{Spectral Classifications} 
\label{sect:class}
Ultimately, 80\% of our targets turned out to be H-rich, of which we classified 37 
as DA; 1 as DAe; 1 as DAe; 1 as DA UHE; 1 as DA UHE; 11 as DAO;
1 as DAO UHE; 1 as an O(H) star; and 2 as sdO stars. 
The remaining 20\% are H-deficient WDs, of which we classified 6 as 
DO; 2 as DO UHE; 1 as DOA; 4 as DOZ; and 1 as DOZ UHE. 
We list the spectral classification of each star in Table~\ref{table:A1}.\\ 
In \fg{fig:cmd}, we show the locations of our targets in the \emph{Gaia}
CMD. Stars that do not show any He are indicated in blue, while hybrid stars
showing both H and He are shown in light (if H is the dominant element in the 
atmosphere) or dark (if He is the dominant element in the atmosphere) purple, 
stars that only show He in red, and He-rich objects that also show metals in 
dark red. Also shown in this figure are theoretical cooling sequences for H-rich 
CO-core WDs with masses of 0.2, 0.6, and 1.3\,\Msol\ of the Montreal WD 
Group\footnote{\url{https://www.astro.umontreal.ca/~bergeron/CoolingModels/}}  
\citep{Bedard+2020}.
It can be seen that one sdO and one DAO are located within the hot subdwarf cloud, 
and the O(H) star as well as another DAO lie in the sparsely populated region 
between the hot subdwarf cloud and the top of the WD banana.
Most of our targets have $M_{\mathrm{G}}<6$ and 
lie on the top of the WD banana, or slightly redward 
of it. The latter is most likely due to interstellar reddening, since the stars 
in our sample have distances that are  approximately between $100-2200$\,pc 
\citep{Bailer-Jones+2021}. It is also worth mentioning that several 
of our newly discovered UHE WDs (star symbols) and possible UHE WDs (crosses) 
lie within the absolute \emph{Gaia} magnitude region in which the UHE 
phenomenon occurs (indicated by the black, dashed lines) as 
uncovered by \cite{Reindl+2021}. In addition, we find that 
several of the DOZ and DAO WDs have $M_\mathrm{G}$ below the upper $MG$ limit 
reported by \cite{Reindl+2021} for the UHE WDs. The only exception to this rule 
is the DAO WD WDJ210110.17$-$052751.14. However, this objects appears to be a 
former DO WD that is currently transforming into a pure DA WD (see \se{sect:summary}).

\begin{figure*}[ht]
 \centering  
 \includegraphics[width=\textwidth]{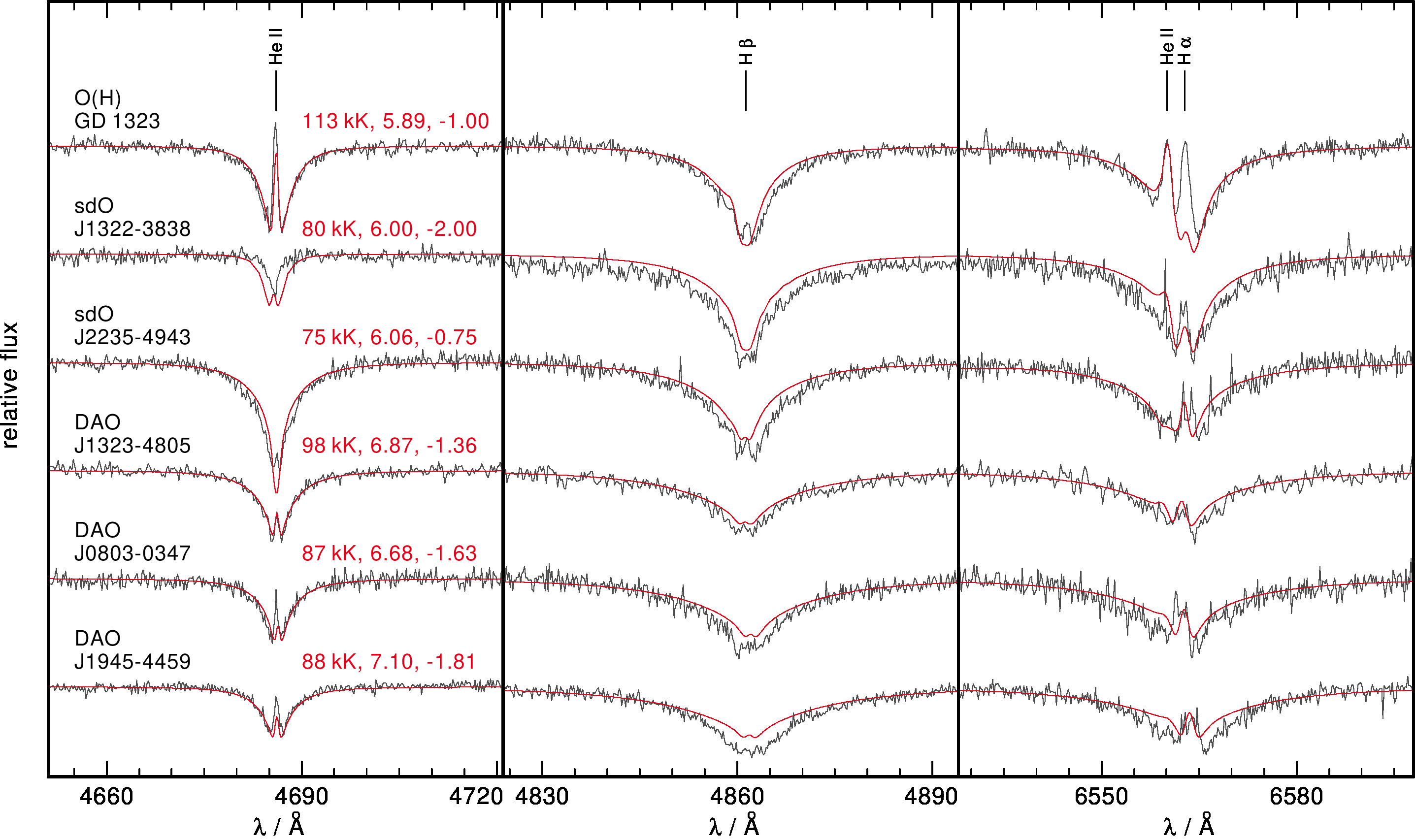}
  \caption{Fits to He\,{\sc ii} 4686\,\AA, H\,$\beta$, and H\,$\alpha$ observed in the X-Shooter spectra of a selection of our O(H), sdO, and DAO WD stars. Over plotted in red are the best fit TMAP models. The spectral classifications, names of the stars, the derived effective temperatures, surface gravities, and logarithmic He/H ratios (by number) are indicated. The Balmer line problem is more or less evident in each star.}
\label{fig:dao}
\end{figure*}

\begin{figure*}[ht]
 \centering  
 \includegraphics[width=\textwidth]{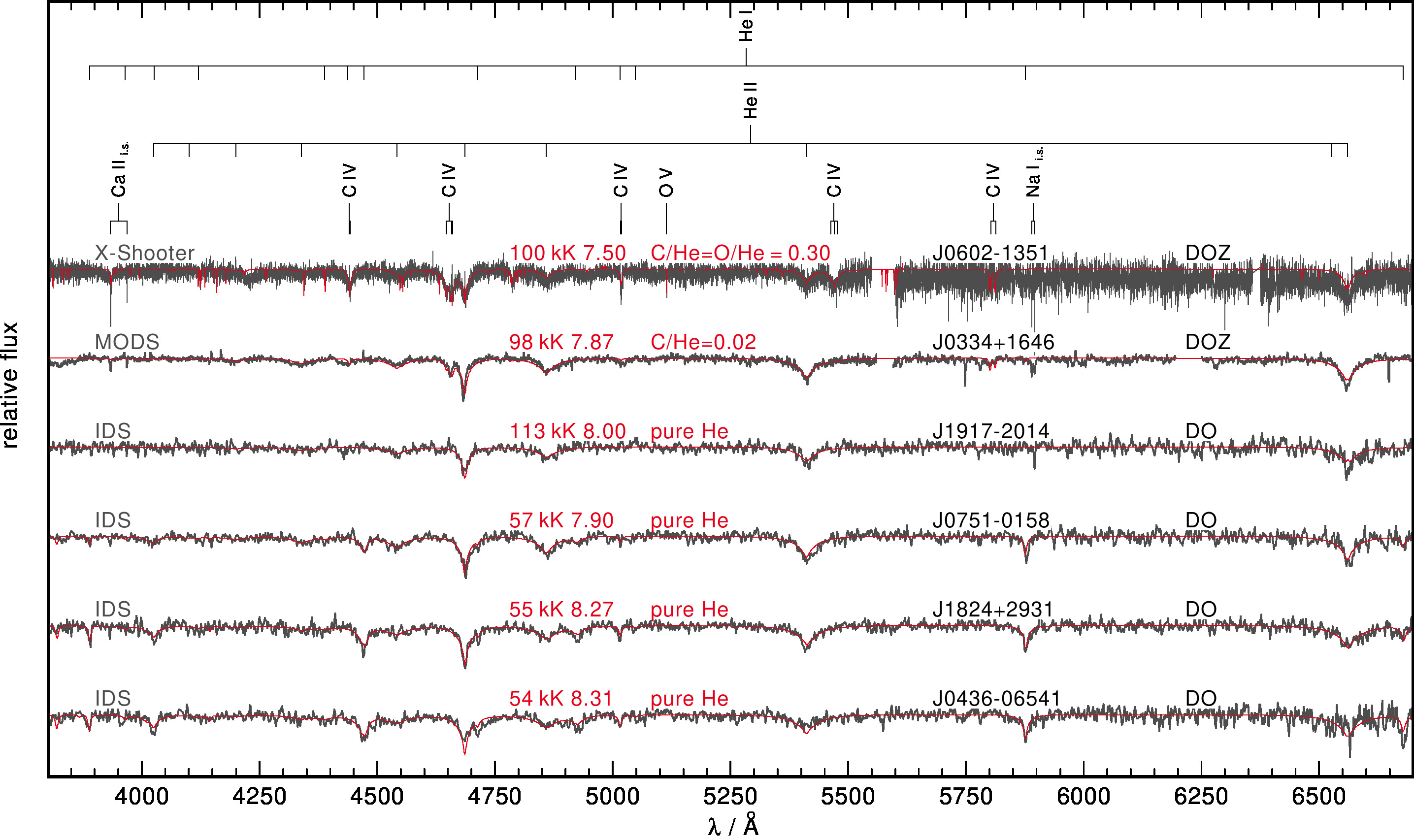}
  \caption{Spectra of some of the DOZ and DO WDs (gray) overplotted with the best fit TMAP models (red). The positions of photospheric lines are marked. Effective temperatures, surface gravities, and chemical compositions (in number ratios), as determined in this work as well as the spectrograph used for the observation are also indicated.}
\label{fig:do}
\end{figure*}

\section{Spectral Analysis}
\label{sec:spec}

\subsection{Model grids}
\label{sec:models}
We computed plane-parallel, non-local thermodynamic equilibrium (NLTE) model atmospheres in radiative and hydrostatic equilibrium using the T{\"u}bingen NLTE Model-Atmosphere Package (\textsc{TMAP}\footnote{\url{http://astro.uni-tuebingen.de/~TMAP}}) along with atomic data that were obtained from the T{\"u}bingen Model Atom Database (\textsc{TMAD}\footnote{\url{http://astro.uni-tuebingen.de/~TMAD}}, \citealt{rauchdeetjen2003, werner+2003, tmap2012}). 
To calculate synthetic line profiles, we used Stark line-broadening tables
provided by \cite{TremblayBergeron2009} for \ion{H}{I}, \cite{Barnard1969} for \Ionww{He}{1}{4026, 4388, 4471, 4921}, 
\cite{Barnard1974} for \Ionw{He}{1}{4471} and \cite{Griem1974} for all other \ion{He}{I} lines, and for \ion{He}{II}.

The pure H model grid covers \Teffw{20-200} (1\,kK steps for \Teff$<40$\,kK, 5\,kK steps 
for \Teff$<100$\,kK, and 10\,kK steps for \Teff$>100$\,kK) and \loggw{6.0-9.0} (in steps of 0.5\,dex).

Our pure He model grid spans from \Teffw{40-200} (in steps of 5\,kK for \Teff$<100$\,kK, 
10\,kK for \Teff$>100$\,kK) and covers surface gravities from \loggw{6.0-9.0} (in steps of 0.5\,dex). Models above the Eddington limit (i.e. \Teffw{190, 200} for \loggw{6.0})
were not calculated.

For the DOZ WDs, we chose two different model sets, depending on the carbon abundance. For stars with low C/He ratios ($\leq 0.03$ by number) we computed models composed of helium and carbon. They were described in detail by \cite{Werner+2014}. For the two DOZs with high carbon abundances (i.e., the PG\,1159 stars), we used models composed of He, C, and O, plus H in order to enable an estimate of an upper hydrogen abundance limit. Nitrogen was included and treated as a trace element, meaning that NLTE line formation iterations for N were performed by keeping fixed the atmospheric structure. This model type was introduced by \cite{WernerRauch2014}. 

For DAO WDs we computed a grid which considers opacities from both H and He. It covers 
the same parameter space and step sizes as the pure He model grid and was calculated 
for six different He abundances ($\log$\,(He/H) $= 0, -1, -2, -3, -4, -5$, logarithmic 
number ratios). Models above the Eddington limit (i.e. \Teffw{190, 200} for \loggw{6.0} 
and $\log$\,(He/H) $= 0, -1, -2, -3, -4, -5,$ and \Teffw{180} for \loggw{6.0} and 
$\log$\,(He/H) $=-2, -3, -4, -5$) were not calculated. For models with \Teff$<100$\,kK, we considered 15, 103, and 20 levels in NLTE for \ion{H}{I} \ion{He}{I}, and \ion{He}{II}, 
respectively. For models with \Teff$\geq 100$\,kK we considered 15, 5, and 32 levels in 
NLTE for \ion{H}{I} \ion{He}{I}, and \ion{He}{II}, respectively. For stars that turned out to be O(H) or sdO stars, we employed the model grid computed by 
\cite{Reindl+2016}.

\subsection{Spectral fitting}
If more than one spectrum of a star -- taken with the same instrument and with a 
similar signal-to-noise ratio (S/N) -- was available, we performed a fit to the 
co-added the spectrum. If the S/N of one spectrum was significantly worse, we 
only performed a fit to the higher S/N spectrum. Five of our INT targets and the 
one SOAR target were later re-observed with X-Shooter for the LBT/MODS. For those, we adopted the 
atmospheric parameters derived in the analysis of the X-Shooter spectra.

To derive the effective temperatures and surface gravities for the DA WDs
we fit the Balmer lines in an automated procedure by means of 
$\chi^2$ minimization using the FITSB2 routine \citep{Napiwotzki1999} 
and calculated the statistical 1\,$\sigma$ errors. Each fit was then inspected 
visually to ensure the quality of the analysis. In some cases, for instance, when a 
certain Balmer line was affected by a cosmic, we excluded this line from the 
fit. In \fg{fig:da}, we show some examples of these fits and in 
Table~\ref{table:DA} we summarize the derived effective temperatures and surface 
gravities for the DA WDs.

For determining \Teff, \logg, and the He abundances of the DAO WDs, O(H) and 
sdO stars, we performed global $\chi^2$  spectral fits to consider 
several absorption lines of H and He. Poor-quality regions of the 
spectra and interstellar lines have been excluded from the fit. 
The Balmer line problem is more or less evident in each of these 
stars as can be seen in \fg{fig:dao}, where we show our best 
fits to He\,{\sc ii} 4686\,\AA\, H\,$\beta$, and H\,$\alpha$ in the 
X-Shooter spectra of some of our DAO WDs, O(H), and sdO stars.
The Balmer line problem describes the failure to achieve a consistent 
fit to all Balmer (and He\,{\sc ii}) lines simultaneously, meaning that 
for a particular object different \Teff\ follow from fits to different 
Balmer line series members. As already mentioned by \cite{Bedard+2020},
the Balmer line problem can easily hide in low S/N 
spectra, and for most of our sdO, O(H), and DAO stars, we have higher 
S/N X-Shooter spectra available. This explains also we find the Balmer 
line problem more frequent on those stars, compared to the majority of DA WDs,
for which we have only INT/IDS spectra. 
Thanks to the high resolution of X-Shooter, the NLTE line core 
emissions, whose strength is very sensitive to \Teff, are clearly 
resolved in the spectra of our sdO, O(H), and DAO stars. We provide an 
overview of the derived atmospheric parameters of 
the DAO WDs, O(H) and sdO stars in Table~\ref{table:DAO}.

The same global $\chi^2$  spectral fitting approach as mentioned 
before has been used to derive the effective temperatures and 
surface gravities for the DO WDs. In the case of DOZ WDs with low C abundances,
we also excluded \ion{C}{iv} lines from the fitting. 
With these parameters, we subsequently computed models including He and 
C by varying the C abundance to obtain a good by-eye fit to the observed
\ion{C}{iv} lines in the DOZ stars. We show the fits to a selection of 
our DO and DOZ WDs in Fig.~\ref{fig:do}. A summary of the derived 
effective temperatures, surface gravities, and C abundances is given 
in Table~\ref{table:DO}.

For DOZ WDs with higher C abundances (PG1159 stars), UHE WDs, as well as 
one of the sdO that shows a particularly strong version of the Balmer line 
problem, we performed the \Teff, \logg, and abundance determination by-eye. 
For a more in-depth descripton of  the fits to each object, we refer to Sect.\,\ref{sect:notes}, .

\begin{figure*}[ht]
 \centering  
 \includegraphics[width=\textwidth]{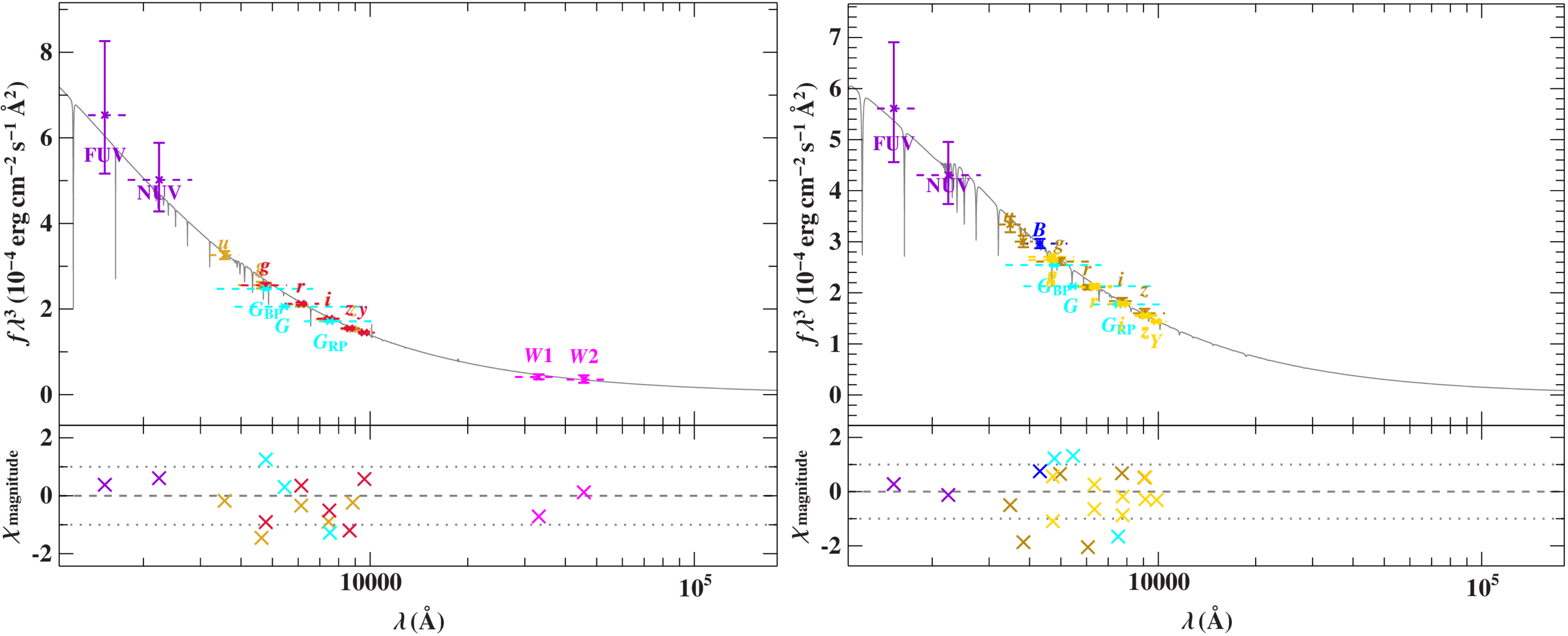}
  \caption{Exemplary SED fits to the post-RGB candidate DAO central star of PN WDJ120728.43$+$540129.16 (left) and the DO  WDJ211532.62$-$615849.50 (right). Top panels: Filter-averaged fluxes converted from observed magnitudes are shown in different colors. The respective full width at tenth maximum are shown as dashed horizontal lines. The best-fitting model, degraded to a spectral resolution of 6\,\AA\ is plotted in gray. To reduce the steep SED slope, the flux is multiplied by the wavelength cubed. Bottom panel: Difference between synthetic and observed magnitudes. The following color code is employed for the various photometric systems: GALEX (violet, \citealt{Bianchi+2017}), Pan-STARRS1 (dark red, \citealt{Chambers+2016}), Johnson (blue, \citealt{Henden+2015}), Gaia (cyan, \citealt{Gaia+2021}), SDSS (golden, \citealt{Alam+2015}), Skymapper (golden, \citealt{Wolf+2018}), Dark Energy Survey (yellow, \citealt{Abdalla+2018}), and WISE (magenta, \citealt{Schlafly+2019}).}
\label{fig:sed}
\end{figure*}

\section{SED fitting}
\label{sec:sed}
We performed fits to the SEDs in order to determine the radius, $R$, and 
luminosity, $L$, of each star assuming our previously derived effective 
temperatures, surface gravities, and He-abundances\footnote{We note that 
for DOZ WDs we neglected the opacity of C, as thus far, we have only 
implemented grids containing H and/or He for WDs in the SED fitting program.}.
In addition, we relied on \emph{Gaia} eDR3 \citep{Gaia+2021, Gaia2022} parallaxes, which were 
corrected for the zeropoint bias using the Python code provided by 
\cite{Lindegren+2021}\footnote{\url{https://gitlab.com/icc-ub/public/gaiadr3_zeropoint}},
as well as photometry from various catalogs and archives.
The $\chi ^2$ SED fitting routine is described in \cite{Heber+2018} and 
\cite{Irrgang+2021}. It allows us to derive the angular diameter (defined 
as $\Theta = 2 R/(\varpi$) of each source by performing a fit of the model
spectrum to the observed SED. The fits account for the effect of interstellar 
reddening by using the reddening law of \cite{Fitzpatrick+2019}. 
We kept the atmospheric parameters fixed, and let the angular diameter, 
$\Theta$, and the color excess, $E(44-55)$\footnote{\cite{Fitzpatrick+2019} employs $E(44-55)$, which is the monochromatic equivalent of usual $E(B-V)$, using 
the wavelengths 4400\,\AA\ and 5500\,\AA, respectively. For high effective 
temperatures such as for the stars in our sample $E(44-55)$ is identical 
to $E(B-V)$.} vary freely. Two exemplary SED fits are shown in \fg{fig:sed}. The radius was then calculated from the angular diameter, via $R = \Theta/(2\varpi)$, 
and the luminosity from the radius and the $T_\mathrm{eff}$ from our spectral fitting via 
$L/L_\odot = (R/R_\odot)^2(T_\mathrm{eff}/T_{\mathrm{eff},\odot})^4$. 
The derived radii and luminosities\footnote{The numbers given are the median 
and the highest density interval with probability 0.6827 
\citep[see][for details on this measure of uncertainty]{Bailer-Jones+2021}.} 
are listed in Tables~\ref{table:DA}-\ref{table:DO}.

\section{Masses}
\label{sec:masses}

\begin{figure*}[ht]
 \centering  
 \includegraphics[width=\textwidth]{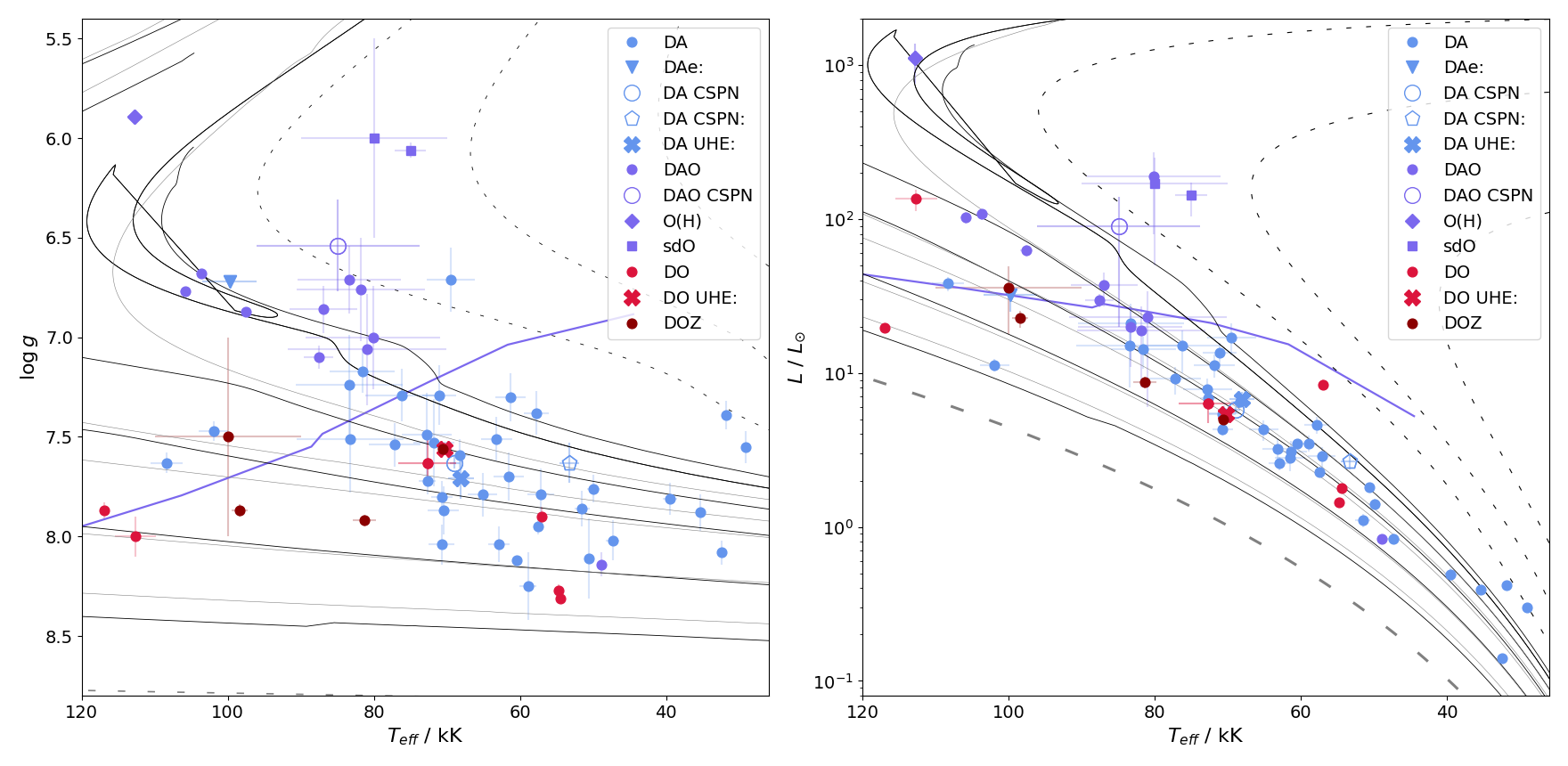}
  \caption{Locations of our targets in the Kiel diagram (left) and HRD (right). The black dashed and solid lines are stellar evolutionary tracks for He-, CO-core WD, respectively \citep{Renedo+2010, Hall+2013},  corresponding to masses of 0.306, 0.378, 0.452, 0.524, 0.570, 0.632, 0.767, and 0.934\,M$_{\odot}$. The gray solid lines are VLTP post-AGB evolutionary tracks form \cite{Althaus+2009} for masses of 0.514, 0.565, 0.609, 0.741, and 0.869\,M$_{\odot}$. The gray, thick, dashed line is a 1.10\,M$_{\odot}$ H-deficient ONe-Core WD track from \cite{Camisassa+2019}. The purple solid line 
  indicates where the He abundances should have decreased down to $\log(\mathrm{He/H})=-3$ according to predictions of \cite{UnglaubBues2000}.} 
\label{fig:hrd}
\end{figure*}

For each star in our sample for which we carried out a formal 
spectroscopic fit, we determined its mass using three 
different methods. For one, we derived the Kiel mass, $M_{\mathrm{Kiel}}$, 
from the Kiel diagram (\fg{fig:hrd}, left) and HRD mass, 
$M_{\mathrm{HRD}}$, from the HRD (\fg{fig:hrd}, right). For 
this we employed theoretical evolutionary sequences 
and the griddata function in \textsc{PYTHON} with the 
rescale option, that rescales the data points of the grid to unit cube 
before the interpolation is performed. The uncertainties 
on the Kiel and HRD masses  were estimated using a Monte 
Carlo method. For H-rich objects, we employed 
evolutionary tracks from \cite{Renedo+2010} for CO-core WDs with $Z=0.01$ 
and for He-core WDs we used tracks from \cite{Hall+2013}. For He-rich 
objects we relied on tracks from \cite{Althaus+2009} for CO-core WDs
and one track from \cite{Camisassa+2019} for a He-rich ONe-core WD.\\
Finally, we also calculated the gravity mass via 
$M_{\mathrm{grav}} = g R^2/G$ from the radius and the surface gravity as 
determined from the SED fitting and spectral analysis.
We list the derived masses in Table~\ref{table:DA}-\ref{table:DO}.

\section{Light curves}
\label{sec:lc}

\begin{table}
\caption{\label{tab:lc} Names, spectral types, bands of the light curves, periods, and peak-to-peak amplitudes of the light curve variation for the newly discovered photometrically variable stars in our sample.}
\begin{tabular}{llccc}
\hline\hline
\noalign{\smallskip}
Name & Spectral & Band & $P$ & Amplitude  \\
(short) & Type &      & [d] & [mag]  \\
\noalign{\smallskip}
\hline
\noalign{\smallskip}
J0143+5841      &       DO      &       g       &       0.298152        &       0.086   \\
                &               &       r       &       0.298152        &       0.088   \\
                &               &       i       &       0.297917        &       0.087   \\
J0334+1646      &       DOZ     &       g       &       0.377464        &       0.038   \\
                &               &       r       &       0.377476        &       0.036   \\
J0600$-$1014 &  DA      &       TESS    &       0.957692        &               \\
J0645+5659      &       DAe:    &       g       &       1.693849        &       0.052   \\
                &                   &   r       &       1.693875        &       0.048   \\
J0702+0514      &       DOZ\,UHE        &       g       &       0.597626        &       0.083   \\
                &               &       r       &       0.597602        &               \\
                &               &       i       &       0.597348        &       0.083   \\
                &               &       TESS    &       0.598129        &               \\
J0706+6133      &       DA\,UHE &       g       &       0.373759        &       0.035   \\
                &               &       r       &       0.373758        &       0.038   \\
                &               &       TESS    &       0.373758        &               \\
J0751+1059      &       DAe     &       g       &       0.276724        &       0.055   \\
                &               &       r       &       0.276729        &       0.127   \\
                &               &       TESS    &       0.276728        &               \\
J0805+4109      &       DA      &       g       &       0.526120        &       0.041   \\
                &               &       r       &       0.526115        &       0.024   \\
                &               &       TESS    &       0.526873        &               \\
J0935+6852      &       DO\,UHE &       g       &       1.344009        &       0.028   \\
                &               &       r       &       1.343939        &       0.032   \\
                &               &       TESS    &       1.344018        &               \\
J1256+7753      &       DA\,UHE:        &       g       &       0.145490        &       0.037   \\
                &               &       r       &       0.145489        &       0.032   \\
                &               &       i       &       0.145491        &       0.043   \\
                &               &       TESS    &       0.145489        &               \\
J2101$-$0527 &  DAO & TESS    &   0.537862    & \\
J2101+1356      &       DAO\,UHE        &       g       &       1.221712        &       0.026   \\
                &               &       r       &       1.221783        &       0.024   \\
                &               &       TESS    &       1.218041        &               \\
J2115$-$6158    &       DO UHE: &       TESS    &       0.620826        &               \\
J2202+2750      &       DA      &       g       &       0.815598        &       0.031   \\
                &               &       r       &       0.815570        &       0.028   \\
J2221+5011      &       DA      &       g       &       2.272771        &       0.018   \\
                &               &       r       &       2.272896        &       0.019   \\
\noalign{\smallskip}
\hline\hline
\end{tabular}
\end{table}

\begin{figure*}[ht]
 \centering  
 \includegraphics[width=0.99\textwidth]{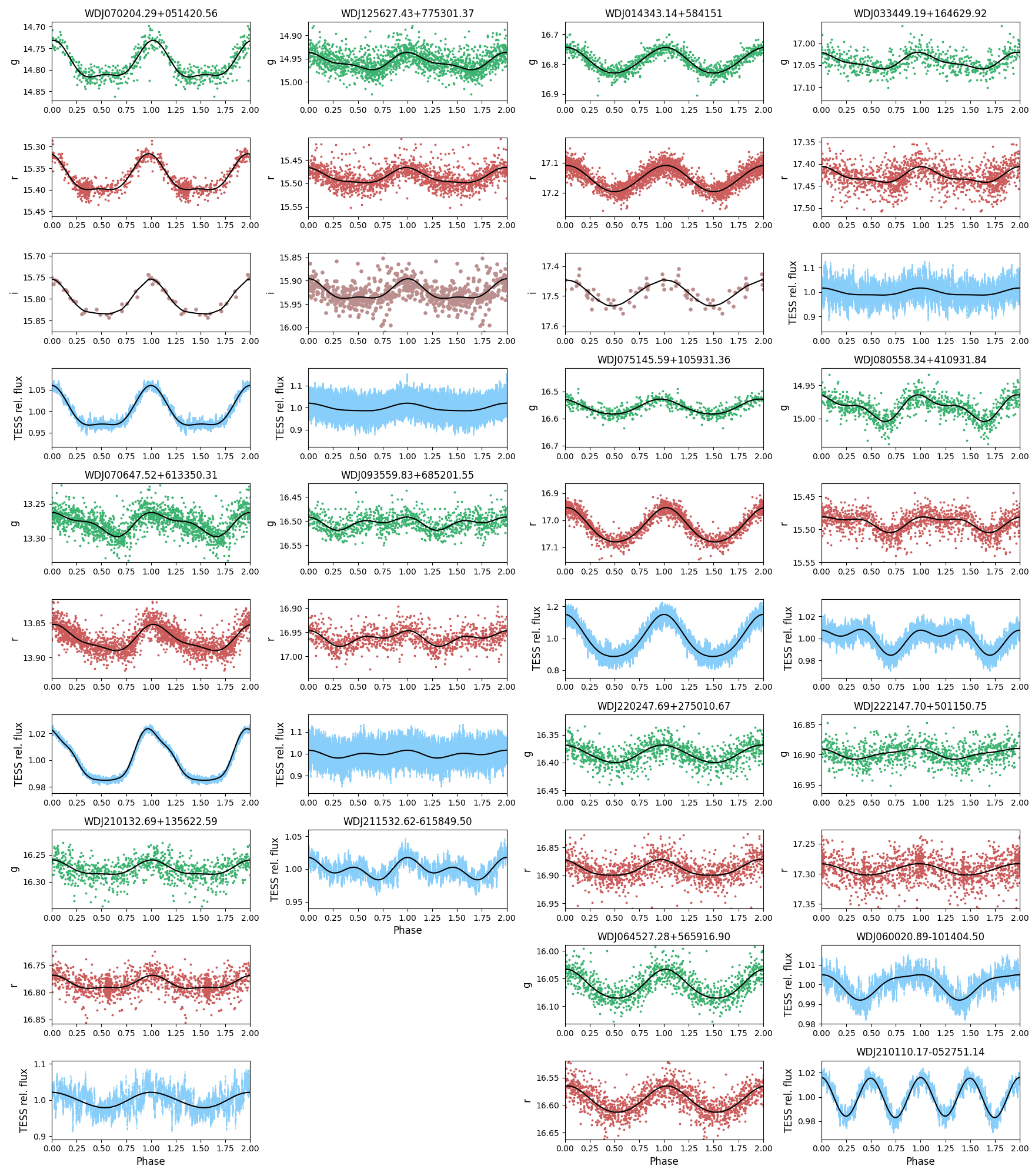}
 \caption{ZTF and phase-averaged TESS light curves of the photometrically variable stars in our sample. In the first and second column the light curves of (possible) UHE WDs are shown, and in the last two columns the light curves of non-UHE WDs. For each star, the light curves are shown one below the other, and the beginning of a set of light curves for a particular star is marked by the title, which contains the name of the variable star.}
 \label{fig:lc}
\end{figure*}

One goal of our observing campaign was to increase the number 
of UHE WDs. The discovery that the majority ($\approx75\%$) of these 
peculiar WDs are photometrically variable \citep{Reindl+2021} provides 
an important observational constraint that can be used to detect more 
of these objects.\\
Before our observing campaigns and (again) after our spectroscopic 
analysis, we inspected light curves from the Zwicky Transient Facility 
(ZTF, \citealt{Bellm+2019, Masci+2019}) survey  which provides 
photometry in the g and r bands, and (with less frequency) in the 
i-band.
In addition, we searched for periodic signals in the 
2\,min cadence and 20\,s cadence light curves obtained
with the Transiting Exoplanet Survey Satellite (TESS).
We downloaded the TESS target pixel files (TPF) of each object from MAST
as FITS format. The FITS files were already processed based
on the Pre-Search Data Conditioning Pipeline \citep{Jenkins+2016}
from  which we  extracted the barycentric corrected dynamical Julian days
("BJD $-$ 2457000", a  time  system  that  is  corrected  by  leap  seconds;
see \citealt{Eastman+2010}) and the pre-search Data Conditioning
Simple Aperture Photometry flux ("PDCSAP  FLUX") for which long-term
trends have been removed using the co-trending basis vectors.  We used 
the PDC light curves and converted the fluxes to fractional
variations from the mean (i.e., differential intensity). 

For the analyses of the light curves, we used the \textsc{VARTOOLS} program
\citep{HartmanBakos2016} to perform a generalized Lomb-Scargle (LS) search
\citep{ZechmeisterKuerster2009, Press1992} for periodic sinusoidal signals.
We consider objects variable that show a periodic signal with a false alarm 
probability (FAP) of $\log(FAP)\leq-4$.
In cases where we found more than one significant period, we whitened the light
curve by removing the strongest periodic signal (including its harmonics and
subharmonics) from the light curve. The periodogram was then recomputed
to check whether or not the FAP of the next strongest signal still remains above our
variability threshold ($\log(FAP)\leq-4$). This whitening procedure was repeated
until no more significant periodic signals could be found. 
By fitting a harmonic series (black lines in \fg{fig:lc}) to each 
light curve, we determined the peak-to-peak amplitude of the light curve, 
which we define as the difference between the maximum and minimum of the fit.

Due to the poor spatial resolution (one detector pixel corresponds to 21\,arcsec 
on the sky), TESS photometry suffers from crowding issues. Since our targets are 
typically faint considering the TESS detection limit, a detected periodic signal 
of a given target may actually originate from a neighboring object. Therefore we 
carefully checked for blends with close-by stars using the tpfplotter code 
\citep{Aller+2020} and by checking the crowdsap parameter, which gives the fraction 
of the flux in the photometric aperture that comes from our target. In case a star showed variability in the TESS data, but was found to be in a crowed region and no 
ZTF data was available, we checked the literature for close-by variable stars. 
Finally, we also employed the open-source Python package \emph{TESS-localize}\footnote{\url{https://github.com/Higgins00/TESS-Localize}} \citep{Higgins+2023}
which allows us to check the probability of a \emph{Gaia} source being the source of variability given TESS pixel data and a set of observed frequencies of variability.
We discuss such cases in more detail in Sect.~\ref{sect:notes}.\\
In Table~\ref{tab:lc}, we summarize the 15 newly discovered photometrically variable 
stars in our sample, list their spectral type, and the band used for the period search, as 
well as the amplitude of the light curve variation in each band. We do not list the 
amplitudes of the TESS light curves as they cannot be considered trustworthy for our 
faint stars and the large pixel size implies that an accurate background 
subtraction is very complicated, particularly in crowded fields.

\section{Notes on individual objects}
\label{sect:notes}

\subsection{DA WDs}
\label{sect:DA}

\paragraph{WDJ060020.89$-$101404.50} The TESS light curve of this 63.2\,kK hot DA WD 
indicates a period of 22.97\,h. The star is located in a 
crowded field (crowdsap$=0.25$), and no ZTF or other archival data 
are available. However, the Python package \emph{TESS-localize} predicts a  
likelihood of 1.0 that the WD is indeed the source of observed variability.
The shape of the light-curve is asymmetric, with an extended maximum. 

\paragraph{WDJ061906.92$-$082807.15}
This is the second hottest DA WD in our sample, we derive \Teffw{102\pm2} 
and \loggw{7.47\pm0.05}. Fitting H$\epsilon\,-\,\beta$ from low resolution 
spectra with pure H NLTE models, \cite{Vennes+1999} derived a significantly
lower \Teffw{76.3\pm0.2} and higher \loggw{8.05\pm0.15}. Since this star only displays 
a very mild Balmer line problem, the exclusion of H\,$\alpha$ from the fit 
is unlikely the origin of this discrepancy. When we exclude H\,$\alpha$ from the
fit, we find an slightly higher \Teffw{103} and that the surface gravity decreases 
by 0.09\,dex. We also note that results from fitting the IDS spectrum 
(\Teffw{92\pm9} and \loggw{7.41\pm0.18}) agree within the error limits 
with our results from the analysis of the X-Shooter spectrum. Thus, also the 
higher resolving power of X-Shooter (that is capable of resolving the NLTE 
emission cores) cannot explain the much lower \Teff\ derived by \cite{Vennes+1999}.
One might speculate, that the outdated line broadening tables used in the analysis 
of \cite{Vennes+1999} could be responsible for the mismatch in \Teff.

\paragraph{WDJ062145.32$+$065239.25}
This is the hottest DA WD in our sample and we find \Teffw{108\pm2} 
and \loggw{7.63\pm0.05}. Just like WDJ061906.92-082807.15 its \Teff\ is 
in excess of 100\,kK and the star displays only a very mild version
of the Balmer line problem, and which only slightly becomes noticeable 
with the higher resolution and S/N of the X-Shooter spectrum. 
In the fit to the low-resolution IDS spectrum, we derived a slightly 
lower temperature (\Teffw{91\pm6}), 
but the surface gravity (\loggw{7.47\pm0.18}) agrees within the error limits 
with the fits from the X-Shooter spectrum. 
We note that the TESS light curve of this star displays photometric variability at 1.91\,h.
The ZTF light curves of this star contain only 66 and 47 data points in the g and r band, respectively and no periodic signal can be found. 
We obtained 3\,h of optical high speed photometry using the 0.6\,m 
telescope at Mt. Cuba Astronomical Observatory. The optical data were 
reduced as outlined in \cite{Provencal+2012}.
From this photometry, we concluded that the variability does not originate from the WD but instead comes from a nearby $\delta$\, Scuti star. In addition, the  \emph{TESS-localize} tool also rules out that the WD is the source of the variability.  

\paragraph{WDJ080029.42$-$015039.82} 
The TESS light curve of this 76.2\,kK hot DA indicates a period of 4.92\,h and 
the crowdsap value is 0.53, indicating that at least half of the 
flux comes from the WD. Yet, \cite{Heinze+2018} reports a nearby 
(1 arcmin away) star, \object{ATO\,J120.1220-01.8600}, which they 
classify as contact or near-contact eclipsing binary with a period 
that is exactly twice the period  we found. Thus, we conclude 
that the variability comes from the nearby star -- and not the WD.

\paragraph{WDJ080558.34$+$410931.84}
We derived \Teffw{50.6\pm0.2} and \loggw{8.11\pm0.02} for this DA WD and 
find that it shows the Balmer line problem. Fitting H\,$\zeta-\beta$
\cite{Gianninas+2011} derived a slightly higher temperature and lower gravity 
(\Teffw{53.2\pm1} and \loggw{7.68\pm0.06}). This is likely because they did 
not have H\,$\alpha$ available in the fit. When we exclude H\,$\alpha$
from our fit, we also end up with a slightly higher \Teff\ and lower
\logg.\\
This star is photmetrically variable with a period of $0.53$\,d. The ZTF 
g- and r-band, as well as the TESS light curves, show two uneven minima. 
Interestingly, the ZTF g band light curve shows two uneven maxima, while 
the maxima in the ZTF r and TESS band seem similar. We note that the 
light curves resemble that of the cool (12\,kK) and apparently non-magnetic 
($B<70$\,kG) WD \object{SDSS\,J152934.98+292801.9}. \cite{Kilic+2015} 
concluded for the latter that it must have a dark spot. In addition,
the TESS light curve of the potential young post-merger hot subdwarf, 
\object{TIC\,220490049}, displays a similar shape and \cite{Vos+2021}
concluded, that the variability is explained by a spot on the surface of 
the star.

\paragraph{WDJ142557.80$-$034139.92 and WDJ165053.99+774844.88} are the 
coolest stars in our sample and both appear to be low-mass -- and 
potentially He-core -- WDs based on the derived gravity, HRD, and Kiel masses.
We speculate that these stars are the outcome of binary evolution.
WDJ142557.80$-$034139.92 is variable in TESS, however, there is 
a 7\,mag brighter star, \object{Gaia DR3 3642987175954122880,} nearby (1\,arcmin) 
and the crowdsap parameter is only 0.05. We also note that \emph{TESS-localize} predicts that 
the variability comes form \object{Gaia DR3 3642987175954122880} and not the WD.
The periodograms of the TESS light curve of WDJ165053.99+774844.88 show several 
significant peaks at $P>1$\,d. However, \emph{TESS-localize} suggest that these
originate from neighbouring stars. This is also in line with the ZTF data, in
which we cannot find a significant variability.

\paragraph{WDJ160152.16$+$380455.24}
\cite{Perez-Fernandez+2016} reported that it is a sdB candidate that 
shows an IR excess based on a poor quality \emph{2MASS} H band magnitude. 
We neither confirm the sdB nature of the star nor the IR excess. The \emph{2MASS} H and K band magnitudes are merely upper limits and we do not see an excess in the \emph{WISE} W1 or W2 magnitudes. We find \Teffw{83\pm7} and \loggw{7.51\pm0.27}, which clearly confirms the WD nature of this star.

\paragraph{WDJ182849.94$+$343649.94}
We derived \Teffw{57.8} and \loggw{7.38} for this DA WD. 
The periodograms of the TESS light curves shows a multitude of significant peaks at 
$P>0.6$\,d. The crowdsap value is only 0.062 and \emph{TESS-localize} suggests 
that one of these periods at 2.96\,d originates from the WD with a probability of
99.8\%. However, we cannot confirm this period based on the ZTF light curves 
which have 961, 1190, and 82 data points in the g, r, and i bands, respectively. 
Thus, we classify this star as possibly variable only.

\paragraph{WDJ220247.69$+$275010.67}
Fitting the MODS spectra of this pure DA WD, we derived 
\Teffw{58.9\pm0.5} and \loggw{8.25\pm0.17}. The star shows 
the Balmer line problem and the gravity, HRD, and Kiel 
masses disagree with each other (see Table~\ref{table:DA}). 
Furthermore, we found this star to be
periodically variable in the ZTF g- and r-band with a period
of 0.82\,d. The light curves look approximately sinusoidal, 
however, the ZTF r-band light curve possibly shows an extended minimum (see 
Fig.~\ref{fig:lc}). There is no significant difference in the 
amplitudes in both bands and there are no hints of a possible
companion in the spectrum. 

\paragraph{WDJ222147.70+501150.75} This DA WD is the second 
faintest ($G=17.01$\,mag) star in our sample. A fit to the MODS 
spectra suggests \Teffw{57.5\pm0.5} and \loggw{7.95\pm0.04}. The 
Balmer line problem is apparent in this WD, but within the error limits
the gravity, HRD, and Kiel masses agree with each other.
The ZTF g- and r-band light curves of this DA WD indicate a period 
of 2.3\,d. The amplitudes in both bands are the same and we do not 
see any hints of emission lines arising from the irradiated side of 
a hypothetical secondary. 

\begin{figure}[ht]
\centering
\includegraphics[width=\hsize]{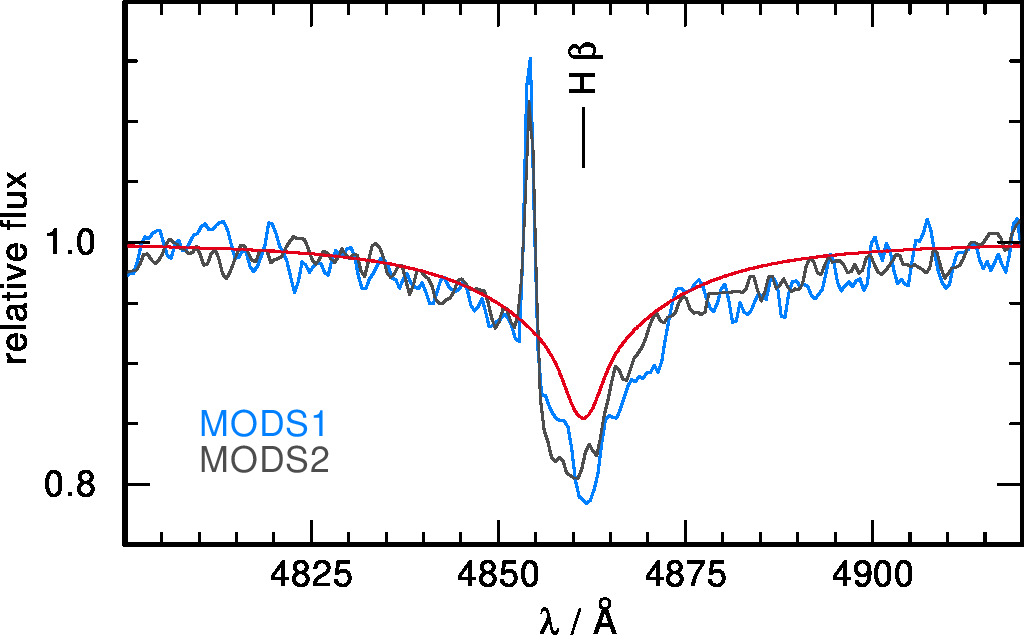}
\caption{LBT/MODS1 (blue) and MODS2 (gray) spectra of the DAe: WD WDJ064527.28$+$565916.90 taken simultaneously  and overplotted with the best fit TMAP model (red). The origin of the emission line located on the blue wing of H\,$\beta$ remains unclear. No emission lines are visible at the other Balmer line members.}
\label{fig:dae2}
\end{figure}

\subsection{DAe WDs}

\paragraph{WDJ064527.28$+$565916.90}
We classify this star as DAe: WD. Fitting the MODS spectra with 
pure H TMAP models, we found \Teffw{99.7\pm3.7} and \loggw{6.72\pm0.03}. 
Moreover, we uncovered a strong Balmer line Problem, which 
could be related to a cool companion which adds 
significant flux to the spectrum. 
The star shows no emission in H$\alpha$, however, both MODS 
spectra show a clear emission line located on the blue wing 
of H$\beta$ (see \fg{fig:dae2}). The relative RV shift of this 
emission line and the H$\beta$ absorption line from the WD is 
estimated to be 450\,km/s.\\
The ZTF g- and r-band light curves indicate a photometric 
variability of 1.69\,d and the light curve looks sinusoidal, 
with no significant difference in the amplitudes in both 
bands. The star also has TESS 2\,min and 20\,s cadence light curves, predicting a period of 2.54\,h. However, we believe that 
this signal comes from \object{Gaia\,DR3\,1001088333316304512,} which 
is located only 7\,arcsec away. This star is reported to be a pulsating 
variable in the \emph{Gaia} DR3 Part 4 Variability catalog with the same
period found by us in the \emph{TESS} data. This is also supported
by \emph{TESS-localize}.\\
Assuming the emission line stems from an irradiated 
companion and the 1.69\,d periodicity found in the ZTF lightcurves
reflects the orbital period, 
then the relative RV shift would be unrealistically high.
It is also unlikely that the emission line is caused 
by a cosmic, since it is visible in both (simultaneously taken)
MODS spectra. Furthermore, we note that immediately after the 
spectra of WDJ064527.28$+$565916.90 were taken, another
four WDs in our sample have been observed within the same 
night (Table~\ref{table:A1}). 
None those spectra show this emission line blue-ward of H$\beta$, 
which rules out a data reduction artefact or a sky line. 

\paragraph{WDJ075145.59$+$105931.36}
is a DAe WD. The whole Balmer line series as well as \Ionw{He}{1}{5876} 
is seen in emission (see \fg{fig:dae}). The relative RV shift of the 
WD absorption lines and the emission lines is estimated to be 200\,km/s. 
Based on ZTF light curves we find a period of 6.64\,h. The amplitude in the
ZTF g band is found to be 0.05\,mag, much less than in the ZTF r band 
(0.13\,mag), indicating a reflection effect. 
Since the companion is expected to add significant flux 
in the optical spectrum, we refrain from a formal fit.

\begin{figure*}[ht]
 \centering  
 \includegraphics[width=\textwidth]{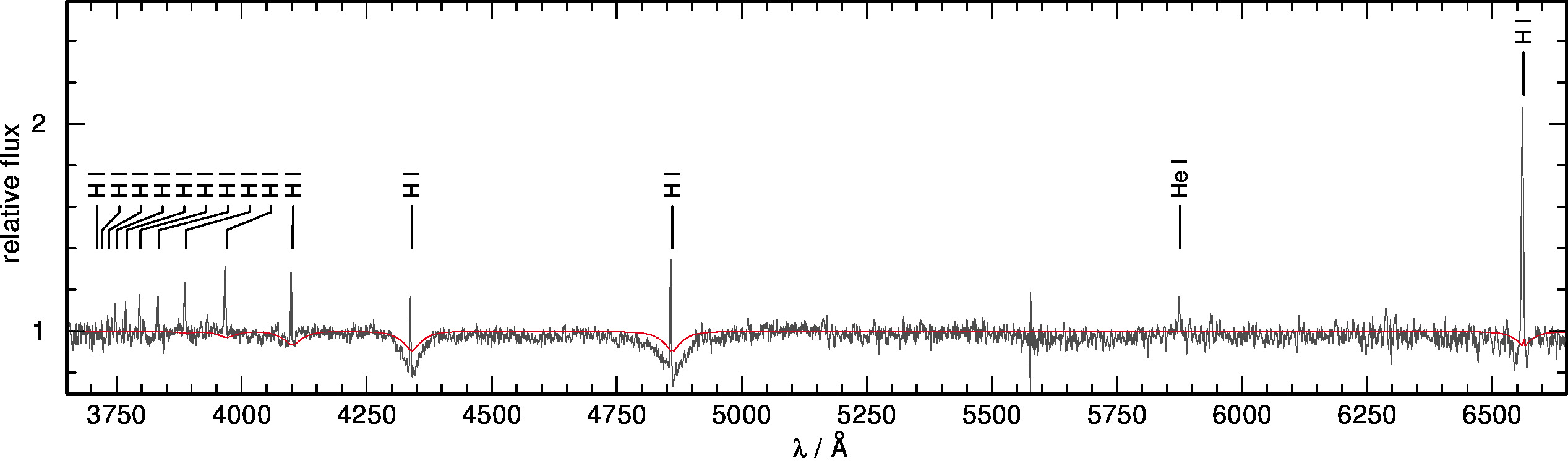}
  \caption{LBT/MODS spectrum of the newly discovered DAe WD WDJ075145.59$+$105931.36. The whole Balmer series as well as He\,{\sc i} 5876\,\AA\ is seen in emission.  A pure H TMAP model (red) is overplotted with \Teffw{100} and \loggw{8.00} (note:\ this is not a formal fit but shown for illustration only).}
\label{fig:dae}
\end{figure*}

\begin{figure}[ht]
\centering
\includegraphics[width=\hsize]{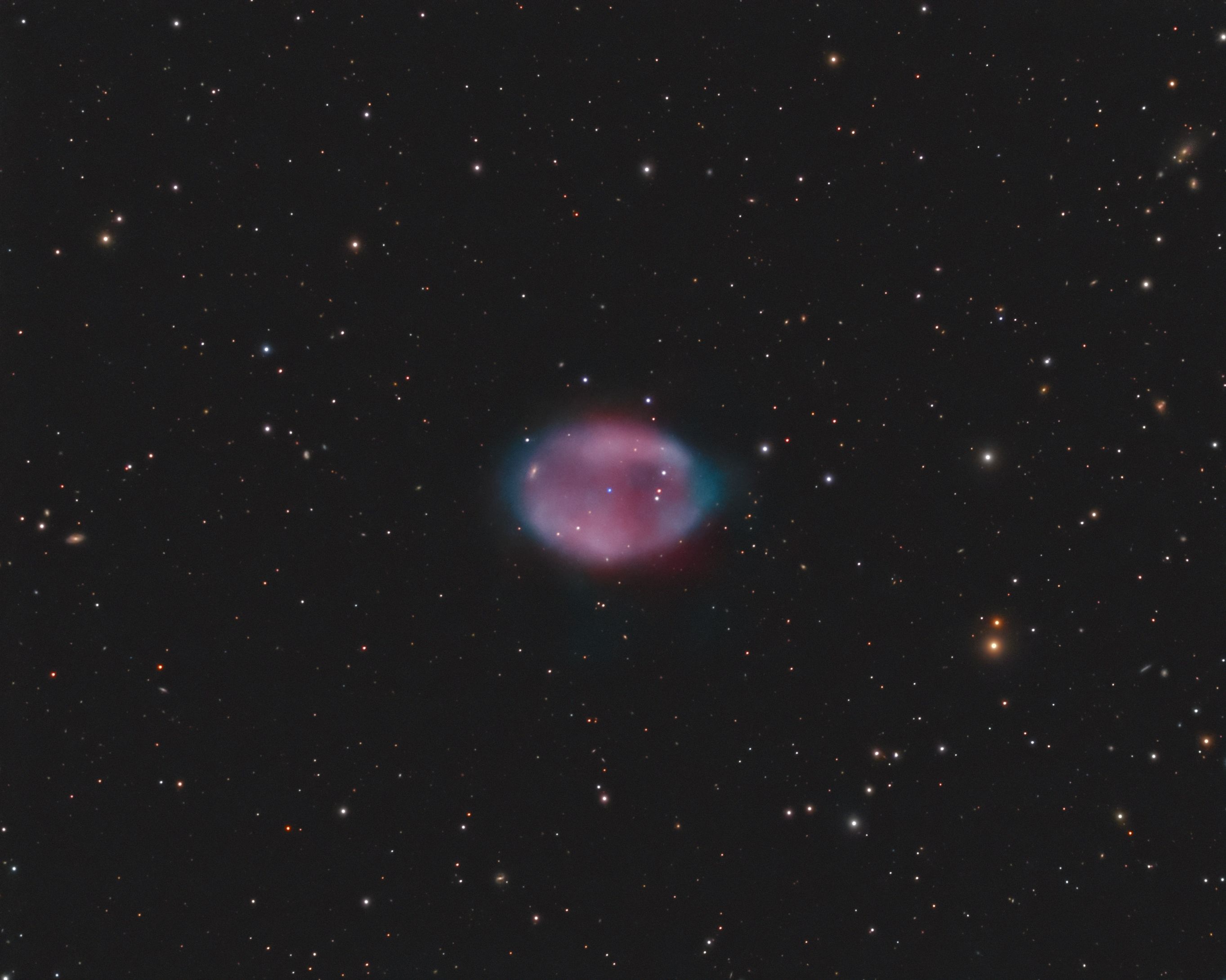}
\caption{Composite R, G, B, [O\,{\sc iii}], and H\,$\alpha$ image of the PN\,G136.7+61.9. The blue star at the center of the PN is the DAO WDJ120728.43$+$540129.16. The size of the image is 2879x2302\,arsec. Image credit: Boris Chausov.}
\label{fig:pn}
\end{figure}

\subsection{Central stars of (possible) PN}
\label{sect:CSPN}

\paragraph{WDJ120728.43$+$540129.16}
is the DAO type central star of the planetary nebula \object{PN\,G136.7+61.9}.
It was first listed as candidate PN by \cite{Yuan+2013}, and its true PN 
nature has been confirmed in 2021 by deep amateur imagery \citep{LeDu+2022}.
In \fg{fig:pn}, we show a R, G, B, [O\,{\sc iii}], and H\,$\alpha$ composite 
image of the PN as obtained by the amateur astronomer 
Boris Chausov using a Starlight XPress Trius-SX694 CCD camera mounted on 
a Celestron Schmidt-Cassegrain telescope. The total integration time of 
the image is 32.5\,h. 
Assuming a distance of 1173\,pc from \cite{Bailer-Jones+2021} as well as  
an expansion velocity of 20\,km/s, we find a radius of 1.2\,pc and an
age of 58\,kyrs.\\
Based on the INT/IDS spectrum, we derived, for the first time, the atmospheric 
parameters of the CS and find \Teffw{84.9\pm11.2} and \loggw{6.54\pm0.23}. 
From the SED fit, (see Fig.~\ref{fig:sed}, left) we derived a radius of 
0.0450\,\Rsol, and a luminosity of $90^{+70}_{-50}$\,\Lsol. The gravity 
($0.26^{+0.20}_{-0.12}$\,\Msol), HRD ($0.49\pm0.09$\,\Msol), and Kiel 
($0.44\pm0.07$\,\Msol) masses all suggests that this star could be a  
candidate for post-RGB CSPNe. Adopting the uncertainty on \Teff, we find 
that the 0.452\,\Msol\ track from \cite{Hall+2013} suggests a post-RGB age of 
59--63\,kyrs while the 0.532\,\Msol\ post-AGB track for $Z=0.01$ of 
\cite{MillerBertolami2016} predicts an post-AGB age of 78-431\,kyrs. 
Since PNe are visible for typically only a few 10\,kyrs, this 
implies that it should actually be more likely to detect a PN around a 
high-mass post-RGB star, than a low-mass post-AGB star.
Unfortunately, we did not detect any significant light curve variations 
in the TESS or ZTF light curves that could indicate a close companion.

\paragraph{WDJ182440.85$-$031959.52}
This DA WD is the CS of the very faint and large (the diameter is 
1900\,arcsec, \citealt{Frew+2016}) planetary nebula \object{PNG\,026.9$+$04.4}. 
Using the distance of 195\,pc from \cite{Bailer-Jones+2021} and assuming 
an expansion velocity of 20\,km/s, a radius of 0.9\,pc and a kinematic 
age of 44\,kyrs can be estimated. Fitting the IDS spectrum with pure 
H models, we derived \Teffw{68.9\pm0.6} and \loggw{7.63\pm0.03} and 
by fitting the SED of the star, we found a radius of 
0.0168\,\Rsol, and a luminosity of $5.74$\,\Lsol. 
The HRD ($0.63\pm0.05$\,\Msol) and Kiel ($0.58\pm0.04$\,\Msol) 
masses agree with each other, but the gravity mass ($0.44\pm0.04$\,\Msol) 
is significantly lower than the former two. We also note that the post-AGB 
age in the Kiel diagram and HRD ($\approx300-400$\,kyrs) predicted by 
the \cite{MillerBertolami2016} tracks is one order of magnitude higher 
than what is expected from the kinematic age. 

\paragraph{WDJ191231.47$-$033131.86}
could be the central star of the possible PN \object{FP\,J1912$-$0331}. 
We derived \Teffw{53.3\pm1.3} and \loggw{7.63\pm0.10}. The gravity 
($0.57^{+0.14}_{-0.12}$\,\Msol), HRD ($0.54\pm0.04$\,\Msol), and Kiel 
($0.55\pm0.05$\,\Msol) masses agree with each other. From the 
 \cite{MillerBertolami2016} tracks, we estimate a post-AGB age of 
 $1.2-1.5$\,Myrs, which makes it seem very unlikely that a PNe 
 should still be visible around such a relatively cool WD.

\subsection{DAO and DOA WDs}
\label{sect:hybrid}

\begin{figure}
\centering
\includegraphics[width=\hsize]{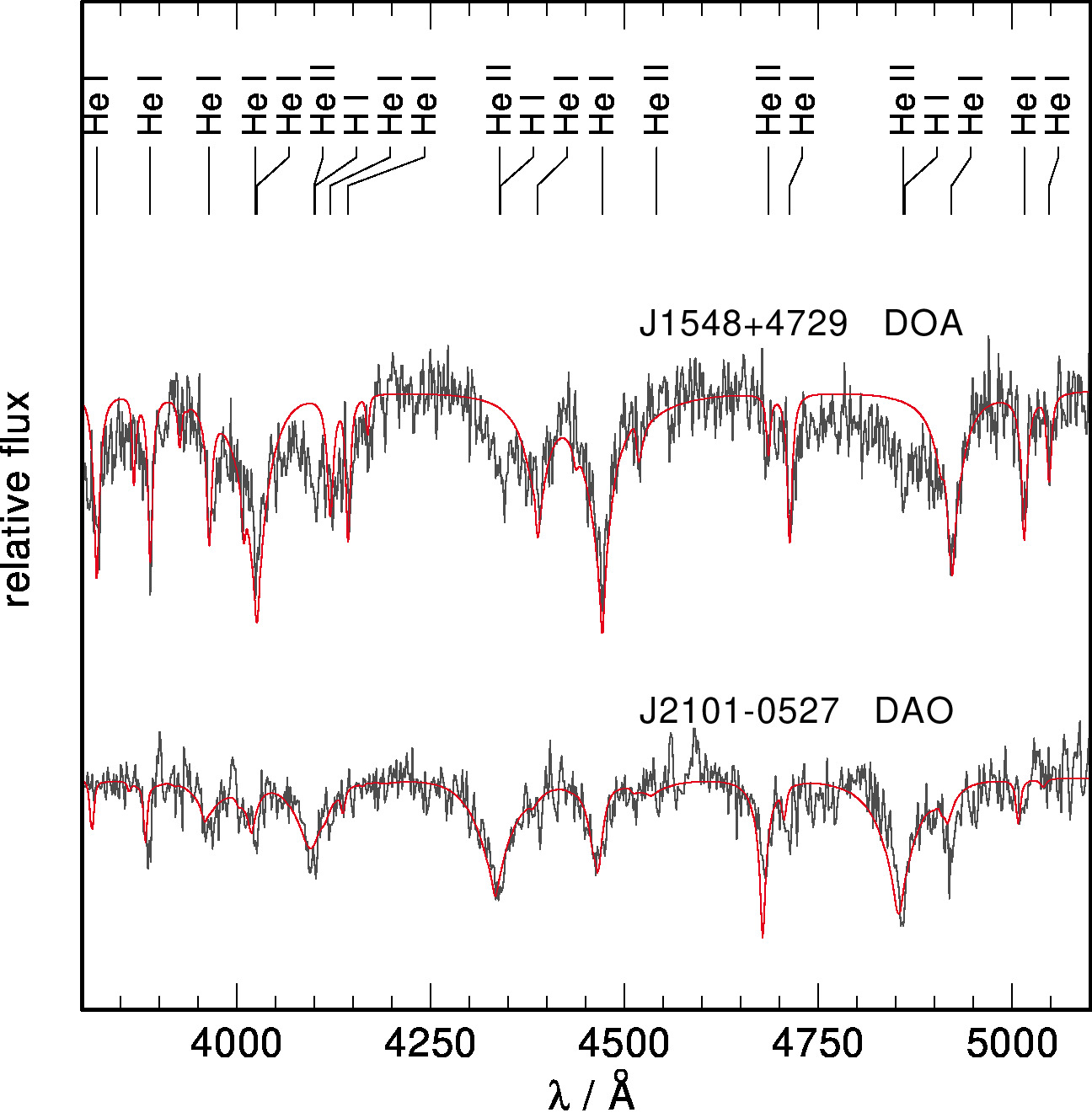}
\caption{INT/IDS spectra (gray) of the DOA WD WDJ154843.31+472936.11 (top, overplotted 
with a pure He model with \Teffw{40} and \loggw{7.5}) and the DAO WD WDJ210110.17$-$052751.14
(bottom, overplotted with a H+He model with \Teffw{48.9}, \loggw{8.14}, and $\log$\,(He/H) $=-0.58$.}
\label{fig:daob}
\end{figure}

\paragraph{WDJ070322.15$-$340331.94 and WDJ080950.28$-$364740.45}
are the hottest DAO WDs in our sample and can be considered as twins
not only due to their very similar atmospheric parameters, also the Gaia
magnitudes and parallaxes are almost the same. For WDJ070322.15$-$340331.94
we find  \Teffw{105.8}, \loggw{6.77}, and $\log(\mathrm{He/H})=-1.14$ (by number), and
for WDJ080950.28$-$364740.45 we derived \Teffw{103.6}, \loggw{6.68}, and 
$\log(\mathrm{He/H})=-1.18$. Their Kiel (0.53 and 0.52\,\Msol, respectively) and HRD 
(0.58 and 0.56\,\Msol) masses agree within the error limits, suggesting 
that they are both post-AGB stars.

\paragraph{WDJ080326.15$-$034746.11}
From the X-Shooter spectrum, we derived \Teffw{86.9\pm4.6}, \loggw{6.86\pm0.12},
and $\log(\mathrm{He/H})=-1.63$. 
From the INT spectrum, we derived a somewhat lower \Teff of $69\pm4.2$\,kK and 
higher \loggw{7.19\pm0.19}.
Like several of our DAO WDs also this 
star shows the Balmer line problem (see Fig.~\ref{fig:dao}). The HRD 
($0.56\pm0.06$\,\Msol) and Kiel ($0.50\pm0.04$\,\Msol) masses agree with 
each other and suggest that the star is a low-mass post-AGB or post-EHB star. 
However, the gravity mass ($0.19^{+0.07}_{-0.05}$\,\Msol) is conspicuously 
lower than the former two. 
The TESS light curve of this DAO WD indicates a period of 5.6\,h and the 
crowdsap value is 0.17. Just as in the case of WDJ080029.42$-$015039.82, 
\cite{Heinze+2018} reported a nearby (0.8\,arcmin) contact or near-contact 
eclipsing binary system with a period that is exactly twice the period we found. Finally, also \emph{TESS-localize} confirms that the variability
does not come from the WD. 

\paragraph{WDJ154843.31$+$472936.11}
Comparing a pure He model with \Teffw{40} and \loggw{7.5,} we found 
that the \ion{He}{I} lines as well as \Ionw{He}{2}{4686} 
line are well reproduced. However, the observed Balmer lines are 
stronger than predicted by our model, suggesting that there is 
some H in the atmosphere of this star (see \fg{fig:daob}).
Thus, we classify it as DOA WD. Since the star lies at the edge
of our grid and most likely has a stratified atmosphere as its 
\Teff\ is below 45\,kK \citep{Bedard+2020}, we refrained from carrying out a formal 
fit.\\
This star was also observed several time with TESS. The TESS light curve 
indicates a period of 0.19\,d and the crowdsap value is 0.52. However, 
we cannot confirm this periodicity from ZTF g-, r-, and i-band photometry.
The observed period is typical for $\delta$\,Scuti stars, and also 
\emph{TESS-localize} confirms that the variability does not come from the WD.

\paragraph{WDJ210110.17$-$052751.14}
This DAO WD shows besides strong Balmer lines also prominent lines of
\ion{He}{ii} and \ion{He}{i}, indicating it is one of the rare cooler 
hybrid WDs. We derived  \Teffw{48.9}, \loggw{8.14}, and $\log$\,(He/H) $=-0.58$.
As can be seen in \fg{fig:daob}, most lines are well reproduced 
with our model, however, \Ionw{He}{2}{4686} is too strong 
compared to the observation. This is a tell-tale sign that this object 
is likely to have a stratified atmosphere \citep{Manseau+2016, Bedard+2020}. 
On the other hand, we do not see a problem with \Ionw{He}{1}{4471}, which is 
 broader and shallower in stratified models. The analysis should, therefore, 
be repeated using stratified models as well, in order to test if a better fit 
can be achieved in this way. Accordingly, we note that these 
parameters should be treated with caution, because neglecting a stratified atmosphere can 
lead to large systematic errors. For example, for the hybrid WD 
\object{SDSS\,J003343.05+142251.4} \cite{Bedard+2020} found \Teffw{49.2} 
and \loggw{8.26} with homogeneous model atmospheres. When assuming a 
stratified atmosphere, the derived values dropped down
significantly to \Teffw{40.6} and \loggw{7.77}.\\
The 2\,min and 20\,s cadence TESS light curves of WDJ210110.17$-$052751.14 
indicate a photometric variability with a period of 12.91\,h. The 
crowdsap value is only 0.38. Yet, based on ATLAS c- and o-band 
light curves, \cite{Heinze+2018} reports exactly twice the period that we 
found in the TESS data, and \emph{TESS-localize} predicts a  likelihood of 
1.0 that the variability originates from the WD. Interestingly, when 
folding the TESS light curve to twice 
the period found in the LS search (1.08\,d, i.e., the period reported 
by \citealt{Heinze+2018}), we found that the depth of the minima differ 
slightly. In the 2\,min and 20\,s cadence TESS light curve the 
difference in the depth of the minima is 0.2\% and 0.5\%, respectively.
Since the star is already very compact (0.0128\,\Rsol), relatively 
massive (HRD, Kiel, and gravity masses all suggest $M>0.72$\,\Msol), 
and with a relatively long period, it seems unlikely that the 
difference in the depth of the minima could by caused by ellipsoidal 
modulation.

\paragraph{WDJ210824.97$+$275049.80}
We derived for this DAO WD \Teffw{80.1}, \loggw{7.00}, and $\log(\mathrm{He/H})=-2.01$. 
The HRD ($0.42\pm0.07$\,\Msol) and Kiel ($0.50\pm0.06$\,\Msol) masses agree 
and make it seem likely that it is a post-EHB star. 
The gravity mass of this star is again very low ($0.25^{+0.22}_{-0.12}$\,\Msol), 
but agrees within the error limits with the Kiel and HRD masses.

\subsection{O(H) and sdO stars}

\paragraph{GD\,1323}
We derived \Teffw{112.7}, \loggw{5.89}, and $\log(\mathrm{He/H})=-1.00$. The Kiel and HRD masses agree and suggest that GD\,1323 
is a low-mass post-AGB star (Fig.~\ref{fig:hrd}), therefore, we considered 
it as a rare naked (i.e., no PN visible) O(H) star. 
However, the gravity mass of this star seems again too low. Using 
the \cite{MillerBertolami2016} tracks, we estimated a post-AGB age of 
27\,kyr and 61\,kyr from the Kiel diagram and the HRD, respectively. 
This could be enough time for a former PN to fade into the interstellar
medium.

\paragraph{WDJ132259.60-383813.13 and WDJ223522.88-494350.80}
The Kiel and HRD masses of both stars are smaller than 0.50\,\Msol, suggesting that they are either post-RGB or post-EHB stars. 
We classify these two stars as sdO stars based on their position in 
the Kiel diagram. When looking into the HRD, however, it could be
the case that both stars have recently entered the WD cooling stage and 
consequently could already be WDs.
WDJ132259.60-383813.13 shows a particularly strong version of the BP. 
Automated fitting attempts showed that the parameters lie at the edge 
of our DAO grid in terms of \logg, and out of the \cite{Reindl+2016} grid
in terms of He abundances. We estimate \Teff$=80\pm10$\,K, 
\loggw{6.00\pm0.50}, and $X_{\mathrm{He}} = -2.00\pm0.50$. 
Based on the sub-solar He abundance one could speculate, that gravitational
settling of He has already started. 
For WDJ223522.88-494350.80 we find \Teffw{75.0\pm2.2}, \loggw{6.06\pm0.04,}
and that the He abundance is slightly enhanced with respect to the solar value.
Also, this star shows the Balmer line problem and the observed 
\Ionw{He}{2}{4686} shows a central emission, which is not reproduced 
by our best-fit model (Fig.~\ref{fig:dao}).

\begin{figure*}[t]
 \centering  
 \includegraphics[width=\textwidth]{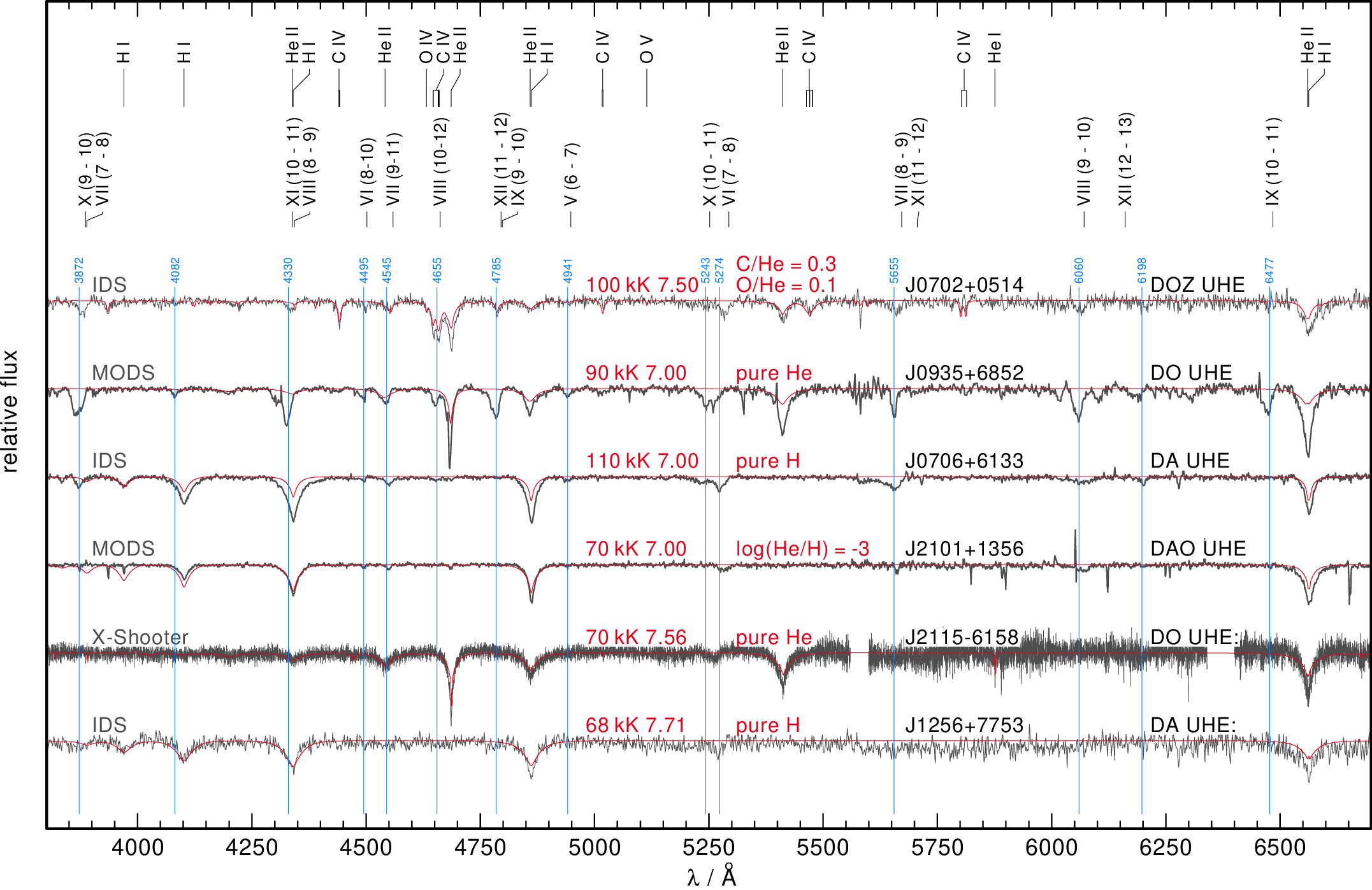}
  \caption{Spectra of the newly discovered UHE WDs. The positions of photospheric
  lines (H\,{\sc i}, He\,{\sc i}, He\,{\sc ii,} and C\,{\sc iv}), $\alpha$ and
  $\beta$ transitions between Rydberg states ($n - n'$) of the ionization
  stages {\sc v}$-${\sc x}, and approximate line positions of the UHE features
  (blue) are marked. Over plotted in red are TMAP models and the effective
  temperatures, surface gravities, and chemical compositions (in number
  ratios) as determined in this work. The spectrograph used for the observation is indicated in gray.}
\label{fig:uhe}
\end{figure*}

\subsection{(Possible) UHE WDs}
\label{sect:UHE}

\paragraph{WDJ070204.29$+$051420.56}
is the second brightest UHE WD known thus far. 
Based on the IDS spectrum of this star, we  estimate \Teffw{100}, \loggw{7.50}, 
and $\mathrm{C/He}=0.3$ (number ratio). It is worth noticing that it is the first UHE 
WD that clearly shows lines of \Ionw{O}{4}{4632} and \Ionw{O}{5}{5114} and 
we estimated $\mathrm{O/He} = 0.1$ (number ratio). 
The TESS and ZTF light curves indicate a period of 0.60\,d and the 
amplitudes in ZTF g-, r-, and i-band light curves do not differ significantly. 
The shapes of the light curves are typical for what is observed for about one 
third of the UHE WDs, namely an extended minimum with a possibly second maximum 
\citep{Reindl+2021}. 

\paragraph{WDJ070647.52$+$613350.31}
is a rare DA-type UHE WD and one of the highlight discoveries of this survey. With 
$G=13.49$\,mag it is about two orders of magnitude brighter than all the other known 
UHE WDs. The star displays a strong Balmer line problem. H\,$\gamma$, H\,$\beta$ 
and H\,$\alpha$ are best reproduced with \Teffw{40} and \loggw{7.00}, while the 
higher order Balmer line series members suggest \Teffw{110} and \loggw{7.00}.
Light curves obtained by ZTF and TESS reveal a photometric period of 8.97\,h, 
an amplitude of 0.04\,mag, and -- again -- an extended minimum. Interestingly, the 
ascents and descents to and from the photometric maximum are clearly asymmetrical.

\paragraph{WDJ093559.83$+$685201.55}
We discovered this DO UHE WD at the LBT. The UHE lines are particularly strong
in this object and we did not detect any lines of \ion{He}{i} nor the 
\Ionww{C}{4}{5803, 5814} doublet. Bas on this,   we estimated a lower limit on \Teff\
of 90\,kK, \loggw{7.00}, and an upper limit on the C abundance of 
$\mathrm{C/He} < 0.002$ (by number).
The star was observed within ZTF and TESS, and the data from both surveys 
suggest a period of 1.34\,d. 
The shapes of the light curves resemble that of the DO UHE WD HS\,0158+2335.
The light curves of this latter object show two maxima, with the first 
one being at phase 0.0, and the second at approximately phase 0.6, and the
minimum is located around phase 0.3 \citep{Reindl+2021}. 

\paragraph{WDJ125627.43$+$775301.37}
We classified this star as DA UHE: based on the IDS spectrum, which
shows possibly a UHE line near 5274\,\AA. Fitting the spectrum with 
H models, we found \Teffw{68.1} and \loggw{7.71} and that object shows 
the Balmer line problem. Based on TESS and ZTF g-, r-, and 
i-data, we found the star is variable with a period of 3.49\,h. The shape
of the light curves appears to vary from band to band. In the g-band 
the light curve looks asymmetrical, while in the i-band it looks 
more symmetrical with an extended minimum. 

\paragraph{WDJ210132.69+135622.59}
Based on the MODS spectrum, we classified this star as DAO UHE. Just as 
WDJ070647.52$+$613350.31, it displays a strong Balmer line problem.
The UHE features are particularly weak in this star (as in the case 
of, e.g.,  DAO UHE WD \object{HS\,2115+1148} or DO UHE WD 
\object{HS\,0158+2335}) and, thus, could have easily been missed in a 
lower S/N spectrum (LBT/MODS provided us with a S/N of 182). 
The star is variable in ZTF and TESS with a period of 1.22\,d.
The ZTF g- and r-band light curve exhibit extended minima.

\paragraph{WDJ211532.62$-$615849.50}
was first observed with SOAR and later with X-Shooter. The star 
shows the \ion{He}{ii} line problem and possibly a UHE line 
near 5274\,\AA, thus, we classified it a DO UHE: WD.
The 2\,min and 20\,s TESS cadence light curves indicate two significant periods at
7.45\,h and at $2\times 7.45 = 14.90$\,h
The crowdsap value is with 0.70 relatively high and \emph{TESS-localize}
predicts a likelihood of 1.0 that the WD is indeed the source of observed 
variability. 
When folding the light curves to the 7.45\,h period, the shapes look 
approximately sinusoidal, with a possible extended minimum. Yet, when 
folding the light curves to the 14.90\,h period, the light curve is
clearly asymmetric with two uneven maxima and minima. Thus, we conclude
that the 14.90\,h period is likely the rotational period of the star.

\subsection{DO(Z) WDs}
\label{sect:DO}

\paragraph{WDJ014343.14$+$584151.39 and WDJ191750.16$-$201409.55}
are the hottest DO WDs in our sample with effective temperatures in excess
of 110\,kK. We were not able to detect lines of C or any 
other metal in either of them. According to calculations of \cite{UnglaubBues2000}, who 
studied the chemical evolution of hot WDs in the presence of diffusion
and mass loss, both stars should still display at least half of their 
original C abundances (meaning the C abundance with which they entered
the WD cooling sequence and before it got affected by gravitational 
settling). Based on this, we can speculate that the H-deficiency in these two 
DO WDs was not cause by a (very) late thermal pulse, during which large 
amounts of C are brought to the surface. Instead, these two DO WDs might 
be the outcome of a double He-WD merger, which predicts only about 1\% of atmospheric carbon (by mass). Therefore, they could be the slightly 
more massive analogs to the DO WDs \object{PG\,1034+001} and \object{PG\,0038+199} 
\citep{Zhang+2012, Reindl+2014b, Werner+2017}.\\
We note, however, that the gravity mass of WDJ191750.16$-$201409.55 is 
unrealistically high ($3.40^{+1.00}_{-0.80}$\Msol), pointing towards 
larger systematic uncertainties on the derived parameters. 
The S/N of the spectrum of WDJ191750.16$-$201409.55 is 
only 35, thus it is difficult to judge if the star belongs to the DOs 
showing the \ion{He}{ii} line problem, which could be responsible for 
the unrealistically high gravity mass.

\paragraph{WDJ060244.99$-$135103.57}
The analysis of the INT spectrum of this pulsating PG1159 (GW Vir)
star was presented in \cite{Uzundag+2021}. The authors derived \Teffw{120\pm10}, 
\loggw{7.5\pm0.5}, He $=0.75^{+0.12}_{-0.08}$, and C $=0.25^{+0.08}_{-0.12}$ 
(number fractions). After analyzing the newly obtained X-shooter spectrum of this star, 
we confirmed \loggw{7.5\pm0.5}, but with a lower \Teffw{100\pm10}.
Thanks to higher resolution of the X-Shooter spectrum, we were able to provide better 
constraints on the He and C abundances (He $=0.72\pm0.25$, C $=0.21\pm0.08$),
tderive an upper limit on the H and N abundances (H $<0.25$, N $<0.0005$), and (for 
the first time) to derive an O abundance (O $=0.07\pm0.04$) of this star (all values are number fractions).

\paragraph{WDJ033449.19$+$164629.92}
Based on the MODS spectrum of this DOZ WD, we derived \Teffw{98.3}, 
\loggw{7.87}, and $\mathrm{C/He}=0.02$ (by number). The star displays a mild  
\ion{He}{ii} line problem (the observed \Ionw{He}{2}{6560} line is too broad and deep compared to our best fit model, see Fig.~\ref{fig:do}), but we find that the Kiel, HRD, and gravity masses agree. The ZTF and TESS light curves indicate a period of 9.06\,h. The shapes of the ZTF light curves are clearly asymmetrical and there is no significant difference in the amplitudes in both bands. 

\paragraph{WDJ043619.17$-$065412.61, WDJ075134.71$-$015807.00, and WDJ182432.46+293115.88}
Based on INT/IDS spectra, we found effective temperatures below 60\,kK for these three DO WDs. It is worth noting that their gravity masses do not agree with the Kiel and HRD masses. In addition, the HRD and Kiel masses of these stars are inconsistent with each other. We speculate that it could be possible that these stars already have stratified atmospheres.

\section{Summary and discussion}
\label{sect:summary}
We carried out a spectroscopic survey targeting 71 bright ($G=13.5-17.2$\,mag) 
and hot (pre-)WD candidates from the \emph{Gaia} DR2 and eDR3 catalogs of hot subdwarfs 
\citep{Geier+2019, Culpan+2022} and WDs \citep{GentileFusillo+2019, GentileFusillo+2021}. Our main motivation
was to increase the number of bright and rare object types (e.g., 
short-lived pre-WDs in the post-AGB region, UHE WDs, extremely hot, i.e., \Teff$>60$\,kK, 
WDs, DOZ WDs or photometrically variable WDs) that are suitable for 
detailed follow-up investigation for instance using high-resolution (UV) spectroscopy, 
spectropolarimetry, or time-resolved photometry.\\
Out of the 71 stars in our sample, 68 are new discoveries, while for the 
remaining 3 objects, we obtained spectra with a higher wavelength coverage and/or higher resolution. This allowed us to almost double the number of the hottest
(\Teff$>60$\,kK) WDs that are brighter than 
$G=16$\,mag\footnote{The Montreal WD database lists 63 WDs with 
\Teff$>60$\,kK and \logg$<6.5$ that are brighter than $G=16$\,mag.}.\\
Using the Gaia eDR3 WD catalog from \cite{GentileFusillo+2021}, we found that 
66\% of all WD candidates that are brighter than 16\,mag and that have 
$6<M_{\mathrm{G}}$/mag $<9$ and bp\_rp\,$<-0.4$\,mag have been spectroscopically confirmed 
and analyzed (either because they have been analyzed in this work or because they are 
listed with atmospheric parameters in the Montreal White Dwarf Database). In the 
northern hemisphere (DEC$>0$) this percentage is higher (82\%) than in the southern 
hemisphere (48\%). Thus, a future survey targeting southern hot WD candidates promises 
further interesting discoveries. Moreover, we note that additional hot WDs may be hidden 
at locations of higher reddening or because an unresolved cool companion adds additional 
flux to the optical or because a cooler companion dominates the optical flux completely.
\\
Using NLTE models we derived the atmospheric parameters of our stars, 
and by fitting their spectral energy distribution, we derived their radii, 
luminosities, and gravity masses. In addition, we derived their masses in the 
Kiel and Hertzsprung–Russell diagram. Furthermore, we have searched 
for periodic signals in the ZTF and TESS light curves of our stars.

\subsection{UHE WDs}
Our surveys resulted in the discovery of four, possibly six new UHE WDs (\fg{fig:uhe}). Previously, only 16 UHE WDs were known \citep{Reindl+2021}, thus, we increased the number of known UHE WDs by 20\%. We stress that we also discovered the two brightest UHE WDs known and increased the number of H-rich UHE WDs by 67\%. With five of the now known twenty UHE WDs being H-rich, we derived the fraction of H-rich UHE WDs as $25.0^{+7}_{-12}$\%. As mentioned earlier, the empirically derived upper $M_{\mathrm{G}}$ limit for UHE WDs seems to coincide with that of DOZ and DAO WDs. Assuming that weak stellar winds keep up C and He in the atmospheres of DOZ and DAO WDs, then it also stands to reason that the UHE phenomenon winds up once the stellar wind has abated. Stellar winds could fade earlier for H-rich objects \citep{UnglaubBues2000}, by that the relatively low percentage of H-rich UHE WDs could be understood. Yet, we are still a long way from understanding what could trigger the beginning of the UHE phenomenon.\\

\subsection{Pure H WDs}
Slightly more than a half of the stars in our sample are pure DA WDs and they are found to have effective temperatures ranging from 29\,kK to 108\,kK. The two coolest DAs in our sample both appear to be low-mass and could be either He-core WDs or post-EHB stars. As a result, they are  likely to be the outcome of binary evolution, although we did not find evidence for photometric variability in these two stars. Four of the pure DA WDs were found to be variable, with none of them displaying a hint for a close companion based on the current data. Furthermore, we found one DAe and one DAe: WD, which are both photometrically variable. The latter only shows an emission line located on the blue wind of H\,$\beta$, and the origin of this line remains unclear. The DAe WD is certainly a close binary system composed of a low-mass companion that is strongly irradiated by the hot DA WD.\\ 

\subsection{H- and He-rich (pre-)WDs}
Our sample includes 16 objects that show both H and He in their spectra. One of them is a rare, naked O(H) star, GD\,1323, that is located in the post-AGB region but does not show evidence for a PN. Only 11 of these naked O(H) stars are known \citep{Heber1986, HeberHunger1987, Chayer+2015, BauerHusfeld1995, Fontaine2008, Reindl+2016, Moehler+2019, Jeffery+2023}.
In the case of GD\,1323, we concluded that its post-AGB age is probably long enough
for a former PN to fade into the interstellar medium.
The Kiel and HRD masses of the two sdO stars in our sample are smaller than 0.50\,\Msol, suggesting that they are either post-RGB or post-EHB stars.\\
As can be assumed based on \fg{fig:hrd}, the mean Kiel mass of our DAO WDs ($\langle M_{\mathrm{Kiel}} \rangle=0.52$\,\Msol, $\sigma = 0.01$\,\Msol) is lower than the mean Kiel mass of the DA WDs ($\langle M_{\mathrm{Kiel}} \rangle=0.60$\,\Msol, $\sigma = 0.01$\,\Msol). This difference is lower but still noticeable in the HRD where we find $\langle M_{\mathrm{HRD}} \rangle=0.62$\,\Msol ($\sigma = 0.01$\,\Msol) for the DA WDs and $\langle M_{\mathrm{HRD}} \rangle=0.57$\,\Msol ($\sigma = 0.01$\,\Msol) for the DAO WDs. It is possible to speculate that this may be related to the striking paucity of H-rich WDs relative to their H-deficient counterparts at the very hot end of the WD cooling sequence \citep{Werner+2019}. This is because more massive DAO WDs should have \Teff in excess of 100\,kK.\\
Several of our DAO WDs are found to have surface gravities of above 7.10\,dex, Kiel masses below $\leq0.55$\,\Msol, and HRD masses $\leq0.59$\,\Msol. The only exception is WDJ210110.17$-$052751.14 were all three mass determination methods suggest $M\geq0.72$\,\Msol. It is also the only DAO WD that lies within the error limits past the thick purple line in Fig.~\ref{fig:hrd}, which indicates where the He abundances should have decreased down to $\log(\mathrm{He/H})=-3$ according to predictions of \cite{UnglaubBues2000}. The star, however, shows the highest He abundance of all DAO WDs in our sample, and -- as mentioned earlier -- it shows signs of a stratified atmosphere. Thus, we suggest it is a former DO WD that is currently transforming into a DA WD.\\
Another possible candidate of a transforming WD is the DOA WDJ154843.31$+$472936.11. It has a \Teff\ around 40\,kK and lies at the edge of our grid which is why we did not perform a formal fit. Yet comparing a pure He model to the observed spectrum, we conclude it shows evidence of H and we encourage a fit with stratified model atmospheres.

\subsection{Central stars of (possible) PNe}
Three of the our targets are central stars of (possible) PNe. One of them, WDJ120728.43$+$540129.16, is a DAO WD, which we consider to be candidate for post-RGB CSPNe. Comparing the predicted post-RGB and post-AGB times of this star with calculations from \cite{Hall+2013} and \cite{MillerBertolami2016}, respectively, we found that it actually should be more likely to discover a PN around a post-RGB star than a low-mass post-AGB star. However, we could not reveal any photometric variability (neither in the TESS nor the ZTF light curves) that could hint towards a close companion, which would support the post-RGB nature of this object.\\
Furthermore, our study revealed the nature of the nucleus of the confirmed PN \object{PNG\,026.9$+$04.4}. We found the central star, WDJ182440.85$-$031959.52, is a pure DA 
WD and that the post-AGB age ($\approx300-400$\,kyrs) predicted by the \cite{MillerBertolami2016} tracks is one order of magnitude higher than what is expected from the kinematic age.\\
Finally, our sample also included the pure DA WD WDJ191231.47$-$033131.86, which could be the central star of the possible PN \object{FP\,J1912$-$0331}. Based on the derived atmospheric parameters, we estimated a cooling age around $1.2-1.5$\,Myrs, which makes it seem very unlikely that a PNe could still be visible.\\

\subsection{He-rich WDs}
A fifth of our targets turned out to be H-deficient WDs, of which 13\% (and possibly even 19\%) show UHE lines. The two hottest (\Teff$>110$\,kK) DO WDs (WDJ014343.14$+$584151.39 and WDJ191750.16$-$201409.55), do not show signs of C, and we suggest that they could be the outcome of a double He-WD merger \citep{Zhang+2012a} and, thus, they could be the slightly 
more massive analogs to the DO WDs \object{PG\,1034+001} and \object{PG\,0038+199}.
Five of the H-deficient WDs in our sample are DOZ WDs, with one of them, WDJ060244.99$-$135103.57, shown to be a previously known pulsating PG1159 (GW Vir) star \citep{Uzundag+2021}. Thanks to the higher resolution of the X-Shooter spectrum presented here, we were able to provide better constrains on its \Teff\ ($=100\,000\pm10\,000$\,K),  to derive its O abundances (O $=0.17\pm0.10$) and upper limits on the H and N abundances (H $<0.05$, N $<0.001$). Besides, WDJ060244.99$-$135103.57, which is known to be variable due to non-radial $g$-mode pulsations, our work has revealed four newly discovered variable H-deficient WDs, with two of them showing UHE lines.\\
The mean Kiel mass of our H-deficient WDs ($\langle M_{\mathrm{Kiel}} \rangle=0.65$\,\Msol, $\sigma = 0.01$\,\Msol) is slightly higher than the mean Kiel mass of the DA WDs, however, the mean HRD mass of DA and DO WDs ($\langle M_{\mathrm{HRD}} \rangle=0.63$\,\Msol, $\sigma = 0.01$\,\Msol) are in agreement. Our mean Kiel mass for DO WDs also agrees with what was previously reported for DO WDs from the SDSS \citep{Bedard+2020}.

\subsection{Mass determinations}
It is important to note that our three mass determinations are not independent
from each other. All three methods rely on the results of the spectroscopic 
analysis. The HRD and Kiel masses additionally depend on stellar evolutionary 
computations (e.g., core composition and thickness of the H- or He-envelope).
In addition, the HRD and gravity masses are both dependent on the 
\emph{Gaia} parallax and the zero point bias.\\
We find that Kiel and HRD masses agree for 81\% of the stars, while the 
gravity mass agrees with the former two only in half of the cases. 
This is not unexpected, as the atmospheric parameters of hot (pre-)WDs are known to be prone to systematic effects. To some degree it might be related to the neglect of metal opacities, which can have a large impact on the theoretical Balmer and \ion{He}{II} lines and, consequently, on the derived atmospheric parameters -- and, thus, the masses. For example, systematic errors on \Teff\ of $10-30$\,kK are not unusual for hot (pre-)WDs (e.g., \citealt{Rauch+2007, Gianninas+2010, Reindl+2014b, Werner+2018a, Werner+2018b, Werner+2019}). However, it should be noted that even if models including metal opacities are used, the discrepancy in the derived masses may persist \citep{Preval+2019, Herrero+2020, Werner+2022a}. In addition the assumption of 
a homogeneous model atmosphere for a star that actually has a stratified atmosphere 
can lead to large systematic errors (e.g., \citealt{Bedard+2020}).\\
We note that the mismatch in the three mass determination methods is not more
common in H-rich objects than in He-rich objects or vice-versa. Furthermore,
(and quite interestingly) the mismatch is also not more frequent in 
photometrically variable objects. However, it is striking that the H+He-rich objects in our sample, namely, sdO, O(H), and DAOs, often have a very low gravity mass ($\approx 0.2$\,\Msol). In addition, we find that for all DO WDs with \Teff$<60$\,kK the three mass
determinations are inconsistent with each other and that their gravity masses are unusually high. We also find that Kiel, HRD, and gravity masses disagree for several stars with \Teff\ in excess of 100\,kK. Finally, the discrepancy in the three masses is more common in high S/N spectra, meaning  that  we found the three masses disagree if the S/N is higher than 50 in
75\% of cases,.\\
A quantitative investigation of this problem is well beyond the scope of this 
paper, however, we would like to briefly discuss possible causes using the example of WDJ080326.15$-$034746.11. For this star, we found a HRD mass of $0.56\pm0.06$\,\Msol, a Kiel mass of ($0.50\pm0.04$\,\Msol), and unrealistically low gravity mass of $0.19^{+0.07}_{-0.05}\pm0.04$\,\Msol. Decreasing \Teff\ by 30\,kK would increase the gravity mass only by $+0.10$\,\Msol. We note that such a low \Teff\ for this star is unrealistic, since the X-Shooter spectrum clearly revealed the NLTE line emission cores in H\,$\alpha$ and \Ionw{He}{2}{4686} (see Fig.~\ref{fig:dao}), which would not be expected at \Teffw{57}. When we assume that the surface gravity has been underestimated by 0.45\,dex, then we obtain a gravity mass of $0.54^{+0.19}_{-0.14}$\,\Msol. Since the surface gravity has barley an impact on the derived radius obtained from the SED fit, also the derived luminosity and, hence, the HRD mass stays essentially the same. On the other hand, with a 0.45\,dex higher surface gravity the Kiel mass of the star would increase to 0.56\,\Msol, namely, that it is identical with the HRD mass. However, whether the neglect of metal opacities alone can be responsible for such a significantly higher surface gravity is debatable, for example \cite{Gianninas+2010}  only obtains a slightly higher values of \logg\ of $\approx0.1$\,dex when models containing CNO are used. In principle, neglecting radiation driven winds can also lead to an underestimation of the surface gravity and, hence, gravity mass. However, given the small mass-loss rates expected in hot WDs, a significant impact on \logg\ cannot be expected \citep{Sander+2015}. Finally, the neglect of metal opacities in our SED fits could also lead to an underestimation of the mass. In models that include metal opacities flux is redistributed from the UV towards longer wavelengths and can lead to an underestimation of the optical (which is our main photometry source for the SED fits) flux from a few to up to 30\% \citep{Reindl+2018a, Reindl+2021}. The derived radius relates to the model flux, $F_{\mathrm{model}}$, via $R \propto \sqrt{F_{\mathrm{model}}}$. If we assume that the our models underestimate the optical flux by 10\%, then the radius would be underestimated by 5\%. In case of WDJ080326.15$-$034746.11, this would then increase the gravity mass to 0.27\,\Msol, while increasing the luminosity to 41\,\Lsol, and decreasing the HRD mass to 0.55\,\Msol.

Besides systematic uncertainties on the atmospheric parameters, which are 
caused (but not limited to) the neglect of metal opacities,  parallax
measurements from the \emph{Gaia} mission could also be responsible. In particular,
the determination of the parallax bias is non-trivial since it depends at 
least on the magnitude, colour, and Ecliptic latitude of the source 
\citep{Lindegren+2021}. We note that some of our targets had 
either pseudo-colours or effective wavenumbers, $\nu_{\mathrm{eff}}$, that 
were outside the  parallax correction recipe provided by \cite{Lindegren+2021}.
However, for those targets we do not find that the mismatch of the
mass determination methods is more common.\\
We do, however, find there is a slightly higher probability that the derived 
masses do not agree for lower Ecliptic latitudes. For instance, for 
$70^{+11}_{-\phantom{0}8}$\% of the stars the gravity, HRD, and Kiel 
mass do not agree if the Ecliptic latitude is lower than zero, while for 
Ecliptic latitudes larger than zero, the mismatch of the three masses is 
only $43^{+7}_{-8}$\%.\\ 
Generally, the larger the parallax of a star, the less likely it becomes that the zero point determination is the main source of discrepancy of the derived masses. To test the impact of the zero point bias on the derived masses we take again WDJ080326.15$-$034746.11 as an example which has a relatively low parallax ($1.45\pm0.05$\,mas) compared to the rest of our sample. If we assume that the parallax bias has been underestimated by 0.1\,mas, then we obtain a 10\% larger radius. This, in turn, leads to a higher gravity mass of 0.23\,\Msol, a higher luminosity of 45\,\Lsol, and a lower HRD mass of 0.55\,\Rsol.\\

In conclusion, likely a combination of systematic errors on the derived atmospheric parameters that are caused by (but not limited to) the neglect of possibly stratified atmospheres and metal opacities in our spectroscoic and SED fits, as well as the uncertainties of the parallax zero point determination are responsible for the discrepancy of the Kiel, HRD, and gravity masses in some of our stars. Therefore, we would like to appeal to the reader that the parameters of the stars derived in this paper should be treated with caution. They can serve as an first estimate of the atmospheric and stellar parameters, but especially for hot (\Teff$>50$\,kK) H-rich WDs and stars showing the \ion{He}{ii} problem, it is only with UV spectroscopy that the metal abundances and reliable \Teff\ can be determined, and, thereby, the trustworthy masses. 

\subsection{Photometric variability}
For 62 out of the 71 stars in our sample, there were either TESS and/or ZTF light 
curves available. Of these 16 (26\%) were found to be photometrically 
variable\footnote{We note that our sample is slightly biased towards 
variable stars, since our LBT sample (ten stars) contained mainly targets 
(eight stars) that were known to be variable before the 
spectroscopic follow-up.}. One of these objects has been previously reported 
to be a GW Vir star \citep{Uzundag+2021}, but for the remaining 15 we discovered
their variability for the first time. Six of the variable stars are (possible)
UHE WDs and only one is a clear irradiation effect system consisting of a hot 
DA WD and an irradiated low-mass companion.\\
Strikingly, the majority (nine out of 15) of the variable stars exhibit 
non-sinusoidal light-curve shapes, with seven showing asymmetrical light curves, 
and two showing flat and extended minima. This includes all, but is not 
restricted to (possible) UHE WDs. Such light-curve shapes can unlikely 
be explained in terms of close binary systems. Instead we propose, that a 
significant fraction of all stars develop spots at a certain point when 
entering the WD cooling phase.\\
Also worth emphasizing is the photometric variability in the DAO 
WDJ210110.17$-$052751.14, which is likely currently transforming into a pure 
DA WD. The light curve looks sinusoidal, but depth of the minima in the TESS 
light curves of this already very compact and relatively massive star differ 
slightly, which could be related to two somewhat different sized spots on the 
surface of the WD.\\ 
In three of the stars  displaying asymmetrical light curves, we find that the 
shape (but not the amplitude) of the light curves seems to change from band 
to band. Such changes are typical for spots and can be explained with different
elements which cause their maximum in the light curve at different phases
\citep{Krticka+2020b}. Furthermore, in the case of metal-enriched spots, the 
band-to-band amplitude variation is expected typically small. 
\cite{Reindl+2021} predicts band-to-band amplitude variations $<0.04$\,mag from the g- to z-band for C, O, and iron-group elements. 
Only in the (near-) UV the difference in the amplitudes can become noticeable.
The typical uncertainty of the ZTF light curves is $0.01-0.02$\,mag, thus, 
the non-detection of amplitude variations from band to band could still be in 
line with spots on the surfaces of these stars.\\
We also note that non-sinusoidal light curves also look very similar to what is 
observed in magnetic chemically peculiar $\alpha^2$ Canum Venaticorum
variables (ACVs, \citealt{Huemmerich+2016, Jagelka+2019}). 
Like hot WDs, ACVs possess calm radiative atmospheres and their peculiar abundance 
patterns are thought to be produced by selective processes, such as
radiative levitation and gravitational settling \citep{Richer+2000}. Thus, we can 
speculate that the underlying mechanism of the photometric variability is the same,
namely, a magnetic field that interacts with photospheric atoms diffusing
under the competitive effects of gravity and radiative levitation \citep{Alecian2017}.
In fact, based on the discovery of spots on extreme horizontal branch (EHB) stars 
in Galactic globular clusters, \cite{Momany+2020} proposed that similar magnetic 
field induced variability phenomena should take place in all radiative-enveloped 
hot-stars.\\
While we did not detect any Zeeman splitting in the spectra of our variable stars,
below the detection threshold, field strengths could still be sufficient to enable the
production of spots. We highly encourage spectropolarimetric and/or higher 
resolution follow-up of our variable stars in order to place constraints on the magnetic field strengths.\\
Yet, the presence of a magnetic field and a radiative atmosphere alone does not seem to be sufficient for the spot production on a hot star. For example,  four magnetic ($B\approx 350$\,kG) hot subdwarfs were recently discovered, and none of them display photometric variability \citep{Dorsch+2022, Pelisoli+2022}. Therefore, at least a third parameter likely determines whether or not a WD develops spots. We speculate whether this could be related to the onset of diffusion.

\begin{acknowledgements}
We thank Boris Chausov for providing us with the image of PN\,G136.7+61.9.
We thank Keaton Bell, JJ Hermes, Simon Murphy, and Gerald Handler for useful discussion.
NR is supported by the Deutsche Forschungsgemeinschaft (DFG) through grant GE2506/17-1. 
IP acknowledges funding by the UK's Science and Technology Facilities Council (STFC), grant ST/T000406/1, and from a Warwick Astrophysics prize post-doctoral fellowship, made possible thanks to a generous philanthropic donation.
Based on observations with the Isaac Newton Telescopes operated by the Isaac Newton Group at the Observatorio del Roque de los Muchachos of the Instituto de Astrofisica de Canarias on the island of La Palma, Spain.
This paper uses data taken with the MODS spectrographs built with funding
from NSF grant AST-9987045 and the NSF Telescope System Instrumentation
Program (TSIP), with additional funds from the Ohio Board of Regents and the
Ohio State University Office of Research. 
IRAF is distributed 
by the National Optical Astronomy Observatory, which is operated by the 
Association of Universities for Research in Astronomy
(AURA) under a cooperative agreement with the National Science Foundation.
This research has made use of the HASH PN database at \url{http://hashpn.space}.
This research made use of TOPCAT, an interactive graphical viewer and editor for tabular data Taylor (\cite{Taylor2005}). This research made use of the SIMBAD database, operated at CDS, Strasbourg, France; the VizieR catalogue access tool, CDS, Strasbourg, France. This research has made use of the services of the ESO Science Archive Facility.
This work has made use of data from the European Space Agency (ESA) mission {\it Gaia} (https://www.cosmos.esa.int/gaia), processed by the {\it Gaia} Data Processing and Analysis Consortium (DPAC, https://www.cosmos.esa.int/web/gaia/dpac/consortium). Funding for the DPAC has been provided by national institutions, in particular the institutions participating in the {\it Gaia} Multilateral Agreement.
Based on observations obtained with the Samuel Oschin 48-inch Telescope
at the Palomar Observatory as part of the Zwicky Transient Facility
project. ZTF is supported by the National Science Foundation under Grant
No. AST-1440341 and a collaboration including Caltech, IPAC, the
Weizmann Institute for Science, the Oskar Klein Center at Stockholm
University, the University of Maryland, the University of Washington,
Deutsches Elektronen-Synchrotron and Humboldt University, Los Alamos
National Laboratories, the TANGO Consortium of Taiwan, the University of
Wisconsin at Milwaukee, and Lawrence Berkeley National Laboratories.
Operations are conducted by COO, IPAC, and UW.
This paper includes data collected by the TESS mission. Funding for the
TESS mission is provided by the NASA Explorer Program.
This work made use of \texttt{tpfplotter} by J. Lillo-Box (publicly
available in \url{www.github.com/jlillo/tpfplotter}), which also made
use of the python packages \texttt{astropy}, \texttt{lightkurve},
\texttt{matplotlib} and \texttt{numpy}.

\end{acknowledgements}

\bibliographystyle{aa}
\bibliography{BB}

\begin{thebibliography}{137}
\expandafter\ifx\csname natexlab\endcsname\relax\def\natexlab#1{#1}\fi

\bibitem[{{Abbott} {et~al.}(2018){Abbott}, {Abdalla}, {Allam}, {Amara},
  {Annis}, {Asorey}, {Avila}, {Ballester}, {Banerji}, {Barkhouse}, {Baruah},
  {Baumer}, {Bechtol}, {Becker}, {Benoit-L{\'e}vy}, {Bernstein}, {Bertin},
  {Blazek}, {Bocquet}, {Brooks}, {Brout}, {Buckley-Geer}, {Burke}, {Busti},
  {Campisano}, {Cardiel-Sas}, {Carnero Rosell}, {Carrasco Kind}, {Carretero},
  {Castander}, {Cawthon}, {Chang}, {Chen}, {Conselice}, {Costa}, {Crocce},
  {Cunha}, {D'Andrea}, {da Costa}, {Das}, {Daues}, {Davis}, {Davis}, {De
  Vicente}, {DePoy}, {DeRose}, {Desai}, {Diehl}, {Dietrich}, {Dodelson},
  {Doel}, {Drlica-Wagner}, {Eifler}, {Elliott}, {Evrard}, {Farahi}, {Fausti
  Neto}, {Fernandez}, {Finley}, {Flaugher}, {Foley}, {Fosalba}, {Friedel},
  {Frieman}, {Garc{\'\i}a-Bellido}, {Gaztanaga}, {Gerdes}, {Giannantonio},
  {Gill}, {Glazebrook}, {Goldstein}, {Gower}, {Gruen}, {Gruendl}, {Gschwend},
  {Gupta}, {Gutierrez}, {Hamilton}, {Hartley}, {Hinton}, {Hislop}, {Hollowood},
  {Honscheid}, {Hoyle}, {Huterer}, {Jain}, {James}, {Jeltema}, {Johnson},
  {Johnson}, {Kacprzak}, {Kent}, {Khullar}, {Klein}, {Kovacs}, {Koziol},
  {Krause}, {Kremin}, {Kron}, {Kuehn}, {Kuhlmann}, {Kuropatkin}, {Lahav},
  {Lasker}, {Li}, {Li}, {Liddle}, {Lima}, {Lin}, {L{\'o}pez-Reyes}, {MacCrann},
  {Maia}, {Maloney}, {Manera}, {March}, {Marriner}, {Marshall}, {Martini},
  {McClintock}, {McKay}, {McMahon}, {Melchior}, {Menanteau}, {Miller},
  {Miquel}, {Mohr}, {Morganson}, {Mould}, {Neilsen}, {Nichol}, {Nogueira},
  {Nord}, {Nugent}, {Nunes}, {Ogando}, {Old}, {Pace}, {Palmese},
  {Paz-Chinch{\'o}n}, {Peiris}, {Percival}, {Petravick}, {Plazas}, {Poh},
  {Pond}, {Porredon}, {Pujol}, {Refregier}, {Reil}, {Ricker}, {Rollins},
  {Romer}, {Roodman}, {Rooney}, {Ross}, {Rykoff}, {Sako}, {Sanchez}, {Sanchez},
  {Santiago}, {Saro}, {Scarpine}, {Scolnic}, {Serrano}, {Sevilla-Noarbe},
  {Sheldon}, {Shipp}, {Silveira}, {Smith}, {Smith}, {Smith}, {Soares-Santos},
  {Sobreira}, {Song}, {Stebbins}, {Suchyta}, {Sullivan}, {Swanson}, {Tarle},
  {Thaler}, {Thomas}, {Thomas}, {Troxel}, {Tucker}, {Vikram}, {Vivas},
  {Walker}, {Wechsler}, {Weller}, {Wester}, {Wolf}, {Wu}, {Yanny}, {Zenteno},
  {Zhang}, {Zuntz}, {DES Collaboration}, {Juneau}, {Fitzpatrick}, {Nikutta},
  {Nidever}, {Olsen}, {Scott}, \& {NOAO Data Lab}}]{Abdalla+2018}
{Abbott}, T.~M.~C., {Abdalla}, F.~B., {Allam}, S., {et~al.} 2018, \apjs, 239,
  18

\bibitem[{{Alam} {et~al.}(2015){Alam}, {Albareti}, {Allende Prieto}, {Anders},
  {Anderson}, {Anderton}, {Andrews}, {Armengaud}, {Aubourg}, {Bailey}, {Basu},
  {Bautista}, {Beaton}, {Beers}, {Bender}, {Berlind}, {Beutler}, {Bhardwaj},
  {Bird}, {Bizyaev}, {Blake}, {Blanton}, {Blomqvist}, {Bochanski}, {Bolton},
  {Bovy}, {Shelden Bradley}, {Brandt}, {Brauer}, {Brinkmann}, {Brown},
  {Brownstein}, {Burden}, {Burtin}, {Busca}, {Cai}, {Capozzi}, {Carnero
  Rosell}, {Carr}, {Carrera}, {Chambers}, {Chaplin}, {Chen}, {Chiappini},
  {Chojnowski}, {Chuang}, {Clerc}, {Comparat}, {Covey}, {Croft}, {Cuesta},
  {Cunha}, {da Costa}, {Da Rio}, {Davenport}, {Dawson}, {De Lee}, {Delubac},
  {Deshpande}, {Dhital}, {Dutra-Ferreira}, {Dwelly}, {Ealet}, {Ebelke},
  {Edmondson}, {Eisenstein}, {Ellsworth}, {Elsworth}, {Epstein}, {Eracleous},
  {Escoffier}, {Esposito}, {Evans}, {Fan}, {Fern{\'a}ndez-Alvar}, {Feuillet},
  {Filiz Ak}, {Finley}, {Finoguenov}, {Flaherty}, {Fleming}, {Font-Ribera},
  {Foster}, {Frinchaboy}, {Galbraith-Frew}, {Garc{\'\i}a},
  {Garc{\'\i}a-Hern{\'a}ndez}, {Garc{\'\i}a P{\'e}rez}, {Gaulme}, {Ge},
  {G{\'e}nova-Santos}, {Georgakakis}, {Ghezzi}, {Gillespie}, {Girardi},
  {Goddard}, {Gontcho}, {Gonz{\'a}lez Hern{\'a}ndez}, {Grebel}, {Green},
  {Grieb}, {Grieves}, {Gunn}, {Guo}, {Harding}, {Hasselquist}, {Hawley},
  {Hayden}, {Hearty}, {Hekker}, {Ho}, {Hogg}, {Holley-Bockelmann}, {Holtzman},
  {Honscheid}, {Huber}, {Huehnerhoff}, {Ivans}, {Jiang}, {Johnson},
  {Kinemuchi}, {Kirkby}, {Kitaura}, {Klaene}, {Knapp}, {Kneib}, {Koenig},
  {Lam}, {Lan}, {Lang}, {Laurent}, {Le Goff}, {Leauthaud}, {Lee}, {Lee},
  {Licquia}, {Liu}, {Long}, {L{\'o}pez-Corredoira}, {Lorenzo-Oliveira},
  {Lucatello}, {Lundgren}, {Lupton}, {Mack}, {Mahadevan}, {Maia}, {Majewski},
  {Malanushenko}, {Malanushenko}, {Manchado}, {Manera}, {Mao}, {Maraston},
  {Marchwinski}, {Margala}, {Martell}, {Martig}, {Masters}, {Mathur},
  {McBride}, {McGehee}, {McGreer}, {McMahon}, {M{\'e}nard}, {Menzel},
  {Merloni}, {M{\'e}sz{\'a}ros}, {Miller}, {Miralda-Escud{\'e}}, {Miyatake},
  {Montero-Dorta}, {More}, {Morganson}, {Morice-Atkinson}, {Morrison},
  {Mosser}, {Muna}, {Myers}, {Nandra}, {Newman}, {Neyrinck}, {Nguyen},
  {Nichol}, {Nidever}, {Noterdaeme}, {Nuza}, {O'Connell}, {O'Connell},
  {O'Connell}, {Ogando}, {Olmstead}, {Oravetz}, {Oravetz}, {Osumi}, {Owen},
  {Padgett}, {Padmanabhan}, {Paegert}, {Palanque-Delabrouille}, {Pan},
  {Parejko}, {P{\^a}ris}, {Park}, {Pattarakijwanich}, {Pellejero-Ibanez},
  {Pepper}, {Percival}, {P{\'e}rez-Fournon}, {P{\'e}rez-R{\`a}fols},
  {Petitjean}, {Pieri}, {Pinsonneault}, {Porto de Mello}, {Prada}, {Prakash},
  {Price-Whelan}, {Protopapas}, {Raddick}, {Rahman}, {Reid}, {Rich}, {Rix},
  {Robin}, {Rockosi}, {Rodrigues}, {Rodr{\'\i}guez-Torres}, {Roe}, {Ross},
  {Ross}, {Rossi}, {Ruan}, {Rubi{\~n}o-Mart{\'\i}n}, {Rykoff},
  {Salazar-Albornoz}, {Salvato}, {Samushia}, {S{\'a}nchez}, {Santiago},
  {Sayres}, {Schiavon}, {Schlegel}, {Schmidt}, {Schneider}, {Schultheis},
  {Schwope}, {Sc{\'o}ccola}, {Scott}, {Sellgren}, {Seo}, {Serenelli}, {Shane},
  {Shen}, {Shetrone}, {Shu}, {Silva Aguirre}, {Sivarani}, {Skrutskie},
  {Slosar}, {Smith}, {Sobreira}, {Souto}, {Stassun}, {Steinmetz}, {Stello},
  {Strauss}, {Streblyanska}, {Suzuki}, {Swanson}, {Tan}, {Tayar}, {Terrien},
  {Thakar}, {Thomas}, {Thomas}, {Thompson}, {Tinker}, {Tojeiro}, {Troup},
  {Vargas-Maga{\~n}a}, {Vazquez}, {Verde}, {Viel}, {Vogt}, {Wake}, {Wang},
  {Weaver}, {Weinberg}, {Weiner}, {White}, {Wilson}, {Wisniewski},
  {Wood-Vasey}, {Ye`che}, {York}, {Zakamska}, {Zamora}, {Zasowski}, {Zehavi},
  {Zhao}, {Zheng}, {Zhou}, {Zhou}, {Zou}, \& {Zhu}}]{Alam+2015}
{Alam}, S., {Albareti}, F.~D., {Allende Prieto}, C., {et~al.} 2015, \apjs, 219,
  12

\bibitem[{{Alecian} \& {Stift}(2017)}]{Alecian2017}
{Alecian}, G. \& {Stift}, M.~J. 2017, \mnras, 468, 1023

\bibitem[{{Aller} {et~al.}(2020){Aller}, {Lillo-Box}, {Jones}, {Miranda}, \&
  {Barcel{\'o} Forteza}}]{Aller+2020}
{Aller}, A., {Lillo-Box}, J., {Jones}, D., {Miranda}, L.~F., \& {Barcel{\'o}
  Forteza}, S. 2020, \aap, 635, A128

\bibitem[{{Althaus} {et~al.}(2009){Althaus}, {Panei}, {Miller Bertolami},
  {Garc{\'{\i}}a-Berro}, {C{\'o}rsico}, {Romero}, {Kepler}, \&
  {Rohrmann}}]{Althaus+2009}
{Althaus}, L.~G., {Panei}, J.~A., {Miller Bertolami}, M.~M., {et~al.} 2009,
  \apj, 704, 1605

\bibitem[{{Bailer-Jones} {et~al.}(2021){Bailer-Jones}, {Rybizki}, {Fouesneau},
  {Demleitner}, \& {Andrae}}]{Bailer-Jones+2021}
{Bailer-Jones}, C.~A.~L., {Rybizki}, J., {Fouesneau}, M., {Demleitner}, M., \&
  {Andrae}, R. 2021, \aj, 161, 147

\bibitem[{{Bainbridge} {et~al.}(2017){Bainbridge}, {Barstow}, {Reindl},
  {Tchang-Brillet}, {Ayres}, {Webb}, {Barrow}, {Hu}, {Holberg}, {Preval},
  {Ubachs}, {Dzuba}, {Flambaum}, {Dumont}, \& {Berengut}}]{Bainbridge+2017}
{Bainbridge}, M., {Barstow}, M., {Reindl}, N., {et~al.} 2017, Universe, 3, 32

\bibitem[{{Barnard} {et~al.}(1969){Barnard}, {Cooper}, \&
  {Shamey}}]{Barnard1969}
{Barnard}, A.~J., {Cooper}, J., \& {Shamey}, L.~J. 1969, \aap, 1, 28

\bibitem[{{Barnard} {et~al.}(1974){Barnard}, {Cooper}, \&
  {Smith}}]{Barnard1974}
{Barnard}, A.~J., {Cooper}, J., \& {Smith}, E.~W. 1974, \jqsrt, 14, 1025

\bibitem[{{Barstow} {et~al.}(2014){Barstow}, {Barstow}, {Casewell}, {Holberg},
  \& {Hubeny}}]{Barstow+2014}
{Barstow}, M.~A., {Barstow}, J.~K., {Casewell}, S.~L., {Holberg}, J.~B., \&
  {Hubeny}, I. 2014, \mnras, 440, 1607

\bibitem[{{Bauer} \& {Husfeld}(1995)}]{BauerHusfeld1995}
{Bauer}, F. \& {Husfeld}, D. 1995, \aap, 300, 481

\bibitem[{{B{\'e}dard} {et~al.}(2022){B{\'e}dard}, {Bergeron}, \&
  {Brassard}}]{Bedrad+2022}
{B{\'e}dard}, A., {Bergeron}, P., \& {Brassard}, P. 2022, \apj, 930, 8

\bibitem[{{B{\'e}dard} {et~al.}(2023){B{\'e}dard}, {Bergeron}, \&
  {Brassard}}]{Bedard+2023}
{B{\'e}dard}, A., {Bergeron}, P., \& {Brassard}, P. 2023, arXiv e-prints,
  arXiv:2302.05424

\bibitem[{{B{\'e}dard} {et~al.}(2020){B{\'e}dard}, {Bergeron}, {Brassard}, \&
  {Fontaine}}]{Bedard+2020}
{B{\'e}dard}, A., {Bergeron}, P., {Brassard}, P., \& {Fontaine}, G. 2020, \apj,
  901, 93

\bibitem[{{Bellm} {et~al.}(2019){Bellm}, {Kulkarni}, {Graham}, {Dekany},
  {Smith}, {Riddle}, {Masci}, {Helou}, {Prince}, {Adams}, {Barbarino},
  {Barlow}, {Bauer}, {Beck}, {Belicki}, {Biswas}, {Blagorodnova}, {Bodewits},
  {Bolin}, {Brinnel}, {Brooke}, {Bue}, {Bulla}, {Burruss}, {Cenko}, {Chang},
  {Connolly}, {Coughlin}, {Cromer}, {Cunningham}, {De}, {Delacroix}, {Desai},
  {Duev}, {Eadie}, {Farnham}, {Feeney}, {Feindt}, {Flynn}, {Franckowiak},
  {Frederick}, {Fremling}, {Gal-Yam}, {Gezari}, {Giomi}, {Goldstein},
  {Golkhou}, {Goobar}, {Groom}, {Hacopians}, {Hale}, {Henning}, {Ho}, {Hover},
  {Howell}, {Hung}, {Huppenkothen}, {Imel}, {Ip}, {Ivezi{\'c}}, {Jackson},
  {Jones}, {Juric}, {Kasliwal}, {Kaspi}, {Kaye}, {Kelley}, {Kowalski},
  {Kramer}, {Kupfer}, {Landry}, {Laher}, {Lee}, {Lin}, {Lin}, {Lunnan},
  {Giomi}, {Mahabal}, {Mao}, {Miller}, {Monkewitz}, {Murphy}, {Ngeow},
  {Nordin}, {Nugent}, {Ofek}, {Patterson}, {Penprase}, {Porter}, {Rauch},
  {Rebbapragada}, {Reiley}, {Rigault}, {Rodriguez}, {van Roestel}, {Rusholme},
  {van Santen}, {Schulze}, {Shupe}, {Singer}, {Soumagnac}, {Stein}, {Surace},
  {Sollerman}, {Szkody}, {Taddia}, {Terek}, {Van Sistine}, {van Velzen},
  {Vestrand}, {Walters}, {Ward}, {Ye}, {Yu}, {Yan}, \& {Zolkower}}]{Bellm+2019}
{Bellm}, E.~C., {Kulkarni}, S.~R., {Graham}, M.~J., {et~al.} 2019, \pasp, 131,
  018002

\bibitem[{{Berengut} {et~al.}(2013){Berengut}, {Flambaum}, {Ong}, {Webb},
  {Barrow}, {Barstow}, {Preval}, \& {Holberg}}]{Berengut+2013}
{Berengut}, J.~C., {Flambaum}, V.~V., {Ong}, A., {et~al.} 2013, Physical Review
  Letters, 111, 010801

\bibitem[{{Bianchi} {et~al.}(2017){Bianchi}, {Shiao}, \&
  {Thilker}}]{Bianchi+2017}
{Bianchi}, L., {Shiao}, B., \& {Thilker}, D. 2017, \apjs, 230, 24

\bibitem[{{Boffin} \& {Jones}(2019)}]{Boffin+Jones+2019}
{Boffin}, H. M.~J. \& {Jones}, D. 2019, {The Importance of Binaries in the
  Formation and Evolution of Planetary Nebulae} (Berlin, New York,
  Springer-Verlag)

\bibitem[{{Bohlin} {et~al.}(2020){Bohlin}, {Hubeny}, \& {Rauch}}]{Bohlin+2020}
{Bohlin}, R.~C., {Hubeny}, I., \& {Rauch}, T. 2020, \aj, 160, 21

\bibitem[{{Camisassa} {et~al.}(2019){Camisassa}, {Althaus}, {C{\'o}rsico}, {De
  Ger{\'o}nimo}, {Miller Bertolami}, {Novarino}, {Rohrmann}, {Wachlin}, \&
  {Garc{\'\i}a-Berro}}]{Camisassa+2019}
{Camisassa}, M.~E., {Althaus}, L.~G., {C{\'o}rsico}, A.~H., {et~al.} 2019,
  \aap, 625, A87

\bibitem[{{Chambers} {et~al.}(2016){Chambers}, {Magnier}, {Metcalfe},
  {Flewelling}, {Huber}, {Waters}, {Denneau}, {Draper}, {Farrow}, {Finkbeiner},
  {Holmberg}, {Koppenhoefer}, {Price}, {Rest}, {Saglia}, {Schlafly}, {Smartt},
  {Sweeney}, {Wainscoat}, {Burgett}, {Chastel}, {Grav}, {Heasley}, {Hodapp},
  {Jedicke}, {Kaiser}, {Kudritzki}, {Luppino}, {Lupton}, {Monet}, {Morgan},
  {Onaka}, {Shiao}, {Stubbs}, {Tonry}, {White}, {Ba{\~n}ados}, {Bell},
  {Bender}, {Bernard}, {Boegner}, {Boffi}, {Botticella}, {Calamida},
  {Casertano}, {Chen}, {Chen}, {Cole}, {Deacon}, {Frenk}, {Fitzsimmons},
  {Gezari}, {Gibbs}, {Goessl}, {Goggia}, {Gourgue}, {Goldman}, {Grant},
  {Grebel}, {Hambly}, {Hasinger}, {Heavens}, {Heckman}, {Henderson}, {Henning},
  {Holman}, {Hopp}, {Ip}, {Isani}, {Jackson}, {Keyes}, {Koekemoer}, {Kotak},
  {Le}, {Liska}, {Long}, {Lucey}, {Liu}, {Martin}, {Masci}, {McLean}, {Mindel},
  {Misra}, {Morganson}, {Murphy}, {Obaika}, {Narayan}, {Nieto-Santisteban},
  {Norberg}, {Peacock}, {Pier}, {Postman}, {Primak}, {Rae}, {Rai}, {Riess},
  {Riffeser}, {Rix}, {R{\"o}ser}, {Russel}, {Rutz}, {Schilbach}, {Schultz},
  {Scolnic}, {Strolger}, {Szalay}, {Seitz}, {Small}, {Smith}, {Soderblom},
  {Taylor}, {Thomson}, {Taylor}, {Thakar}, {Thiel}, {Thilker}, {Unger},
  {Urata}, {Valenti}, {Wagner}, {Walder}, {Walter}, {Watters}, {Werner},
  {Wood-Vasey}, \& {Wyse}}]{Chambers+2016}
{Chambers}, K.~C., {Magnier}, E.~A., {Metcalfe}, N., {et~al.} 2016, arXiv
  e-prints, arXiv:1612.05560

\bibitem[{{Chayer} {et~al.}(2015){Chayer}, {Dixon}, {Fullerton},
  {Ooghe-Tabanou}, \& {Reid}}]{Chayer+2015}
{Chayer}, P., {Dixon}, W.~V., {Fullerton}, A.~W., {Ooghe-Tabanou}, B., \&
  {Reid}, I.~N. 2015, \mnras, 452, 2292

\bibitem[{{Clemens} {et~al.}(2004){Clemens}, {Crain}, \&
  {Anderson}}]{clemens2004}
{Clemens}, J.~C., {Crain}, J.~A., \& {Anderson}, R. 2004, in Society of
  Photo-Optical Instrumentation Engineers (SPIE) Conference Series, Vol. 5492,
  Ground-based Instrumentation for Astronomy, ed. A.~F.~M. {Moorwood} \&
  M.~{Iye}, 331--340

\bibitem[{{Croxall} \& {Pogge}(2019)}]{Croxall+2019}
{Croxall}, K.~V. \& {Pogge}, R.~W. 2019, {rwpogge/modsIDL: modsIDL Binocular
  Release}, Zenodo

\bibitem[{{Culpan} {et~al.}(2022){Culpan}, {Geier}, {Reindl}, {Pelisoli},
  {Gentile Fusillo}, \& {Vorontseva}}]{Culpan+2022}
{Culpan}, R., {Geier}, S., {Reindl}, N., {et~al.} 2022, \aap, 662, A40

\bibitem[{{De Marco}(2009)}]{DeMarco+2009}
{De Marco}, O. 2009, \pasp, 121, 316

\bibitem[{{Dorsch} {et~al.}(2019){Dorsch}, {Latour}, \& {Heber}}]{Dorsch+2019}
{Dorsch}, M., {Latour}, M., \& {Heber}, U. 2019, \aap, 630, A130

\bibitem[{{Dorsch} {et~al.}(2022){Dorsch}, {Reindl}, {Pelisoli}, {Heber},
  {Geier}, {Istrate}, \& {Justham}}]{Dorsch+2022}
{Dorsch}, M., {Reindl}, N., {Pelisoli}, I., {et~al.} 2022, \aap, 658, L9

\bibitem[{{Eastman} {et~al.}(2010){Eastman}, {Siverd}, \&
  {Gaudi}}]{Eastman+2010}
{Eastman}, J., {Siverd}, R., \& {Gaudi}, B.~S. 2010, \pasp, 122, 935

\bibitem[{{Fantin} {et~al.}(2021){Fantin}, {C{\^o}t{\'e}}, {McConnachie},
  {Bergeron}, {Cuillandre}, {Dufour}, {Gwyn}, {Ibata}, \&
  {Thomas}}]{Fantin+2021}
{Fantin}, N.~J., {C{\^o}t{\'e}}, P., {McConnachie}, A.~W., {et~al.} 2021, \apj,
  913, 30

\bibitem[{{Fitzpatrick} {et~al.}(2019){Fitzpatrick}, {Massa}, {Gordon},
  {Bohlin}, \& {Clayton}}]{Fitzpatrick+2019}
{Fitzpatrick}, E.~L., {Massa}, D., {Gordon}, K.~D., {Bohlin}, R., \& {Clayton},
  G.~C. 2019, \apj, 886, 108

\bibitem[{{Fleming} {et~al.}(1986){Fleming}, {Liebert}, \&
  {Green}}]{Fleming+1986}
{Fleming}, T.~A., {Liebert}, J., \& {Green}, R.~F. 1986, \apj, 308, 176

\bibitem[{{Fontaine} {et~al.}(2008){Fontaine}, {Chayer}, {Oliveira},
  {Wesemael}, \& {Fontaine}}]{Fontaine2008}
{Fontaine}, M., {Chayer}, P., {Oliveira}, C.~M., {Wesemael}, F., \& {Fontaine},
  G. 2008, \apj, 678, 394

\bibitem[{{Frew} {et~al.}(2016){Frew}, {Parker}, \&
  {Boji{\v{c}}i{\'c}}}]{Frew+2016}
{Frew}, D.~J., {Parker}, Q.~A., \& {Boji{\v{c}}i{\'c}}, I.~S. 2016, \mnras,
  455, 1459

\bibitem[{{Gaia Collaboration} {et~al.}(2021){Gaia Collaboration}, {Brown},
  {Vallenari}, {Prusti}, {de Bruijne}, {Babusiaux}, {Biermann}, {Creevey},
  {Evans}, {Eyer}, {Hutton}, {Jansen}, {Jordi}, {Klioner}, {Lammers},
  {Lindegren}, {Luri}, {Mignard}, {Panem}, {Pourbaix}, {Randich}, {Sartoretti},
  {Soubiran}, {Walton}, {Arenou}, {Bailer-Jones}, {Bastian}, {Cropper},
  {Drimmel}, {Katz}, {Lattanzi}, {van Leeuwen}, {Bakker}, {Cacciari},
  {Casta{\~n}eda}, {De Angeli}, {Ducourant}, {Fabricius}, {Fouesneau},
  {Fr{\'e}mat}, {Guerra}, {Guerrier}, {Guiraud}, {Jean-Antoine Piccolo},
  {Masana}, {Messineo}, {Mowlavi}, {Nicolas}, {Nienartowicz}, {Pailler},
  {Panuzzo}, {Riclet}, {Roux}, {Seabroke}, {Sordo}, {Tanga}, {Th{\'e}venin},
  {Gracia-Abril}, {Portell}, {Teyssier}, {Altmann}, {Andrae}, {Bellas-Velidis},
  {Benson}, {Berthier}, {Blomme}, {Brugaletta}, {Burgess}, {Busso}, {Carry},
  {Cellino}, {Cheek}, {Clementini}, {Damerdji}, {Davidson}, {Delchambre},
  {Dell'Oro}, {Fern{\'a}ndez-Hern{\'a}ndez}, {Galluccio}, {Garc{\'\i}a-Lario},
  {Garcia-Reinaldos}, {Gonz{\'a}lez-N{\'u}{\~n}ez}, {Gosset}, {Haigron},
  {Halbwachs}, {Hambly}, {Harrison}, {Hatzidimitriou}, {Heiter},
  {Hern{\'a}ndez}, {Hestroffer}, {Hodgkin}, {Holl}, {Jan{\ss}en}, {Jevardat de
  Fombelle}, {Jordan}, {Krone-Martins}, {Lanzafame}, {L{\"o}ffler}, {Lorca},
  {Manteiga}, {Marchal}, {Marrese}, {Moitinho}, {Mora}, {Muinonen}, {Osborne},
  {Pancino}, {Pauwels}, {Petit}, {Recio-Blanco}, {Richards}, {Riello},
  {Rimoldini}, {Robin}, {Roegiers}, {Rybizki}, {Sarro}, {Siopis}, {Smith},
  {Sozzetti}, {Ulla}, {Utrilla}, {van Leeuwen}, {van Reeven}, {Abbas}, {Abreu
  Aramburu}, {Accart}, {Aerts}, {Aguado}, {Ajaj}, {Altavilla}, {{\'A}lvarez},
  {{\'A}lvarez Cid-Fuentes}, {Alves}, {Anderson}, {Anglada Varela}, {Antoja},
  {Audard}, {Baines}, {Baker}, {Balaguer-N{\'u}{\~n}ez}, {Balbinot}, {Balog},
  {Barache}, {Barbato}, {Barros}, {Barstow}, {Bartolom{\'e}}, {Bassilana},
  {Bauchet}, {Baudesson-Stella}, {Becciani}, {Bellazzini}, {Bernet}, {Bertone},
  {Bianchi}, {Blanco-Cuaresma}, {Boch}, {Bombrun}, {Bossini}, {Bouquillon},
  {Bragaglia}, {Bramante}, {Breedt}, {Bressan}, {Brouillet}, {Bucciarelli},
  {Burlacu}, {Busonero}, {Butkevich}, {Buzzi}, {Caffau}, {Cancelliere},
  {C{\'a}novas}, {Cantat-Gaudin}, {Carballo}, {Carlucci}, {Carnerero},
  {Carrasco}, {Casamiquela}, {Castellani}, {Castro-Ginard}, {Castro Sampol},
  {Chaoul}, {Charlot}, {Chemin}, {Chiavassa}, {Cioni}, {Comoretto}, {Cooper},
  {Cornez}, {Cowell}, {Crifo}, {Crosta}, {Crowley}, {Dafonte}, {Dapergolas},
  {David}, {David}, {de Laverny}, {De Luise}, {De March}, {De Ridder}, {de
  Souza}, {de Teodoro}, {de Torres}, {del Peloso}, {del Pozo}, {Delbo},
  {Delgado}, {Delgado}, {Delisle}, {Di Matteo}, {Diakite}, {Diener},
  {Distefano}, {Dolding}, {Eappachen}, {Edvardsson}, {Enke}, {Esquej}, {Fabre},
  {Fabrizio}, {Faigler}, {Fedorets}, {Fernique}, {Fienga}, {Figueras},
  {Fouron}, {Fragkoudi}, {Fraile}, {Franke}, {Gai}, {Garabato},
  {Garcia-Gutierrez}, {Garc{\'\i}a-Torres}, {Garofalo}, {Gavras}, {Gerlach},
  {Geyer}, {Giacobbe}, {Gilmore}, {Girona}, {Giuffrida}, {Gomel}, {Gomez},
  {Gonzalez-Santamaria}, {Gonz{\'a}lez-Vidal}, {Granvik},
  {Guti{\'e}rrez-S{\'a}nchez}, {Guy}, {Hauser}, {Haywood}, {Helmi}, {Hidalgo},
  {Hilger}, {H{\l}adczuk}, {Hobbs}, {Holland}, {Huckle}, {Jasniewicz},
  {Jonker}, {Juaristi Campillo}, {Julbe}, {Karbevska}, {Kervella}, {Khanna},
  {Kochoska}, {Kontizas}, {Kordopatis}, {Korn}, {Kostrzewa-Rutkowska},
  {Kruszy{\'n}ska}, {Lambert}, {Lanza}, {Lasne}, {Le Campion}, {Le Fustec},
  {Lebreton}, {Lebzelter}, {Leccia}, {Leclerc}, {Lecoeur-Taibi}, {Liao},
  {Licata}, {Lindstr{\o}m}, {Lister}, {Livanou}, {Lobel}, {Madrero Pardo},
  {Managau}, {Mann}, {Marchant}, {Marconi}, {Marcos Santos}, {Marinoni},
  {Marocco}, {Marshall}, {Martin Polo}, {Mart{\'\i}n-Fleitas}, {Masip},
  {Massari}, {Mastrobuono-Battisti}, {Mazeh}, {McMillan}, {Messina},
  {Michalik}, {Millar}, {Mints}, {Molina}, {Molinaro}, {Moln{\'a}r},
  {Montegriffo}, {Mor}, {Morbidelli}, {Morel}, {Morris}, {Mulone}, {Munoz},
  {Muraveva}, {Murphy}, {Musella}, {Noval}, {Ord{\'e}novic}, {Orr{\`u}},
  {Osinde}, {Pagani}, {Pagano}, {Palaversa}, {Palicio}, {Panahi}, {Pawlak},
  {Pe{\~n}alosa Esteller}, {Penttil{\"a}}, {Piersimoni}, {Pineau}, {Plachy},
  {Plum}, {Poggio}, {Poretti}, {Poujoulet}, {Pr{\v{s}}a}, {Pulone}, {Racero},
  {Ragaini}, {Rainer}, {Raiteri}, {Rambaux}, {Ramos}, {Ramos-Lerate}, {Re
  Fiorentin}, {Regibo}, {Reyl{\'e}}, {Ripepi}, {Riva}, {Rixon}, {Robichon},
  {Robin}, {Roelens}, {Rohrbasser}, {Romero-G{\'o}mez}, {Rowell}, {Royer},
  {Rybicki}, {Sadowski}, {Sagrist{\`a} Sell{\'e}s}, {Sahlmann}, {Salgado},
  {Salguero}, {Samaras}, {Sanchez Gimenez}, {Sanna}, {Santove{\~n}a},
  {Sarasso}, {Schultheis}, {Sciacca}, {Segol}, {Segovia}, {S{\'e}gransan},
  {Semeux}, {Shahaf}, {Siddiqui}, {Siebert}, {Siltala}, {Slezak}, {Smart},
  {Solano}, {Solitro}, {Souami}, {Souchay}, {Spagna}, {Spoto}, {Steele},
  {Steidelm{\"u}ller}, {Stephenson}, {S{\"u}veges}, {Szabados}, {Szegedi-Elek},
  {Taris}, {Tauran}, {Taylor}, {Teixeira}, {Thuillot}, {Tonello}, {Torra},
  {Torra}, {Turon}, {Unger}, {Vaillant}, {van Dillen}, {Vanel}, {Vecchiato},
  {Viala}, {Vicente}, {Voutsinas}, {Weiler}, {Wevers}, {Wyrzykowski}, {Yoldas},
  {Yvard}, {Zhao}, {Zorec}, {Zucker}, {Zurbach}, \& {Zwitter}}]{Gaia+2021}
{Gaia Collaboration}, {Brown}, A.~G.~A., {Vallenari}, A., {et~al.} 2021, \aap,
  650, C3

\bibitem[{{Gaia Collaboration} {et~al.}(2016){Gaia Collaboration}, {Prusti},
  {de Bruijne}, {Brown}, {Vallenari}, {Babusiaux}, {Bailer-Jones}, {Bastian},
  {Biermann}, {Evans}, {Eyer}, {Jansen}, {Jordi}, {Klioner}, {Lammers},
  {Lindegren}, {Luri}, {Mignard}, {Milligan}, {Panem}, {Poinsignon},
  {Pourbaix}, {Randich}, {Sarri}, {Sartoretti}, {Siddiqui}, {Soubiran},
  {Valette}, {van Leeuwen}, {Walton}, {Aerts}, {Arenou}, {Cropper}, {Drimmel},
  {H{\o}g}, {Katz}, {Lattanzi}, {O'Mullane}, {Grebel}, {Holland}, {Huc},
  {Passot}, {Bramante}, {Cacciari}, {Casta{\~n}eda}, {Chaoul}, {Cheek}, {De
  Angeli}, {Fabricius}, {Guerra}, {Hern{\'a}ndez}, {Jean-Antoine-Piccolo},
  {Masana}, {Messineo}, {Mowlavi}, {Nienartowicz}, {Ord{\'o}{\~n}ez-Blanco},
  {Panuzzo}, {Portell}, {Richards}, {Riello}, {Seabroke}, {Tanga},
  {Th{\'e}venin}, {Torra}, {Els}, {Gracia-Abril}, {Comoretto},
  {Garcia-Reinaldos}, {Lock}, {Mercier}, {Altmann}, {Andrae}, {Astraatmadja},
  {Bellas-Velidis}, {Benson}, {Berthier}, {Blomme}, {Busso}, {Carry},
  {Cellino}, {Clementini}, {Cowell}, {Creevey}, {Cuypers}, {Davidson}, {De
  Ridder}, {de Torres}, {Delchambre}, {Dell'Oro}, {Ducourant}, {Fr{\'e}mat},
  {Garc{\'\i}a-Torres}, {Gosset}, {Halbwachs}, {Hambly}, {Harrison}, {Hauser},
  {Hestroffer}, {Hodgkin}, {Huckle}, {Hutton}, {Jasniewicz}, {Jordan},
  {Kontizas}, {Korn}, {Lanzafame}, {Manteiga}, {Moitinho}, {Muinonen},
  {Osinde}, {Pancino}, {Pauwels}, {Petit}, {Recio-Blanco}, {Robin}, {Sarro},
  {Siopis}, {Smith}, {Smith}, {Sozzetti}, {Thuillot}, {van Reeven}, {Viala},
  {Abbas}, {Abreu Aramburu}, {Accart}, {Aguado}, {Allan}, {Allasia},
  {Altavilla}, {{\'A}lvarez}, {Alves}, {Anderson}, {Andrei}, {Anglada Varela},
  {Antiche}, {Antoja}, {Ant{\'o}n}, {Arcay}, {Atzei}, {Ayache}, {Bach},
  {Baker}, {Balaguer-N{\'u}{\~n}ez}, {Barache}, {Barata}, {Barbier}, {Barblan},
  {Baroni}, {Barrado y Navascu{\'e}s}, {Barros}, {Barstow}, {Becciani},
  {Bellazzini}, {Bellei}, {Bello Garc{\'\i}a}, {Belokurov}, {Bendjoya},
  {Berihuete}, {Bianchi}, {Bienaym{\'e}}, {Billebaud}, {Blagorodnova},
  {Blanco-Cuaresma}, {Boch}, {Bombrun}, {Borrachero}, {Bouquillon}, {Bourda},
  {Bouy}, {Bragaglia}, {Breddels}, {Brouillet}, {Br{\"u}semeister},
  {Bucciarelli}, {Budnik}, {Burgess}, {Burgon}, {Burlacu}, {Busonero}, {Buzzi},
  {Caffau}, {Cambras}, {Campbell}, {Cancelliere}, {Cantat-Gaudin}, {Carlucci},
  {Carrasco}, {Castellani}, {Charlot}, {Charnas}, {Charvet}, {Chassat},
  {Chiavassa}, {Clotet}, {Cocozza}, {Collins}, {Collins}, {Costigan}, {Crifo},
  {Cross}, {Crosta}, {Crowley}, {Dafonte}, {Damerdji}, {Dapergolas}, {David},
  {David}, {De Cat}, {de Felice}, {de Laverny}, {De Luise}, {De March}, {de
  Martino}, {de Souza}, {Debosscher}, {del Pozo}, {Delbo}, {Delgado},
  {Delgado}, {di Marco}, {Di Matteo}, {Diakite}, {Distefano}, {Dolding}, {Dos
  Anjos}, {Drazinos}, {Dur{\'a}n}, {Dzigan}, {Ecale}, {Edvardsson}, {Enke},
  {Erdmann}, {Escolar}, {Espina}, {Evans}, {Eynard Bontemps}, {Fabre},
  {Fabrizio}, {Faigler}, {Falc{\~a}o}, {Farr{\`a}s Casas}, {Faye}, {Federici},
  {Fedorets}, {Fern{\'a}ndez-Hern{\'a}ndez}, {Fernique}, {Fienga}, {Figueras},
  {Filippi}, {Findeisen}, {Fonti}, {Fouesneau}, {Fraile}, {Fraser}, {Fuchs},
  {Furnell}, {Gai}, {Galleti}, {Galluccio}, {Garabato}, {Garc{\'\i}a-Sedano},
  {Gar{\'e}}, {Garofalo}, {Garralda}, {Gavras}, {Gerssen}, {Geyer}, {Gilmore},
  {Girona}, {Giuffrida}, {Gomes}, {Gonz{\'a}lez-Marcos},
  {Gonz{\'a}lez-N{\'u}{\~n}ez}, {Gonz{\'a}lez-Vidal}, {Granvik}, {Guerrier},
  {Guillout}, {Guiraud}, {G{\'u}rpide}, {Guti{\'e}rrez-S{\'a}nchez}, {Guy},
  {Haigron}, {Hatzidimitriou}, {Haywood}, {Heiter}, {Helmi}, {Hobbs},
  {Hofmann}, {Holl}, {Holland}, {Hunt}, {Hypki}, {Icardi}, {Irwin}, {Jevardat
  de Fombelle}, {Jofr{\'e}}, {Jonker}, {Jorissen}, {Julbe}, {Karampelas},
  {Kochoska}, {Kohley}, {Kolenberg}, {Kontizas}, {Koposov}, {Kordopatis},
  {Koubsky}, {Kowalczyk}, {Krone-Martins}, {Kudryashova}, {Kull}, {Bachchan},
  {Lacoste-Seris}, {Lanza}, {Lavigne}, {Le Poncin-Lafitte}, {Lebreton},
  {Lebzelter}, {Leccia}, {Leclerc}, {Lecoeur-Taibi}, {Lemaitre}, {Lenhardt},
  {Leroux}, {Liao}, {Licata}, {Lindstr{\o}m}, {Lister}, {Livanou}, {Lobel},
  {L{\"o}ffler}, {L{\'o}pez}, {Lopez-Lozano}, {Lorenz}, {Loureiro},
  {MacDonald}, {Magalh{\~a}es Fernandes}, {Managau}, {Mann}, {Mantelet},
  {Marchal}, {Marchant}, {Marconi}, {Marie}, {Marinoni}, {Marrese},
  {Marschalk{\'o}}, {Marshall}, {Mart{\'\i}n-Fleitas}, {Martino}, {Mary},
  {Matijevi{\v{c}}}, {Mazeh}, {McMillan}, {Messina}, {Mestre}, {Michalik},
  {Millar}, {Miranda}, {Molina}, {Molinaro}, {Molinaro}, {Moln{\'a}r},
  {Moniez}, {Montegriffo}, {Monteiro}, {Mor}, {Mora}, {Morbidelli}, {Morel},
  {Morgenthaler}, {Morley}, {Morris}, {Mulone}, {Muraveva}, {Musella},
  {Narbonne}, {Nelemans}, {Nicastro}, {Noval}, {Ord{\'e}novic},
  {Ordieres-Mer{\'e}}, {Osborne}, {Pagani}, {Pagano}, {Pailler}, {Palacin},
  {Palaversa}, {Parsons}, {Paulsen}, {Pecoraro}, {Pedrosa}, {Pentik{\"a}inen},
  {Pereira}, {Pichon}, {Piersimoni}, {Pineau}, {Plachy}, {Plum}, {Poujoulet},
  {Pr{\v{s}}a}, {Pulone}, {Ragaini}, {Rago}, {Rambaux}, {Ramos-Lerate},
  {Ranalli}, {Rauw}, {Read}, {Regibo}, {Renk}, {Reyl{\'e}}, {Ribeiro},
  {Rimoldini}, {Ripepi}, {Riva}, {Rixon}, {Roelens}, {Romero-G{\'o}mez},
  {Rowell}, {Royer}, {Rudolph}, {Ruiz-Dern}, {Sadowski}, {Sagrist{\`a}
  Sell{\'e}s}, {Sahlmann}, {Salgado}, {Salguero}, {Sarasso}, {Savietto},
  {Schnorhk}, {Schultheis}, {Sciacca}, {Segol}, {Segovia}, {Segransan},
  {Serpell}, {Shih}, {Smareglia}, {Smart}, {Smith}, {Solano}, {Solitro},
  {Sordo}, {Soria Nieto}, {Souchay}, {Spagna}, {Spoto}, {Stampa}, {Steele},
  {Steidelm{\"u}ller}, {Stephenson}, {Stoev}, {Suess}, {S{\"u}veges}, {Surdej},
  {Szabados}, {Szegedi-Elek}, {Tapiador}, {Taris}, {Tauran}, {Taylor},
  {Teixeira}, {Terrett}, {Tingley}, {Trager}, {Turon}, {Ulla}, {Utrilla},
  {Valentini}, {van Elteren}, {Van Hemelryck}, {van Leeuwen}, {Varadi},
  {Vecchiato}, {Veljanoski}, {Via}, {Vicente}, {Vogt}, {Voss}, {Votruba},
  {Voutsinas}, {Walmsley}, {Weiler}, {Weingrill}, {Werner}, {Wevers},
  {Whitehead}, {Wyrzykowski}, {Yoldas}, {{\v{Z}}erjal}, {Zucker}, {Zurbach},
  {Zwitter}, {Alecu}, {Allen}, {Allende Prieto}, {Amorim},
  {Anglada-Escud{\'e}}, {Arsenijevic}, {Azaz}, {Balm}, {Beck}, {Bernstein},
  {Bigot}, {Bijaoui}, {Blasco}, {Bonfigli}, {Bono}, {Boudreault}, {Bressan},
  {Brown}, {Brunet}, {Bunclark}, {Buonanno}, {Butkevich}, {Carret}, {Carrion},
  {Chemin}, {Ch{\'e}reau}, {Corcione}, {Darmigny}, {de Boer}, {de Teodoro}, {de
  Zeeuw}, {Delle Luche}, {Domingues}, {Dubath}, {Fodor}, {Fr{\'e}zouls},
  {Fries}, {Fustes}, {Fyfe}, {Gallardo}, {Gallegos}, {Gardiol}, {Gebran},
  {Gomboc}, {G{\'o}mez}, {Grux}, {Gueguen}, {Heyrovsky}, {Hoar}, {Iannicola},
  {Isasi Parache}, {Janotto}, {Joliet}, {Jonckheere}, {Keil}, {Kim},
  {Klagyivik}, {Klar}, {Knude}, {Kochukhov}, {Kolka}, {Kos}, {Kutka}, {Lainey},
  {LeBouquin}, {Liu}, {Loreggia}, {Makarov}, {Marseille}, {Martayan},
  {Martinez-Rubi}, {Massart}, {Meynadier}, {Mignot}, {Munari}, {Nguyen},
  {Nordlander}, {Ocvirk}, {O'Flaherty}, {Olias Sanz}, {Ortiz}, {Osorio},
  {Oszkiewicz}, {Ouzounis}, {Palmer}, {Park}, {Pasquato}, {Peltzer}, {Peralta},
  {P{\'e}turaud}, {Pieniluoma}, {Pigozzi}, {Poels}, {Prat}, {Prod'homme},
  {Raison}, {Rebordao}, {Risquez}, {Rocca-Volmerange}, {Rosen}, {Ruiz-Fuertes},
  {Russo}, {Sembay}, {Serraller Vizcaino}, {Short}, {Siebert}, {Silva},
  {Sinachopoulos}, {Slezak}, {Soffel}, {Sosnowska}, {Strai{\v{z}}ys}, {ter
  Linden}, {Terrell}, {Theil}, {Tiede}, {Troisi}, {Tsalmantza}, {Tur},
  {Vaccari}, {Vachier}, {Valles}, {Van Hamme}, {Veltz}, {Virtanen}, {Wallut},
  {Wichmann}, {Wilkinson}, {Ziaeepour}, \& {Zschocke}}]{Gaia+2016}
{Gaia Collaboration}, {Prusti}, T., {de Bruijne}, J.~H.~J., {et~al.} 2016,
  \aap, 595, A1

\bibitem[{{Gaia Collaboration} {et~al.}(2022){Gaia Collaboration}, {Vallenari},
  {Brown}, {Prusti}, {de Bruijne}, {Arenou}, {Babusiaux}, {Biermann},
  {Creevey}, {Ducourant}, {Evans}, {Eyer}, {Guerra}, {Hutton}, {Jordi},
  {Klioner}, {Lammers}, {Lindegren}, {Luri}, {Mignard}, {Panem}, {Pourbaix},
  {Randich}, {Sartoretti}, {Soubiran}, {Tanga}, {Walton}, {Bailer-Jones},
  {Bastian}, {Drimmel}, {Jansen}, {Katz}, {Lattanzi}, {van Leeuwen}, {Bakker},
  {Cacciari}, {Casta{\~n}eda}, {De Angeli}, {Fabricius}, {Fouesneau},
  {Fr{\'e}mat}, {Galluccio}, {Guerrier}, {Heiter}, {Masana}, {Messineo},
  {Mowlavi}, {Nicolas}, {Nienartowicz}, {Pailler}, {Panuzzo}, {Riclet}, {Roux},
  {Seabroke}, {Sordo{\o}rcit}, {Th{\'e}venin}, {Gracia-Abril}, {Portell},
  {Teyssier}, {Altmann}, {Andrae}, {Audard}, {Bellas-Velidis}, {Benson},
  {Berthier}, {Blomme}, {Burgess}, {Busonero}, {Busso}, {C{\'a}novas}, {Carry},
  {Cellino}, {Cheek}, {Clementini}, {Damerdji}, {Davidson}, {de Teodoro},
  {Nu{\~n}ez Campos}, {Delchambre}, {Dell'Oro}, {Esquej},
  {Fern{\'a}ndez-Hern{\'a}ndez}, {Fraile}, {Garabato}, {Garc{\'\i}a-Lario},
  {Gosset}, {Haigron}, {Halbwachs}, {Hambly}, {Harrison}, {Hern{\'a}ndez},
  {Hestroffer}, {Hodgkin}, {Holl}, {Jan{\ss}en}, {Jevardat de Fombelle},
  {Jordan}, {Krone-Martins}, {Lanzafame}, {L{\"o}ffler}, {Marchal}, {Marrese},
  {Moitinho}, {Muinonen}, {Osborne}, {Pancino}, {Pauwels}, {Recio-Blanco},
  {Reyl{\'e}}, {Riello}, {Rimoldini}, {Roegiers}, {Rybizki}, {Sarro}, {Siopis},
  {Smith}, {Sozzetti}, {Utrilla}, {van Leeuwen}, {Abbas}, {{\'A}brah{\'a}m},
  {Abreu Aramburu}, {Aerts}, {Aguado}, {Ajaj}, {Aldea-Montero}, {Altavilla},
  {{\'A}lvarez}, {Alves}, {Anders}, {Anderson}, {Anglada Varela}, {Antoja},
  {Baines}, {Baker}, {Balaguer-N{\'u}{\~n}ez}, {Balbinot}, {Balog}, {Barache},
  {Barbato}, {Barros}, {Barstow}, {Bartolom{\'e}}, {Bassilana}, {Bauchet},
  {Becciani}, {Bellazzini}, {Berihuete}, {Bernet}, {Bertone}, {Bianchi},
  {Binnenfeld}, {Blanco-Cuaresma}, {Blazere}, {Boch}, {Bombrun}, {Bossini},
  {Bouquillon}, {Bragaglia}, {Bramante}, {Breedt}, {Bressan}, {Brouillet},
  {Brugaletta}, {Bucciarelli}, {Burlacu}, {Butkevich}, {Buzzi}, {Caffau},
  {Cancelliere}, {Cantat-Gaudin}, {Carballo}, {Carlucci}, {Carnerero},
  {Carrasco}, {Casamiquela}, {Castellani}, {Castro-Ginard}, {Chaoul},
  {Charlot}, {Chemin}, {Chiaramida}, {Chiavassa}, {Chornay}, {Comoretto},
  {Contursi}, {Cooper}, {Cornez}, {Cowell}, {Crifo}, {Cropper}, {Crosta},
  {Crowley}, {Dafonte}, {Dapergolas}, {David}, {David}, {de Laverny}, {De
  Luise}, {De March}, {De Ridder}, {de Souza}, {de Torres}, {del Peloso}, {del
  Pozo}, {Delbo}, {Delgado}, {Delisle}, {Demouchy}, {Dharmawardena}, {Di
  Matteo}, {Diakite}, {Diener}, {Distefano}, {Dolding}, {Edvardsson}, {Enke},
  {Fabre}, {Fabrizio}, {Faigler}, {Fedorets}, {Fernique}, {Fienga}, {Figueras},
  {Fournier}, {Fouron}, {Fragkoudi}, {Gai}, {Garcia-Gutierrez},
  {Garcia-Reinaldos}, {Garc{\'\i}a-Torres}, {Garofalo}, {Gavel}, {Gavras},
  {Gerlach}, {Geyer}, {Giacobbe}, {Gilmore}, {Girona}, {Giuffrida}, {Gomel},
  {Gomez}, {Gonz{\'a}lez-N{\'u}{\~n}ez}, {Gonz{\'a}lez-Santamar{\'\i}a},
  {Gonz{\'a}lez-Vidal}, {Granvik}, {Guillout}, {Guiraud},
  {Guti{\'e}rrez-S{\'a}nchez}, {Guy}, {Hatzidimitriou}, {Hauser}, {Haywood},
  {Helmer}, {Helmi}, {Sarmiento}, {Hidalgo}, {Hilger}, {H{\l}adczuk}, {Hobbs},
  {Holland}, {Huckle}, {Jardine}, {Jasniewicz}, {Jean-Antoine Piccolo},
  {Jim{\'e}nez-Arranz}, {Jorissen}, {Juaristi Campillo}, {Julbe}, {Karbevska},
  {Kervella}, {Khanna}, {Kontizas}, {Kordopatis}, {Korn}, {K{\'o}sp{\'a}l},
  {Kostrzewa-Rutkowska}, {Kruszy{\'n}ska}, {Kun}, {Laizeau}, {Lambert},
  {Lanza}, {Lasne}, {Le Campion}, {Lebreton}, {Lebzelter}, {Leccia}, {Leclerc},
  {Lecoeur-Taibi}, {Liao}, {Licata}, {Lindstr{\o}m}, {Lister}, {Livanou},
  {Lobel}, {Lorca}, {Loup}, {Madrero Pardo}, {Magdaleno Romeo}, {Managau},
  {Mann}, {Manteiga}, {Marchant}, {Marconi}, {Marcos}, {Marcos Santos},
  {Mar{\'\i}n Pina}, {Marinoni}, {Marocco}, {Marshall}, {Polo},
  {Mart{\'\i}n-Fleitas}, {Marton}, {Mary}, {Masip}, {Massari},
  {Mastrobuono-Battisti}, {Mazeh}, {McMillan}, {Messina}, {Michalik}, {Millar},
  {Mints}, {Molina}, {Molinaro}, {Moln{\'a}r}, {Monari}, {Mongui{\'o}},
  {Montegriffo}, {Montero}, {Mor}, {Mora}, {Morbidelli}, {Morel}, {Morris},
  {Muraveva}, {Murphy}, {Musella}, {Nagy}, {Noval}, {Oca{\~n}a}, {Ogden},
  {Ordenovic}, {Osinde}, {Pagani}, {Pagano}, {Palaversa}, {Palicio},
  {Pallas-Quintela}, {Panahi}, {Payne-Wardenaar}, {Pe{\~n}alosa Esteller},
  {Penttil{\"a}}, {Pichon}, {Piersimoni}, {Pineau}, {Plachy}, {Plum}, {Poggio},
  {Pr{\v{s}}a}, {Pulone}, {Racero}, {Ragaini}, {Rainer}, {Raiteri}, {Rambaux},
  {Ramos}, {Ramos-Lerate}, {Re Fiorentin}, {Regibo}, {Richards}, {Rios Diaz},
  {Ripepi}, {Riva}, {Rix}, {Rixon}, {Robichon}, {Robin}, {Robin}, {Roelens},
  {Rogues}, {Rohrbasser}, {Romero-G{\'o}mez}, {Rowell}, {Royer}, {Ruz Mieres},
  {Rybicki}, {Sadowski}, {S{\'a}ez N{\'u}{\~n}ez}, {Sagrist{\`a} Sell{\'e}s},
  {Sahlmann}, {Salguero}, {Samaras}, {Sanchez Gimenez}, {Sanna},
  {Santove{\~n}a}, {Sarasso}, {Schultheis}, {Sciacca}, {Segol}, {Segovia},
  {S{\'e}gransan}, {Semeux}, {Shahaf}, {Siddiqui}, {Siebert}, {Siltala},
  {Silvelo}, {Slezak}, {Slezak}, {Smart}, {Snaith}, {Solano}, {Solitro},
  {Souami}, {Souchay}, {Spagna}, {Spina}, {Spoto}, {Steele},
  {Steidelm{\"u}ller}, {Stephenson}, {S{\"u}veges}, {Surdej}, {Szabados},
  {Szegedi-Elek}, {Taris}, {Taylo}, {Teixeira}, {Tolomei}, {Tonello}, {Torra},
  {Torra}, {Torralba Elipe}, {Trabucchi}, {Tsounis}, {Turon}, {Ulla}, {Unger},
  {Vaillant}, {van Dillen}, {van Reeven}, {Vanel}, {Vecchiato}, {Viala},
  {Vicente}, {Voutsinas}, {Weiler}, {Wevers}, {Wyrzykowski}, {Yoldas}, {Yvard},
  {Zhao}, {Zorec}, {Zucker}, \& {Zwitter}}]{Gaia2022}
{Gaia Collaboration}, {Vallenari}, A., {Brown}, A.~G.~A., {et~al.} 2022, arXiv
  e-prints, arXiv:2208.00211

\bibitem[{{Geier} {et~al.}(2019){Geier}, {Raddi}, {Gentile Fusillo}, \&
  {Marsh}}]{Geier+2019}
{Geier}, S., {Raddi}, R., {Gentile Fusillo}, N.~P., \& {Marsh}, T.~R. 2019,
  \aap, 621, A38

\bibitem[{{Gentile Fusillo} {et~al.}(2021){Gentile Fusillo}, {Tremblay},
  {Cukanovaite}, {Vorontseva}, {Lallement}, {Hollands}, {G{\"a}nsicke},
  {Burdge}, {McCleery}, \& {Jordan}}]{GentileFusillo+2021}
{Gentile Fusillo}, N.~P., {Tremblay}, P.~E., {Cukanovaite}, E., {et~al.} 2021,
  \mnras, 508, 3877

\bibitem[{{Gentile Fusillo} {et~al.}(2019){Gentile Fusillo}, {Tremblay},
  {G{\"a}nsicke}, {Manser}, {Cunningham}, {Cukanovaite}, {Hollands}, {Marsh},
  {Raddi}, {Jordan}, {Toonen}, {Geier}, {Barstow}, \&
  {Cummings}}]{GentileFusillo+2019}
{Gentile Fusillo}, N.~P., {Tremblay}, P.-E., {G{\"a}nsicke}, B.~T., {et~al.}
  2019, \mnras, 482, 4570

\bibitem[{{Gianninas} {et~al.}(2010){Gianninas}, {Bergeron}, {Dupuis}, \&
  {Ruiz}}]{Gianninas+2010}
{Gianninas}, A., {Bergeron}, P., {Dupuis}, J., \& {Ruiz}, M.~T. 2010, \apj,
  720, 581

\bibitem[{{Gianninas} {et~al.}(2011){Gianninas}, {Bergeron}, \&
  {Ruiz}}]{Gianninas+2011}
{Gianninas}, A., {Bergeron}, P., \& {Ruiz}, M.~T. 2011, \apj, 743, 138

\bibitem[{{Griem}(1974)}]{Griem1974}
{Griem}, H.~R. 1974, {Spectral line broadening by plasmas} (New York, Academic
  Press, Inc.~Pure and Applied Physics.~Volume 39, 1974.~421)

\bibitem[{{Hall} {et~al.}(2013){Hall}, {Tout}, {Izzard}, \&
  {Keller}}]{Hall+2013}
{Hall}, P.~D., {Tout}, C.~A., {Izzard}, R.~G., \& {Keller}, D. 2013, \mnras,
  435, 2048

\bibitem[{{Hartman} \& {Bakos}(2016)}]{HartmanBakos2016}
{Hartman}, J.~D. \& {Bakos}, G.~{\'A}. 2016, Astronomy and Computing, 17, 1

\bibitem[{{Heber} \& {Hunger}(1987)}]{HeberHunger1987}
{Heber}, U. \& {Hunger}, K. 1987, The Messenger, 47, 36

\bibitem[{{Heber} {et~al.}(2018){Heber}, {Irrgang}, \&
  {Schaffenroth}}]{Heber+2018}
{Heber}, U., {Irrgang}, A., \& {Schaffenroth}, J. 2018, Open Astronomy, 27, 35

\bibitem[{{Heber} \& {Kudritzki}(1986)}]{Heber1986}
{Heber}, U. \& {Kudritzki}, R.~P. 1986, \aap, 169, 244

\bibitem[{{Heinze} {et~al.}(2018){Heinze}, {Tonry}, {Denneau}, {Flewelling},
  {Stalder}, {Rest}, {Smith}, {Smartt}, \& {Weiland}}]{Heinze+2018}
{Heinze}, A.~N., {Tonry}, J.~L., {Denneau}, L., {et~al.} 2018, \aj, 156, 241

\bibitem[{{Henden} {et~al.}(2015){Henden}, {Levine}, {Terrell}, \&
  {Welch}}]{Henden+2015}
{Henden}, A.~A., {Levine}, S., {Terrell}, D., \& {Welch}, D.~L. 2015, in
  American Astronomical Society Meeting Abstracts, Vol. 225, American
  Astronomical Society Meeting Abstracts \#225, 336.16

\bibitem[{{Hermes} {et~al.}(2017){Hermes}, {G{\"a}nsicke}, {Kawaler}, {Greiss},
  {Tremblay}, {Gentile Fusillo}, {Raddi}, {Fanale}, {Bell}, {Dennihy}, {Fuchs},
  {Dunlap}, {Clemens}, {Montgomery}, {Winget}, {Chote}, {Marsh}, \&
  {Redfield}}]{Hermes+2017}
{Hermes}, J.~J., {G{\"a}nsicke}, B.~T., {Kawaler}, S.~D., {et~al.} 2017, \apjs,
  232, 23

\bibitem[{{Herrero} {et~al.}(2020){Herrero}, {Parthasarathy},
  {Sim{\'o}n-D{\'\i}az}, {Hubrig}, {Sarkar}, \& {Muneer}}]{Herrero+2020}
{Herrero}, A., {Parthasarathy}, M., {Sim{\'o}n-D{\'\i}az}, S., {et~al.} 2020,
  \mnras, 494, 2117

\bibitem[{{Higgins} \& {Bell}(2023)}]{Higgins+2023}
{Higgins}, M.~E. \& {Bell}, K.~J. 2023, \aj, 165, 141

\bibitem[{{Hillwig} {et~al.}(2017){Hillwig}, {Frew}, {Reindl}, {Rotter},
  {Webb}, \& {Margheim}}]{Hillwig+2017}
{Hillwig}, T.~C., {Frew}, D.~J., {Reindl}, N., {et~al.} 2017, \aj, 153, 24

\bibitem[{{Hu} {et~al.}(2020){Hu}, {Webb}, {Ayres}, {Bainbridge}, {Barrow},
  {Barstow}, {Berengut}, {Carswell}, {Dumont}, {Dzuba}, {Flambaum}, {Lee},
  {Reindl}, {Preval}, \& {Tchang-Brillet}}]{Hu+2020}
{Hu}, J., {Webb}, J.~K., {Ayres}, T.~R., {et~al.} 2020, arXiv e-prints,
  arXiv:2007.10905

\bibitem[{{H{\"u}mmerich} {et~al.}(2016){H{\"u}mmerich}, {Paunzen}, \&
  {Bernhard}}]{Huemmerich+2016}
{H{\"u}mmerich}, S., {Paunzen}, E., \& {Bernhard}, K. 2016, \aj, 152, 104

\bibitem[{{Irrgang} {et~al.}(2021){Irrgang}, {Geier}, {Heber}, {Kupfer},
  {El-Badry}, \& {Bloemen}}]{Irrgang+2021}
{Irrgang}, A., {Geier}, S., {Heber}, U., {et~al.} 2021, \aap, 650, A102

\bibitem[{{Isern} {et~al.}(2008){Isern}, {Garc{\'{\i}}a-Berro}, {Torres}, \&
  {Catal{\'a}n}}]{Isern2008}
{Isern}, J., {Garc{\'{\i}}a-Berro}, E., {Torres}, S., \& {Catal{\'a}n}, S.
  2008, \apjl, 682, L109

\bibitem[{{Isern} {et~al.}(2022){Isern}, {Torres}, \&
  {Rebassa-Mansergas}}]{Isern+2022}
{Isern}, J., {Torres}, S., \& {Rebassa-Mansergas}, A. 2022, Frontiers in
  Astronomy and Space Sciences, 9, 6

\bibitem[{{Jagelka} {et~al.}(2019){Jagelka}, {Mikul{\'a}{\v{s}}ek},
  {H{\"u}mmerich}, \& {Paunzen}}]{Jagelka+2019}
{Jagelka}, M., {Mikul{\'a}{\v{s}}ek}, Z., {H{\"u}mmerich}, S., \& {Paunzen}, E.
  2019, \aap, 622, A199

\bibitem[{{Jeffery} {et~al.}(2023){Jeffery}, {Werner}, {Kilkenny}, {Miszalski},
  {Monageng}, \& {Snowdon}}]{Jeffery+2023}
{Jeffery}, C.~S., {Werner}, K., {Kilkenny}, D., {et~al.} 2023, \mnras, 519,
  2321

\bibitem[{{Jenkins} {et~al.}(2016){Jenkins}, {Twicken}, {McCauliff},
  {Campbell}, {Sanderfer}, {Lung}, {Mansouri-Samani}, {Girouard}, {Tenenbaum},
  {Klaus}, {Smith}, {Caldwell}, {Chacon}, {Henze}, {Heiges}, {Latham},
  {Morgan}, {Swade}, {Rinehart}, \& {Vanderspek}}]{Jenkins+2016}
{Jenkins}, J.~M., {Twicken}, J.~D., {McCauliff}, S., {et~al.} 2016, in Society
  of Photo-Optical Instrumentation Engineers (SPIE) Conference Series, Vol.
  9913, Software and Cyberinfrastructure for Astronomy IV, ed. G.~{Chiozzi} \&
  J.~C. {Guzman}, 99133E

\bibitem[{{Jones}(2019)}]{Jones2019}
{Jones}, D. 2019, in Highlights on Spanish Astrophysics X, ed. B.~{Montesinos},
  A.~{Asensio Ramos}, F.~{Buitrago}, R.~{Sch{\"o}del}, E.~{Villaver},
  S.~{P{\'e}rez-Hoyos}, \& I.~{Ord{\'o}{\~n}ez-Etxeberria}, 340--345

\bibitem[{{Jones} {et~al.}(2020){Jones}, {Boffin}, {Hibbert}, {Steinmetz},
  {Wesson}, {Hillwig}, {Sowicka}, {Corradi}, {Garc{\'\i}a-Rojas},
  {Rodr{\'\i}guez-Gil}, \& {Munday}}]{Jones+2020}
{Jones}, D., {Boffin}, H.~M.~J., {Hibbert}, J., {et~al.} 2020, \aap, 642, A108

\bibitem[{{Jones} {et~al.}(2022){Jones}, {Munday}, {Corradi},
  {Rodr{\'\i}guez-Gil}, {Boffin}, {Zak}, {Sowicka}, {Parsons}, {Dhillon},
  {Littlefair}, {Marsh}, {Reindl}, \& {Garc{\'\i}a-Rojas}}]{Jones+2022}
{Jones}, D., {Munday}, J., {Corradi}, R. L.~M., {et~al.} 2022, \mnras, 510,
  3102

\bibitem[{{Justham} {et~al.}(2011){Justham}, {Podsiadlowski}, \&
  {Han}}]{Justham+2011}
{Justham}, S., {Podsiadlowski}, P., \& {Han}, Z. 2011, \mnras, 410, 984

\bibitem[{{Kalirai}(2012)}]{Kalirai2012}
{Kalirai}, J.~S. 2012, \nat, 486, 90

\bibitem[{{Kilic} {et~al.}(2019){Kilic}, {Bergeron}, {Dame}, {Hambly},
  {Rowell}, \& {Crawford}}]{Kilic+2019}
{Kilic}, M., {Bergeron}, P., {Dame}, K., {et~al.} 2019, \mnras, 482, 965

\bibitem[{{Kilic} {et~al.}(2015){Kilic}, {Gianninas}, {Bell}, {Curd}, {Brown},
  {Hermes}, {Dufour}, {Wisniewski}, {Winget}, \& {Winget}}]{Kilic+2015}
{Kilic}, M., {Gianninas}, A., {Bell}, K.~J., {et~al.} 2015, \apjl, 814, L31

\bibitem[{{Krti{\v{c}}ka} {et~al.}(2020){Krti{\v{c}}ka}, {Kawka},
  {Mikul{\'a}{\v{s}}ek}, {Fossati}, {Krti{\v{c}}kov{\'a}}, {Prv{\'a}k},
  {Jan{\'\i}k}, {Skarka}, \& {Liptaj}}]{Krticka+2020b}
{Krti{\v{c}}ka}, J., {Kawka}, A., {Mikul{\'a}{\v{s}}ek}, Z., {et~al.} 2020,
  \aap, 639, A8

\bibitem[{{Krzesi{\'n}ski} {et~al.}(2009){Krzesi{\'n}ski}, {Kleinman}, {Nitta},
  {H{\"u}gelmeyer}, {Dreizler}, {Liebert}, \& {Harris}}]{Krzesinski2009}
{Krzesi{\'n}ski}, J., {Kleinman}, S.~J., {Nitta}, A., {et~al.} 2009, \aap, 508,
  339

\bibitem[{{Latour} {et~al.}(2018){Latour}, {Chayer}, {Green}, {Irrgang}, \&
  {Fontaine}}]{Latour+2018}
{Latour}, M., {Chayer}, P., {Green}, E.~M., {Irrgang}, A., \& {Fontaine}, G.
  2018, \aap, 609, A89

\bibitem[{{Le D{\^u}} {et~al.}(2022){Le D{\^u}}, {Mulato}, {Parker}, {Petit},
  {Ritter}, {Drechsler}, {Strottner}, {Patchick}, {Prestgard}, {Garde},
  {Outters}, \& {Raffaelli}}]{LeDu+2022}
{Le D{\^u}}, P., {Mulato}, L., {Parker}, Q.~A., {et~al.} 2022, \aap, 666, A152

\bibitem[{{Lindegren} {et~al.}(2021){Lindegren}, {Bastian}, {Biermann},
  {Bombrun}, {de Torres}, {Gerlach}, {Geyer}, {Hern{\'a}ndez}, {Hilger},
  {Hobbs}, {Klioner}, {Lammers}, {McMillan}, {Ramos-Lerate},
  {Steidelm{\"u}ller}, {Stephenson}, \& {van Leeuwen}}]{Lindegren+2021}
{Lindegren}, L., {Bastian}, U., {Biermann}, M., {et~al.} 2021, \aap, 649, A4

\bibitem[{{L{\"o}bling} {et~al.}(2020){L{\"o}bling}, {Maney}, {Rauch},
  {Quinet}, {Gamrath}, {Kruk}, \& {Werner}}]{Loebling+2020}
{L{\"o}bling}, L., {Maney}, M.~A., {Rauch}, T., {et~al.} 2020, \mnras, 492, 528

\bibitem[{{Manseau} {et~al.}(2016){Manseau}, {Bergeron}, \&
  {Green}}]{Manseau+2016}
{Manseau}, P.~M., {Bergeron}, P., \& {Green}, E.~M. 2016, \apj, 833, 127

\bibitem[{{Masci} {et~al.}(2019){Masci}, {Laher}, {Rusholme}, {Shupe}, {Groom},
  {Surace}, {Jackson}, {Monkewitz}, {Beck}, {Flynn}, {Terek}, {Landry},
  {Hacopians}, {Desai}, {Howell}, {Brooke}, {Imel}, {Wachter}, {Ye}, {Lin},
  {Cenko}, {Cunningham}, {Rebbapragada}, {Bue}, {Miller}, {Mahabal}, {Bellm},
  {Patterson}, {Juri{\'c}}, {Golkhou}, {Ofek}, {Walters}, {Graham}, {Kasliwal},
  {Dekany}, {Kupfer}, {Burdge}, {Cannella}, {Barlow}, {Van Sistine}, {Giomi},
  {Fremling}, {Blagorodnova}, {Levitan}, {Riddle}, {Smith}, {Helou}, {Prince},
  \& {Kulkarni}}]{Masci+2019}
{Masci}, F.~J., {Laher}, R.~R., {Rusholme}, B., {et~al.} 2019, \pasp, 131,
  018003

\bibitem[{{Michaud}(1970)}]{Michaud1970}
{Michaud}, G. 1970, \apj, 160, 641

\bibitem[{{Miller Bertolami}(2014)}]{MillerBertolami2014b}
{Miller Bertolami}, M.~M. 2014, \aap, 562, A123

\bibitem[{{Miller Bertolami}(2016)}]{MillerBertolami2016}
{Miller Bertolami}, M.~M. 2016, \aap, 588, A25

\bibitem[{{Miller Bertolami} {et~al.}(2014){Miller Bertolami}, {Melendez},
  {Althaus}, \& {Isern}}]{MillerBertolami2014a}
{Miller Bertolami}, M.~M., {Melendez}, B.~E., {Althaus}, L.~G., \& {Isern}, J.
  2014, ArXiv e-prints 1406.7712 [\eprint[arXiv]{1406.7712}]

\bibitem[{{Moe} \& {De Marco}(2006)}]{Moe+2006}
{Moe}, M. \& {De Marco}, O. 2006, \apj, 650, 916

\bibitem[{{Moe} \& {De Marco}(2012)}]{Moe+2012}
{Moe}, M. \& {De Marco}, O. 2012, IAU Symposium, 283, 111

\bibitem[{{Moehler} {et~al.}(2019){Moehler}, {Landsman}, {Lanz}, \& {Miller
  Bertolami}}]{Moehler+2019}
{Moehler}, S., {Landsman}, W.~B., {Lanz}, T., \& {Miller Bertolami}, M.~M.
  2019, \aap, 627, A34

\bibitem[{{Momany} {et~al.}(2020){Momany}, {Zaggia}, {Montalto}, {Jones},
  {Boffin}, {Cassisi}, {Moni Bidin}, {Gullieuszik}, {Saviane}, {Monaco},
  {Mason}, {Girardi}, {D'Orazi}, {Piotto}, {Milone}, {Lala}, {Stetson}, \&
  {Beletsky}}]{Momany+2020}
{Momany}, Y., {Zaggia}, S., {Montalto}, M., {et~al.} 2020, Nature Astronomy, 4,
  1092

\bibitem[{{Napiwotzki}(1999)}]{Napiwotzki1999}
{Napiwotzki}, R. 1999, \aap, 350, 101

\bibitem[{{Pelisoli} {et~al.}(2022){Pelisoli}, {Dorsch}, {Heber},
  {G{\"a}nsicke}, {Geier}, {Kupfer}, {N{\'e}meth}, {Scaringi}, \&
  {Schaffenroth}}]{Pelisoli+2022}
{Pelisoli}, I., {Dorsch}, M., {Heber}, U., {et~al.} 2022, \mnras, 515, 2496

\bibitem[{{P{\'e}rez-Fern{\'a}ndez} {et~al.}(2016){P{\'e}rez-Fern{\'a}ndez},
  {Ulla}, {Solano}, {Oreiro}, \& {Rodrigo}}]{Perez-Fernandez+2016}
{P{\'e}rez-Fern{\'a}ndez}, E., {Ulla}, A., {Solano}, E., {Oreiro}, R., \&
  {Rodrigo}, C. 2016, \mnras, 457, 3396

\bibitem[{{Pogge}(2019)}]{Pogge2019}
{Pogge}, R. 2019, {rwpogge/modsCCDRed: v2.0.1}, Zenodo

\bibitem[{{Pogge} {et~al.}(2010){Pogge}, {Atwood}, {Brewer}, {Byard},
  {Derwent}, {Gonzalez}, {Martini}, {Mason}, {O'Brien}, {Osmer}, {Pappalardo},
  {Steinbrecher}, {Teiga}, \& {Zhelem}}]{Pogge+2010}
{Pogge}, R.~W., {Atwood}, B., {Brewer}, D.~F., {et~al.} 2010, in Society of
  Photo-Optical Instrumentation Engineers (SPIE) Conference Series, Vol. 7735,
  Ground-based and Airborne Instrumentation for Astronomy III, ed. I.~S.
  {McLean}, S.~K. {Ramsay}, \& H.~{Takami}, 77350A

\bibitem[{{Press} {et~al.}(1992){Press}, {Teukolsky}, {Vetterling}, \&
  {Flannery}}]{Press1992}
{Press}, W.~H., {Teukolsky}, S.~A., {Vetterling}, W.~T., \& {Flannery}, B.~P.
  1992, {Numerical recipes in C. The art of scientific computing} (New York:
  Cambridge University Press)

\bibitem[{{Preval} {et~al.}(2019){Preval}, {Barstow}, {Bainbridge}, {Reindl},
  {Ayres}, {Holberg}, {Barrow}, {Lee}, {Webb}, \& {Hu}}]{Preval+2019}
{Preval}, S.~P., {Barstow}, M.~A., {Bainbridge}, M., {et~al.} 2019, \mnras,
  487, 3470

\bibitem[{{Provencal} {et~al.}(2012){Provencal}, {Montgomery}, {Kanaan},
  {Thompson}, {Dalessio}, {Shipman}, {Childers}, {Clemens}, {Rosen},
  {Henrique}, {Bischoff-Kim}, {Strickland}, {Chandler}, {Walter}, {Watson},
  {Castanheira}, {Wang}, {Handler}, {Wood}, {Vennes}, {Nemeth}, {Kepler},
  {Reed}, {Nitta}, {Kleinman}, {Brown}, {Kim}, {Sullivan}, {Chen}, {Yang},
  {Shih}, {Jiang}, {Sergeev}, {Maksim}, {Janulis}, {Baliyan}, {Vats}, {Zola},
  {Baran}, {Winiarski}, {Ogloza}, {Paparo}, {Bognar}, {Papics}, {Kilkenny},
  {Sefako}, {Buckley}, {Loaring}, {Kniazev}, {Silvotti}, {Galleti}, {Nagel},
  {Vauclair}, {Dolez}, {Fremy}, {Perez}, {Almenara}, \&
  {Fraga}}]{Provencal+2012}
{Provencal}, J.~L., {Montgomery}, M.~H., {Kanaan}, A., {et~al.} 2012, \apj,
  751, 91

\bibitem[{{Pych}(2004)}]{wojtek2004}
{Pych}, W. 2004, \pasp, 116, 148

\bibitem[{{Rauch} \& {Deetjen}(2003)}]{rauchdeetjen2003}
{Rauch}, T. \& {Deetjen}, J.~L. 2003, in Astronomical Society of the Pacific
  Conference Series, Vol. 288, Stellar Atmosphere Modeling, ed. I.~{Hubeny},
  D.~{Mihalas}, \& K.~{Werner}, 103

\bibitem[{{Rauch} {et~al.}(2015{\natexlab{a}}){Rauch}, {Hoyer}, {Quinet},
  {Gallardo}, \& {Raineri}}]{Rauch+2015b}
{Rauch}, T., {Hoyer}, D., {Quinet}, P., {Gallardo}, M., \& {Raineri}, M.
  2015{\natexlab{a}}, \aap, 577, A88

\bibitem[{{Rauch} {et~al.}(2015{\natexlab{b}}){Rauch}, {Werner}, {Quinet}, \&
  {Kruk}}]{Rauch+2015a}
{Rauch}, T., {Werner}, K., {Quinet}, P., \& {Kruk}, J.~W. 2015{\natexlab{b}},
  \aap, 577, A6

\bibitem[{{Rauch} {et~al.}(2007{\natexlab{a}}){Rauch}, {Ziegler}, {Werner},
  {Kruk}, {Oliveira}, {Vande Putte}, {Mignani}, \& {Kerber}}]{Rauch2007}
{Rauch}, T., {Ziegler}, M., {Werner}, K., {et~al.} 2007{\natexlab{a}}, \aap,
  470, 317

\bibitem[{{Rauch} {et~al.}(2007{\natexlab{b}}){Rauch}, {Ziegler}, {Werner},
  {Kruk}, {Oliveira}, {Vande Putte}, {Mignani}, \& {Kerber}}]{Rauch+2007}
{Rauch}, T., {Ziegler}, M., {Werner}, K., {et~al.} 2007{\natexlab{b}}, \aap,
  470, 317

\bibitem[{{Reindl} {et~al.}(2019){Reindl}, {Bainbridge}, {Przybilla}, {Geier},
  {Prv{\'a}k}, {Krti{\v{c}}ka}, {{\O}stensen}, {Telting}, \&
  {Werner}}]{Reindl+2019}
{Reindl}, N., {Bainbridge}, M., {Przybilla}, N., {et~al.} 2019, \mnras, 482,
  L93

\bibitem[{{Reindl} {et~al.}(2016){Reindl}, {Geier}, {Kupfer}, {Bloemen},
  {Schaffenroth}, {Heber}, {Barlow}, \& {{\O}stensen}}]{Reindl+2016}
{Reindl}, N., {Geier}, S., {Kupfer}, T., {et~al.} 2016, \aap, 587, A101

\bibitem[{{Reindl} {et~al.}(2018){Reindl}, {Geier}, \&
  {{\O}stensen}}]{Reindl+2018a}
{Reindl}, N., {Geier}, S., \& {{\O}stensen}, R.~H. 2018, \mnras, 480, 1211

\bibitem[{{Reindl} {et~al.}(2014{\natexlab{a}}){Reindl}, {Rauch}, {Werner},
  {Kepler}, {G{\"a}nsicke}, \& {Gentile Fusillo}}]{Reindl+2014c}
{Reindl}, N., {Rauch}, T., {Werner}, K., {et~al.} 2014{\natexlab{a}}, \aap,
  572, A117

\bibitem[{{Reindl} {et~al.}(2014{\natexlab{b}}){Reindl}, {Rauch}, {Werner},
  {Kruk}, \& {Todt}}]{Reindl+2014b}
{Reindl}, N., {Rauch}, T., {Werner}, K., {Kruk}, J.~W., \& {Todt}, H.
  2014{\natexlab{b}}, \aap, 566, A116

\bibitem[{{Reindl} {et~al.}(2021){Reindl}, {Schaffenroth}, {Filiz}, {Geier},
  {Pelisoli}, \& {Kepler}}]{Reindl+2021}
{Reindl}, N., {Schaffenroth}, V., {Filiz}, S., {et~al.} 2021, \aap, 647, A184

\bibitem[{{Renedo} {et~al.}(2010){Renedo}, {Althaus}, {Miller Bertolami},
  {Romero}, {C{\'o}rsico}, {Rohrmann}, \& {Garc{\'\i}a-Berro}}]{Renedo+2010}
{Renedo}, I., {Althaus}, L.~G., {Miller Bertolami}, M.~M., {et~al.} 2010, \apj,
  717, 183

\bibitem[{{Richer} {et~al.}(2000){Richer}, {Michaud}, \&
  {Turcotte}}]{Richer+2000}
{Richer}, J., {Michaud}, G., \& {Turcotte}, S. 2000, \apj, 529, 338

\bibitem[{{Saio} \& {Jeffery}(2000)}]{Sion+2000}
{Saio}, H. \& {Jeffery}, C.~S. 2000, \mnras, 313, 671

\bibitem[{{Sander} {et~al.}(2015){Sander}, {Shenar}, {Hainich},
  {G{\'\i}menez-Garc{\'\i}a}, {Todt}, \& {Hamann}}]{Sander+2015}
{Sander}, A., {Shenar}, T., {Hainich}, R., {et~al.} 2015, \aap, 577, A13

\bibitem[{{Schlafly} {et~al.}(2019){Schlafly}, {Meisner}, \&
  {Green}}]{Schlafly+2019}
{Schlafly}, E.~F., {Meisner}, A.~M., \& {Green}, G.~M. 2019, \apjs, 240, 30

\bibitem[{{Schreiber} {et~al.}(2019){Schreiber}, {G{\"a}nsicke}, {Toloza},
  {Hernandez}, \& {Lagos}}]{Schreiber+2019}
{Schreiber}, M.~R., {G{\"a}nsicke}, B.~T., {Toloza}, O., {Hernandez}, M.-S., \&
  {Lagos}, F. 2019, \apjl, 887, L4

\bibitem[{{Taylor}(2005)}]{Taylor2005}
{Taylor}, M.~B. 2005, in Astronomical Society of the Pacific Conference Series,
  Vol. 347, Astronomical Data Analysis Software and Systems XIV, ed.
  P.~{Shopbell}, M.~{Britton}, \& R.~{Ebert}, 29

\bibitem[{{Tremblay} \& {Bergeron}(2009)}]{TremblayBergeron2009}
{Tremblay}, P.-E. \& {Bergeron}, P. 2009, \apj, 696, 1755

\bibitem[{{Unglaub} \& {Bues}(2000)}]{UnglaubBues2000}
{Unglaub}, K. \& {Bues}, I. 2000, \aap, 359, 1042

\bibitem[{{Uzundag} {et~al.}(2021){Uzundag}, {C{\'o}rsico}, {Kepler},
  {Althaus}, {Werner}, {Reindl}, {Bell}, {Higgins}, {da Rosa},
  {Vu{\v{c}}kovi{\'c}}, \& {Istrate}}]{Uzundag+2021}
{Uzundag}, M., {C{\'o}rsico}, A.~H., {Kepler}, S.~O., {et~al.} 2021, \aap, 655,
  A27

\bibitem[{{Vennes}(1999)}]{Vennes+1999}
{Vennes}, S. 1999, \apj, 525, 995

\bibitem[{{Vos} {et~al.}(2021){Vos}, {Pelisoli}, {Budaj}, {Reindl},
  {Schaffenroth}, {Bobrick}, {Geier}, {Hermes}, {Nemeth}, {{\O}stensen},
  {Reding}, {Uzundag}, \& {Vu{\v{c}}kovi{\'c}}}]{Vos+2021}
{Vos}, J., {Pelisoli}, I., {Budaj}, J., {et~al.} 2021, \aap, 655, A43

\bibitem[{{Weidmann} {et~al.}(2020){Weidmann}, {Mari}, {Schmidt}, {Gaspar},
  {Miller Bertolami}, {Oio}, {Guti{\'e}rrez-Soto}, {Volpe}, {Gamen}, \&
  {Mast}}]{Weidmann+2020}
{Weidmann}, W.~A., {Mari}, M.~B., {Schmidt}, E.~O., {et~al.} 2020, \aap, 640,
  A10

\bibitem[{{Werner} {et~al.}(2003){Werner}, {Deetjen}, {Dreizler}, {Nagel},
  {Rauch}, \& {Schuh}}]{werner+2003}
{Werner}, K., {Deetjen}, J.~L., {Dreizler}, S., {et~al.} 2003, in Astronomical
  Society of the Pacific Conference Series, Vol. 288, Stellar Atmosphere
  Modeling, ed. I.~{Hubeny}, D.~{Mihalas}, \& K.~{Werner}, 31

\bibitem[{{Werner} {et~al.}(1995){Werner}, {Dreizler}, {Heber}, {Rauch},
  {Wisotzki}, \& {Hagen}}]{Werner+1995}
{Werner}, K., {Dreizler}, S., {Heber}, U., {et~al.} 1995, \aap, 293, L75

\bibitem[{{Werner} {et~al.}(2012){Werner}, {Dreizler}, \& {Rauch}}]{tmap2012}
{Werner}, K., {Dreizler}, S., \& {Rauch}, T. 2012, {TMAP: T{\"u}bingen NLTE
  Model-Atmosphere Package}, Astrophysics Source Code Library

\bibitem[{{Werner} \& {Herwig}(2006)}]{WernerHerwig2006}
{Werner}, K. \& {Herwig}, F. 2006, \pasp, 118, 183

\bibitem[{{Werner} \& {Rauch}(2014)}]{WernerRauch2014}
{Werner}, K. \& {Rauch}, T. 2014, \aap, 569, A99

\bibitem[{{Werner} {et~al.}(2014){Werner}, {Rauch}, \& {Kepler}}]{Werner+2014}
{Werner}, K., {Rauch}, T., \& {Kepler}, S.~O. 2014, \aap, 564, A53

\bibitem[{{Werner} {et~al.}(2017){Werner}, {Rauch}, \& {Kruk}}]{Werner+2017}
{Werner}, K., {Rauch}, T., \& {Kruk}, J.~W. 2017, \aap, 601, A8

\bibitem[{{Werner} {et~al.}(2018{\natexlab{a}}){Werner}, {Rauch}, \&
  {Kruk}}]{Werner+2018b}
{Werner}, K., {Rauch}, T., \& {Kruk}, J.~W. 2018{\natexlab{a}}, \aap, 609, A107

\bibitem[{{Werner} {et~al.}(2018{\natexlab{b}}){Werner}, {Rauch}, \&
  {Kruk}}]{Werner+2018a}
{Werner}, K., {Rauch}, T., \& {Kruk}, J.~W. 2018{\natexlab{b}}, \aap, 616, A73

\bibitem[{{Werner} {et~al.}(2019){Werner}, {Rauch}, \& {Reindl}}]{Werner+2019}
{Werner}, K., {Rauch}, T., \& {Reindl}, N. 2019, \mnras, 483, 5291

\bibitem[{{Werner} {et~al.}(2022{\natexlab{a}}){Werner}, {Reindl}, {Dorsch},
  {Geier}, {Munari}, \& {Raddi}}]{Werner+2022a}
{Werner}, K., {Reindl}, N., {Dorsch}, M., {et~al.} 2022{\natexlab{a}}, \aap,
  658, A66

\bibitem[{{Werner} {et~al.}(2022{\natexlab{b}}){Werner}, {Reindl}, {Geier}, \&
  {Pritzkuleit}}]{Werner+2022b}
{Werner}, K., {Reindl}, N., {Geier}, S., \& {Pritzkuleit}, M.
  2022{\natexlab{b}}, \mnras, 511, L66

\bibitem[{{Werner} {et~al.}(2020){Werner}, {Reindl}, {L{\"o}bling}, {Pelisoli},
  {Schaffenroth}, {Rebassa-Mansergas}, {Irawati}, \& {Ren}}]{Werner+2020}
{Werner}, K., {Reindl}, N., {L{\"o}bling}, L., {et~al.} 2020, \aap, 642, A228

\bibitem[{{Wolf} {et~al.}(2018){Wolf}, {Onken}, {Luvaul}, {Schmidt}, {Bessell},
  {Chang}, {Da Costa}, {Mackey}, {Martin-Jones}, {Murphy}, {Preston}, {Scalzo},
  {Shao}, {Smillie}, {Tisserand}, {White}, \& {Yuan}}]{Wolf+2018}
{Wolf}, C., {Onken}, C.~A., {Luvaul}, L.~C., {et~al.} 2018, \pasa, 35, e010

\bibitem[{{Yuan} \& {Liu}(2013)}]{Yuan+2013}
{Yuan}, H.~B. \& {Liu}, X.~W. 2013, \mnras, 436, 718

\bibitem[{{Zechmeister} \& {K{\"u}rster}(2009)}]{ZechmeisterKuerster2009}
{Zechmeister}, M. \& {K{\"u}rster}, M. 2009, \aap, 496, 577

\bibitem[{{Zhang} \& {Jeffery}(2012{\natexlab{a}})}]{Zhang+2012b}
{Zhang}, X. \& {Jeffery}, C.~S. 2012{\natexlab{a}}, \mnras, 426, L81

\bibitem[{{Zhang} \& {Jeffery}(2012{\natexlab{b}})}]{Zhang+2012a}
{Zhang}, X. \& {Jeffery}, C.~S. 2012{\natexlab{b}}, \mnras, 419, 452

\bibitem[{{Zhang} {et~al.}(2012){Zhang}, {Fang}, {Chau}, {Hsia}, {Liu}, {Kwok},
  \& {Koning}}]{Zhang+2012}
{Zhang}, Y., {Fang}, X., {Chau}, W., {et~al.} 2012, \apj, 754, 28

\end{thebibliography}

\clearpage
\begin{onecolumn}
\begin{appendix}

\section{Tables}

\begin{longtable}{llrrlcrr}
\caption{\label{table:A1} Overview of the observations. Name, J2000 coordinates of our targets given along with what telescopes and instruments they have been observed. The MJD at the start of the observation, exposure times and the S/N as measured between $5100-5300$\,\AA\ of the observations are listed as well.}\\
\hline\hline
\noalign{\smallskip}
Name & Type & Ra & Dec& Telescope/ & MJD-OBS & $T_{\mathrm{exp}}$ & S/N \\
 & & [deg] & [deg] & Instrument & & [s] & \\
 \noalign{\smallskip}
\hline
\noalign{\smallskip}
\endfirsthead 
\caption{continued}\\
\hline\hline
\noalign{\smallskip}
Name & Type & Ra & Dec& Telescope/ & MJD-OBS & $T_{\mathrm{exp}}$ & S/N \\
 & & [deg] & [deg] & Instrument & & [s] & \\
 \noalign{\smallskip}
\hline
\noalign{\smallskip}
\endhead 
\hline
\endfoot
\hline \hline
\endlastfoot
WDJ014343.14$+$584151.39        &       DO V*   &       25.92974        &       58.69761        &       LBT/MODS        &       59526.31541     &       900     &       127     \\
WDJ020351.33$-$063906.46        &       DA      &       30.96387        &       $-$6.65179      &       LBT/MODS        &       59528.31042     &       1184    &       272     \\
WDJ020351.33$-$063906.46        &       DA      &       30.96402        &       $-$6.65179      &       INT/IDS(R400V)  &       58827.91127     &       1800    &       36      \\
WDJ023722.53$+$583405.41        &       DA      &       39.34390        &       58.56817        &       INT/IDS(R400V)  &       59260.91354     &       1800    &       51      \\
WDJ031437.37$+$313641.91        &       DA      &       48.65573        &       31.61164        &       INT/IDS(R400V)  &       59260.94453     &       900     &       89      \\
WDJ033449.19$+$164629.92        &       DOZ V*  &       53.70498        &       16.77498        &       LBT/MODS        &       59528.33045     &       900     &       125     \\
WDJ043619.17$-$065412.61        &       DO      &       69.07988        &       $-$6.9035       &       INT/IDS(R400V)  &       59262.91673     &       1800    &       39      \\
WDJ051351.78$-$225919.24        &       DA      &       78.46576        &       $-$22.9887      &       INT/IDS(R400V)  &       59261.88301     &       1800    &       57      \\
WDJ051949.17$-$044300.03        &       DO      &       79.95486        &       $-$4.71667      &       INT/IDS(R400V)  &       58825.99324     &       1800    &       22      \\
WDJ053810.41$-$103644.81        &       DA      &       84.54339        &       $-$10.6124      &       INT/IDS(R400V)  &       59261.96735     &       1800    &       59      \\
WDJ060020.89$-$101404.50        &       DA V*   & 90.08705        &       $-$10.2346      &       INT/IDS(R400V)  &       59262.98942     &       1800    &       45      \\
WDJ060244.99$-$135103.57        &       DOZ V*  &       90.68749        &       $-$13.85096     &       VLT/X-Shooter   &       59470.33534     &       1800    &       32      \\
WDJ060244.99$-$135103.57        &       DOZ V*  &       90.68749        &       $-$13.85096     &       INT/IDS(R400V)  &       59260.97692     &       1800    &       45      \\
WDJ060244.99$-$135103.57        &       DOZ V*  &       90.68749        &       $-$13.85096     &       INT/IDS(R400V)  &       59261.91840     &       1800    &       44      \\
WDJ061906.92$-$082807.15        &       DA      &       94.77885        &       $-$8.46865      &       VLT/X-Shooter   &       59314.02969     &       1800    &       60      \\
WDJ061906.92$-$082807.15        &       DA      &       94.77889        &       $-$8.46865      &       INT/IDS(R400V)  &       58828.11719     &       1800    &       17      \\
WDJ062145.32$+$065239.25        &       DA      &       95.43885        &       6.87757 &       VLT/X-Shooter   &       59314.00268     &       1800    &       54      \\
WDJ062145.32$+$065239.25        &       DA      &       95.43883        &       6.87757 &       INT/IDS(R400V)  &       58828.0017      &       1800    &       19      \\
WDJ064527.28$+$565916.90        &       DAe: V* &       101.36365       &       56.98803        &       LBT/MODS        &       59528.28488     &       900     &       134     \\
WDJ070204.29$+$051420.56        &       DOZ UHE V*      &       105.51790       &       5.23905 &       INT/IDS(R400V)  &       59261.02997     &       1800    &       67      \\
WDJ070322.15$-$340331.94        &       DAO     &       105.84231       & -34.05881 &     VLT/X-Shooter   &       60011.02962 & 1200 & 60  \\
WDJ070647.52$+$613350.31        &       DA UHE V*       &       106.69800       &       61.56398        &       INT/IDS(R400V)  &       59261.10608     &       900     &       109     \\
WDJ070647.52$+$613350.31        &       DA UHE V*       &       106.69800       &       61.56398        &       INT/IDS(R400V)  &       59261.08357     &       900     &       75      \\
WDJ070647.52$+$613350.31        &       DA UHE V*       &       106.69800       &       61.56398        &       INT/IDS(R400V)  &       59261.09485     &       900     &       93      \\
WDJ070647.52$+$613350.31        &       DA UHE V*       &       106.69800       &       61.56398        &       INT/IDS(R400V)  &       59262.94971     &       1100    &       102     \\
WDJ073607.90$+$144451.07        &       DOZ     &       114.03290       &       14.74752        &       INT/IDS(R400V)  &       59261.12629     &       900     &       65      \\
WDJ073957.35$+$091439.23        &       DA      &       114.98895       &       9.24423 &       INT/IDS(R400V)  &       58825.14766     &       1200    &       22      \\
WDJ073957.35$+$091439.23        &       DA      &       114.98900       &       9.24423 &       INT/IDS(R400V)  &       59262.05103     &       1800    &       28      \\
WDJ075134.71$-$015807.00        &       DO      &       117.89460       &       $-$1.96861      &       INT/IDS(R400V)  &       59261.99625     &       1800    &       40      \\
WDJ075145.59$+$105931.36        &       DAe     V* &    117.93997       &       10.99204        &       LBT/MODS        &       59551.16308     &       900     &       94      \\
WDJ080029.42$-$015039.82        &       DA      &       120.12258       &       $-$1.8444       &       INT/IDS(R400V)  &       58828.25166     &       1800    &       35      \\
WDJ080326.15$-$034746.11        &       DAO     &       120.85896       &       $-$3.79614      &       VLT/X-Shooter   &       59281.16694     &       1800    &       60      \\
WDJ080326.15$-$034746.11        &       DAO     &       120.85896       &       $-$3.79614      &       INT/IDS(R400V)  &       58828.22633     &       1800    &       32      \\
WDJ080558.34$+$410931.84        &       DA V*   &       121.49307       &       41.15885        &       LBT/MODS        &       59528.36973     &       900     &       241     \\
WDJ080950.28$-$364740.45        &       DAO     &       122.45944       & -36.79449 &     VLT/X-Shooter   &       60014.14361 & 1200 & 50  \\
WDJ085137.48$+$185119.67        &       DA      &       132.90615       &       18.85546        &       INT/IDS(R400V)  &       58825.2317      &       1800    &       36      \\
WDJ093559.83$+$685201.55        &       DO UHE V*       &       143.99930       &       68.86710        &       LBT/MODS        &       59526.33673     &       900     &       118     \\
WDJ120728.43$+$540129.16        &       DAO     &       181.86850       &       54.02477        &       INT/IDS(R400V)  &       59261.20041     &       1800    &       37      \\
WDJ123952.73$+$720546.58        &       DA      &       189.96951       &       72.09627        &       INT/IDS(R400V)  &       59376.92141     &       900     &       36      \\
WDJ125627.43$+$775301.37        &       DA UHE: V*      &       194.11430       &       77.88372        &       INT/IDS(R400V)  &       59261.24365     &       840     &       30      \\
WDJ125627.43$+$775301.37        &       DA UHE: V*      &       194.11430       &       77.88372        &       INT/IDS(R400V)  &       59261.25424     &       840     &       48      \\
WDJ132259.60$-$383813.13        &       sdO     &       200.74833       &       $-$38.63694     &       VLT/X-Shooter   &       59314.22414     &       1800    &       77      \\
WDJ132307.79$-$480542.75        &       DAO     &       200.78246       &       $-$48.09521     &       VLT/X-Shooter   &       59987.32415     &       1200    &       86      \\
WDJ134843.58$+$741545.75        &       DA      &       207.18159       &       74.26271        &       INT/IDS(R400V)  &       58837.25518     &       1800    &       43      \\
WDJ134843.58$+$741545.75        &       DA      &       207.18154       &       74.26264        &       INT/IDS(R400V)  &       59376.89045     &       1800    &       39      \\
WDJ135724.86$+$753723.60        &       DA      &       209.35335       &       75.62321        &       INT/IDS(R400V)  &       59375.99466     &       1800    &       35      \\
WDJ135724.86$+$753723.60        &       DA      &       209.35359       &       75.62322        &       INT/IDS(R400V)  &       58836.23492     &       1800    &       49      \\
WDJ142557.80$-$034139.92        &       DA      &       216.49080       &       $-$3.69442      &       INT/IDS(R400V)  &       59262.18566     &       1800    &       34      \\
WDJ154843.31$+$472936.11        &       DOA     &       237.18036       &       47.49333        &       INT/IDS(R400V)  &       59377.90543     &       1200    &       30      \\
WDJ160152.16$+$380455.24        &       DA      &       240.46730       &       38.08201        &       INT/IDS(R400V)  &       59262.15119     &       1800    &       32      \\
WDJ165053.99$+$774844.88        &       DA      &       252.72426       &       77.81275        &       INT/IDS(R400V)  &       59375.94995     &       120     &       35      \\
WDJ175145.75$+$382015.60        &       DA      &       267.94060       &       38.33767        &       INT/IDS(R400V)  &       59262.22048     &       1800    &       18      \\
WDJ175629.64$-$072248.89        &       DA      &       269.12351       &       $-$7.38025      &       INT/IDS(R400V)  &       59377.03563     &       1800    &       43      \\
WDJ181104.68$+$193658.10        &       DAO     &       272.76950       &       19.61614        &       INT/IDS(R400V)  &       59262.24628     &       1800    &       24      \\
WDJ181104.68$+$193658.10        &       DAO     &       272.76950       &       19.61614        &       INT/IDS(R400V)  &       59262.26786     &       1800    &       32      \\
WDJ182432.46$+$293115.88        &       DO      &       276.13524       &       29.52108        &       INT/IDS(R400V)  &       59376.0252      &       1800    &       48      \\
WDJ182440.85$-$031959.52        &       DA      &       276.17028       &       $-$3.33329      &       INT/IDS(R400V)  &       59378.01071     &       900     &       36      \\
WDJ182849.94$+$343649.94        &       DA      &       277.20809       &       34.61387        &       INT/IDS(R400V)  &       59377.92625     &       1800    &       36      \\
WDJ182849.94$+$343649.94        &       DA      &       277.20809       &       34.61387        &       INT/IDS(R400V)  &       59377.92625     &       1800    &       31      \\
WDJ183429.29$+$181101.71        &       DA      &       278.62208       &       18.18376        &       INT/IDS(R400V)  &       59376.11081     &       900     &       30      \\
WDJ183429.29$+$181101.71        &       DA      &       278.62208       &       18.18376        &       INT/IDS(R400V)  &       59376.11081     &       900     &       27      \\
WDJ184101.20$+$612032.16        &       DA      &       280.25499       &       61.34227        &       INT/IDS(R400V)  &       59376.97473     &       1800    &       37      \\
WDJ190416.13$-$170153.08        &       DAO     &       286.06718       &       $-$17.03141     &       INT/IDS(R400V)  &       59379.07986     &       1500    &       27      \\
WDJ190717.50$+$364000.37        &       DAO     &       286.82291       &       36.66677        &       INT/IDS(R400V)  &       59376.15473     &       900     &       33      \\
WDJ191231.47$-$033131.86        &       DA      &       288.13102       &       $-$3.52566      &       INT/IDS(R400V)  &       59379.01501     &       900     &       43      \\
WDJ191750.16$-$201409.55        &       DO      &       289.45901       &       $-$20.23599     &       INT/IDS(R400V)  &       59377.14625     &       1800    &       35      \\
WDJ192705.33$+$112452.84        &       DA      &       291.77224       &       11.41463        &       INT/IDS(R400V)  &       59377.07346     &       900     &       66      \\
WDJ194344.19$+$521732.66        &       DOZ     &       295.93412       &       52.29241        &       INT/IDS(R400V)  &       59377.0019      &       500     &       65      \\
WDJ194511.31$-$445954.57        &       DAO     &       296.29703       &       $-$44.9985      &       VLT/X-Shooter   &       59427.16143     &       900     &       97      \\
WDJ202547.32$+$355411.38        &       DA      &       306.44716       &       35.90316        &       INT/IDS(R400V)  &       59378.05155     &       900     &       34      \\
WDJ210110.17$-$052751.14        &       DAO V*  &       315.29212       &       $-$5.46432      &       INT/IDS(R400V)  &       59377.22622     &       900     &       24      \\
WDJ210110.17$-$052751.14        &       DAO V*  &       315.29212       &       $-$5.46432      &       INT/IDS(R400V)  &       59378.18389     &       900     &       28      \\
WDJ210132.69$+$135622.59        &       DAO UHE V*      &       315.38619       &       13.93961        &       LBT/MODS        &       59528.34973     &       900     &       182     \\
WDJ210824.97$+$275049.80        &       DAO     &       317.10403 &     27.84717 &       INT/IDS(R400V)  &       59378.15987     &       1500    &       30      \\
WDJ211532.62$-$615849.50        &       DO UHE: V* &    318.88590       &       $-$61.98035     &       VLT/X-Shooter   &       59469.26343     &       1800    &       65      \\
WDJ211532.62$-$615849.50        &       DO UHE: V* &    318.88590       &       $-$61.98035     &       SOAR/GHTS       &       59372.30922     &       800     &       40      \\
WDJ212355.09$+$342017.32        &       DA      &       320.97954       &       34.33815        &       INT/IDS(R400V)  &       59378.06714     &       1200    &       30      \\
WDJ214212.56$+$322052.17        &       DA      &       325.55249       &       32.34777        &       INT/IDS(R400V)  &       59379.2067      &       1500    &       32      \\
WDJ214327.96$+$414318.99        &       DA      &       325.86627       &       41.72174        &       INT/IDS(R400V)  &       59379.05566     &       1500    &       26      \\
WDJ215101.38$-$151839.71        &       DA      &       327.75574       &       $-$15.31103     &       INT/IDS(R400V)  &       59379.17214     &       1800    &       29      \\
WDJ220247.69$+$275010.67        &       DA V*   &       330.69871       &       27.83630        &       LBT/MODS        &       59555.14628     &       900     &       74      \\
WDJ220817.34$+$390711.18        &       DA      &       332.07226       &       39.11977        &       INT/IDS(R400V)  &       59378.11333     &       1500    &       22      \\
WDJ222147.70$+$501150.75        &       DA V*   &       335.44874       &       50.19743        &       LBT/MODS        &       59526.07477     &       900     &       100     \\
WDJ223522.88$-$494350.80        &       sdO     &       338.84547       &       $-$49.73088     &       VLT/X-Shooter   &       59409.3123      &       1800    &       63      \\
WDJ224305.77$+$584336.70        &       DA      &       340.77417       &       58.72693        &       INT/IDS(R400V)  &       59377.17848     &       1500    &       43      \\
WDJ225808.72$+$685751.45        &       DA      &       344.53632       &       68.96429        &       INT/IDS(R400V)  &       58834.90353     &       1800    &       40      \\
WDJ231757.59$+$555026.25        &       DA      &       349.49002       &       55.84066        &       INT/IDS(R400V)  &       59378.13711     &       1500    &       27      \\
GD1323  &       DAO     &       356.52117       &       $-$7.6516       &       VLT/X-Shooter   &       59765.37629     &       1800    &       78      \\
\noalign{\smallskip}
\end{longtable}

\begin{center}
\begin{table*}
\caption{\label{table:DA} Effective temperatures, surface gravities, radii, luminosities, gravity, HRD, and Kiel masses of the DA WDs.}
\renewcommand{\tabcolsep}{1.8mm}
\begin{tabular}{l r@{$\pm$}l r@{$\pm$}l r@{}l r@{}l r@{}l r@{$\pm$}l r@{$\pm$}l} 
\hline\hline
\noalign{\smallskip}
Name &  \multicolumn{2}{l}{\Teff} & \multicolumn{2}{l}{\logg} & \multicolumn{2}{l}{$R$} & \multicolumn{2}{l}{$L$} & \multicolumn{2}{l}{$M_{\mathrm{grav}}$} & \multicolumn{2}{l}{$M_{\mathrm{HRD}}$} & \multicolumn{2}{l}{$M_{\mathrm{Kiel}}$}\\
 (short) & \multicolumn{2}{l}{[K]} & \multicolumn{2}{l}{[cm\,/\,s$^2$]}  & \multicolumn{2}{l}{[\Rsol/100]} &\multicolumn{2}{l}{[\Lsol]} & \multicolumn{2}{l}{[\Msol]} & \multicolumn{2}{l}{[\Msol]} & \multicolumn{2}{l}{[\Msol]} \\
\noalign{\smallskip}
\hline
\noalign{\smallskip}
J0203$-$0639 & 60507 & 300 & 8.12 & 0.02                &$ 1.70 $&$\pm0.03 $&$ 3.5 $&$^{+0.13}_{-0.12}$&$ 1.40 $&$^{+0.09}_{-0.08}$&       0.60 & 0.05 & 0.75 & 0.02   \\ \noalign{\smallskip}
J0237$+$5834 & 70753 & 1820 & 8.04 & 0.1                &$ 1.55 $&$\pm0.04 $&$ 5.4 $&$^{+0.7}_{-0.6}$&$ 0.96 $&$^{+0.26}_{-0.20}$& 0.67 & 0.06 & 0.73 & 0.05  \\ \noalign{\smallskip}
J0314$+$3136 & 72664 & 1158 & 7.72 & 0.07               &$ 1.66 $&$\pm0.02 $&$ 6.9 $&$\pm0.5        $&$ 0.53 $&$^{+0.10}_{-0.08}$& 0.64 & 0.04 & 0.62 & 0.04  \\ \noalign{\smallskip}
J0513$-$2259 & 70702 & 1426 & 7.8 & 0.08                &$ 1.39 $&$\pm0.02 $&$ 4.3 $&$\pm0.4        $&$ 0.44 $&$^{+0.10}_{-0.08}$& 0.72 & 0.04 & 0.64 & 0.04  \\ \noalign{\smallskip}
J0538$-$1036 & 68255 & 1292 & 7.59 & 0.08               &$ 1.82 $&$\pm0.02 $&$ 6.5 $&$^{+0.6}_{-0.5}$&$ 0.47 $&$^{+0.10}_{-0.08}$& 0.60 & 0.05 & 0.57 & 0.05  \\ \noalign{\smallskip}
J0600$-$1014 & 63198 & 2099 & 7.51 & 0.11               &$ 1.50 $&$^{+0.03}_{-0.02}$&$ 3.2 $&$\pm0.5    $&$ 0.26 $&$^{+0.08}_{-0.06}$& 0.67 & 0.05 & 0.54 & 0.04       \\ \noalign{\smallskip}
J0619$-$0828 & 101921 & 1991 & 7.47 & 0.05              &$ 1.08 $&$\pm0.02 $&$ 11.3 $&$^{+1.1}_{-1}$&$ 0.13 $&$^{+0.02}_{-0.02}$&  0.89 & 0.02 & 0.61 & 0.04  \\ \noalign{\smallskip}
J0621$+$0652 & 108314 & 2186 & 7.63 & 0.05              &$ 1.75 $&$\pm0.05 $&$ 38 $&$\pm4   $&$ 0.48 $&$^{+0.07}_{-0.06}$& 0.70 & 0.04 & 0.65 & 0.05       \\ \noalign{\smallskip}
J0645$+$5659 & 99762 & 3672 & 6.72 & 0.03               &$ 1.89 $&$^{+0.11}_{-0.10}$&$ 32 $&$^{+7}_{-6}$&$ 0.07 $&$^{+0.01}_{-0.01}$&  0.66 & 0.06 & 0.51 & 0.06       \\ \noalign{\smallskip}
J0739$+$0914 & 81585 & 4449 & 7.17 & 0.11               &$ 1.89 $&$\pm0.07 $&$ 14.2 $&$^{+3.6}_{-3}$&$ 0.19 $&$^{+0.06}_{-0.05}$&  0.62 & 0.06 & 0.52 & 0.05  \\ \noalign{\smallskip}
J0800$-$0150 & 76169 & 5053 & 7.29 & 0.13               &$ 2.20 $&$^{+0.09}_{-0.08}$&$ 15 $&$^{+5}_{-4}$&$ 0.35 $&$^{+0.13}_{-0.10}$&  0.57 & 0.07 & 0.53 & 0.04       \\ \noalign{\smallskip}
J0805$+$4109 & 50590 & 227 & 8.11 & 0.2                 &$ 1.76 $&$\pm0.02 $&$ 1.82 $&$\pm0.11      $&$ 1.90 $&$\pm0.40     $&     0.56 & 0.05 & 0.73 & 0.08  \\ \noalign{\smallskip}
J0851$+$1851 & 70491 & 2125 & 7.87 & 0.12               &$ 1.57 $&$^{+0.05}_{-0.04}$&$ 5.5 $&$^{+0.8}_{-0.7}$&$ 0.67 $&$^{+0.23}_{-0.17}$&     0.66 & 0.06 & 0.66 & 0.06  \\ \noalign{\smallskip}
J1239$+$7205 & 71049 & 2312 & 7.29 & 0.15               &$ 2.43 $&$^{+0.12}_{-0.11}$&$ 13.6 $&$^{+2.4}_{-2.1}$&$ 0.42 $&$^{+0.19}_{-0.13}$&    0.54 & 0.07 & 0.52 & 0.06  \\ \noalign{\smallskip}
J1256$+$7753 & 68086 & 1723 & 7.71 & 0.1                &$ 1.88 $&$^{+0.03}_{-0.03}$&$ 6.8 $&$^{+0.8}_{-0.7}$&$ 0.66 $&$^{+0.18}_{-0.14}$&     0.59 & 0.06 & 0.61 & 0.05  \\ \noalign{\smallskip}
J1348$+$7415 & 65092 & 2008 & 7.79 & 0.11               &$ 1.64 $&$^{+0.05}_{-0.04}$&$ 4.3 $&$^{+0.7}_{-0.6}$&$ 0.60 $&$^{+0.18}_{-0.14}$&     0.63 & 0.05 & 0.62 & 0.05  \\ \noalign{\smallskip}
J1357$+$7537 & 51512 & 1065 & 7.86 & 0.09               &$ 1.32 $&$^{+0.02}_{-0.02}$&$ 1.11 $&$^{+0.1}_{-0.1}$&$ 0.46 $&$\pm0.10        $&     0.70 & 0.04 & 0.62 & 0.05  \\ \noalign{\smallskip}
J1425$-$0341 & 31795 & 268 & 7.39 & 0.07                &$ 2.14 $&$\pm0.04 $&$ 0.422 $&$^{+0.02}_{-0.019}$&$ 0.41 $&$^{+0.07}_{-0.06}$&    0.47 & 0.03 & 0.46 & 0.03   \\ \noalign{\smallskip}
J1601$+$3804 & 83236 & 7307 & 7.51 & 0.27               &$ 2.22 $&$^{+0.10}_{-0.09}$&$ 21 $&$^{+9}_{-7}$&$ 0.58 $&$^{+0.51}_{-0.27}$&  0.59 & 0.08 & 0.59 & 0.07       \\ \noalign{\smallskip}
J1650$+$7748 & 29104 & 310 & 7.55 & 0.08                &$ 2.13 $&$^{+0.03}_{-0.03}$&$ 0.295 $&$^{+0.016}_{-0.015}$&$ 0.59 $&$^{+0.13}_{-0.11}$&       0.46 & 0.04 & 0.48 & 0.04   \\ \noalign{\smallskip}
J1751$+$3820 & 69474 & 3314 & 6.71 & 0.16               &$ 2.88 $&$\pm0.08 $&$ 17 $&$\pm4   $&$ 0.15 $&$^{+0.07}_{-0.05}$& 0.51 & 0.06 & 0.43 & 0.06       \\ \noalign{\smallskip}
J1756$-$0722 & 62941 & 1459 & 8.04 & 0.09               &$ 1.35 $&$\pm0.02 $&$ 2.59 $&$^{+0.26}_{-0.24}$&$ 0.74 $&$^{+0.17}_{-0.14}$&      0.71 & 0.04 & 0.72 & 0.04   \\ \noalign{\smallskip}
J1824$-$0319 & 68937 & 630 & 7.63 & 0.03                &$ 1.68 $&$\pm0.02 $&$ 5.74 $&$^{+0.26}_{-0.25}$&$ 0.44 $&$\pm0.04  $&     0.63 & 0.05 & 0.58 & 0.04  \\ \noalign{\smallskip}
J1828$+$3436 & 39478 & 927 & 7.81 & 0.08                &$ 1.50 $&$\pm0.02 $&$ 0.49 $&$\pm0.05      $&$ 0.53 $&$^{+0.11}_{-0.10}$& 0.60 & 0.05 & 0.58 & 0.04  \\ \noalign{\smallskip}
J1834$+$1811 & 57790 & 1685 & 7.38 & 0.11               &$ 2.13 $&$\pm0.04 $&$ 4.6 $&$\pm0.6        $&$ 0.40 $&$^{+0.12}_{-0.09}$& 0.52 & 0.05 & 0.51 & 0.06  \\ \noalign{\smallskip}
J1841$+$6120 & 47322 & 897 & 8.02 & 0.1                 &$ 1.36 $&$\pm0.02 $&$ 0.84 $&$\pm0.07      $&$ 0.71 $&$^{+0.18}_{-0.15}$& 0.67 & 0.04 & 0.68 & 0.05  \\ \noalign{\smallskip}
J1912$-$0331 & 53271 & 1316 & 7.63 & 0.1                &$ 1.92 $&$^{+0.02}_{-0.02}$&$ 2.66 $&$^{+0.29}_{-0.27}$&$ 0.57 $&$^{+0.14}_{-0.12}$&  0.54 & 0.04 & 0.55 & 0.05  \\ \noalign{\smallskip}
J1927$+$1124 & 77190 & 3559 & 7.54 & 0.11               &$ 1.69 $&$\pm0.05 $&$ 9.2 $&$^{+2}_{-1.7}$&$ 0.36 $&$^{+0.11}_{-0.09}$&   0.65 & 0.06 & 0.58 & 0.05  \\ \noalign{\smallskip}
J2025$+$3554 & 35343 & 511 & 7.88 & 0.09                &$ 1.66 $&$^{+0.02}_{-0.02}$&$ 0.389 $&$^{+0.026}_{-0.024}$&$ 0.77 $&$^{+0.18}_{-0.15}$&       0.54 & 0.05 & 0.60 & 0.05   \\ \noalign{\smallskip}
J2123$+$3420 & 57145 & 1858 & 7.79 & 0.13               &$ 1.75 $&$\pm0.04 $&$ 2.9 $&$^{+0.5}_{-0.4}$&$ 0.69 $&$^{+0.25}_{-0.18}$& 0.59 & 0.05 & 0.61 & 0.05  \\ \noalign{\smallskip}
J2142$+$3220 & 61309 & 2128 & 7.3 & 0.12                &$ 1.56 $&$^{+0.03}_{-0.03}$&$ 3.1 $&$\pm0.5    $&$ 0.18 $&$^{+0.06}_{-0.05}$& 0.64 & 0.05 & 0.50 & 0.04       \\ \noalign{\smallskip}
J2143$+$4143 & 83390 & 7284 & 7.24 & 0.25               &$ 1.84 $&$^{+0.08}_{-0.07}$&$ 15 $&$^{+7}_{-5}$&$ 0.22 $&$^{+0.17}_{-0.10}$&  0.63 & 0.09 & 0.54 & 0.06       \\ \noalign{\smallskip}
J2151$-$1518 & 61511 & 2087 & 7.7 & 0.12                &$ 1.46 $&$\pm0.04 $&$ 2.8 $&$^{+0.5}_{-0.4}$&$ 0.39 $&$^{+0.13}_{-0.10}$& 0.67 & 0.06 & 0.59 & 0.05  \\ \noalign{\smallskip}
J2202$+$2750 & 58915 & 1135 & 8.25 & 0.17               &$ 1.78 $&$\pm0.07 $&$ 3.5 $&$\pm0.4        $&$ 2.10 $&$^{+1.10}_{-0.70}$& 0.58 & 0.05 & 0.82 & 0.06  \\ \noalign{\smallskip}
J2208$+$3907 & 71826 & 2783 & 7.53 & 0.22               &$ 2.17 $&$\pm0.05 $&$ 11.3 $&$^{+2}_{-1.8}$&$ 0.58 $&$^{+0.39}_{-0.24}$&  0.56 & 0.05 & 0.57 & 0.06  \\ \noalign{\smallskip}
J2221$+$5011 & 57480 & 552 & 7.95 & 0.04                &$ 1.52 $&$\pm0.05 $&$ 2.27 $&$^{+0.17}_{-0.16}$&$ 0.75 $&$^{+0.09}_{-0.08}$&      0.64 & 0.04 & 0.67 & 0.03   \\ \noalign{\smallskip}
J2243$+$5843 & 49918 & 850 & 7.76 & 0.07                &$ 1.59 $&$^{+0.02}_{-0.02}$&$ 1.41 $&$\pm0.11  $&$ 0.52 $&$^{+0.10}_{-0.09}$& 0.60 & 0.05 & 0.58 & 0.05       \\ \noalign{\smallskip}
J2258$+$6857 & 32424 & 290 & 8.08 & 0.06                &$ 1.17 $&$\pm0.01 $&$ 0.137 $&$\pm0.006    $&$ 0.60 $&$^{+0.10}_{-0.08}$& 0.73 & 0.04 & 0.69 & 0.03  \\ \noalign{\smallskip}
J2317$+$5550 & 72758 & 3383 & 7.49 & 0.21               &$ 1.76 $&$\pm0.04 $&$ 7.8 $&$^{+1.7}_{-1.4}$&$ 0.35 $&$^{+0.22}_{-0.14}$& 0.62 & 0.05 & 0.56 & 0.07  \\ \noalign{\smallskip}
\noalign{\smallskip}
\hline\hline
\end{tabular}
\end{table*}
\end{center}

\begin{center}
\begin{table}
\caption{\label{table:DAO} Effective temperatures, surface gravities, logarithmic He/H  abundances ratios (by number), radii, luminosities, gravity, HRD, and Kiel masses of the DAO WDs, sdO, and O(H) stars.}
\renewcommand{\tabcolsep}{1.8mm}
\begin{tabular}{l r@{$\pm$}l r@{$\pm$}l r@{$\pm$}l r@{}l r@{}l r@{}l r@{$\pm$}l r@{$\pm$}l} 
\hline\hline
\noalign{\smallskip}
Name &  \multicolumn{2}{l}{\Teff} & \multicolumn{2}{l}{\logg} & \multicolumn{2}{l}{$\log$\,(He/H)} & \multicolumn{2}{l}{$R$} & \multicolumn{2}{l}{$L$} & \multicolumn{2}{l}{$M_{\mathrm{grav}}$} & \multicolumn{2}{l}{$M_{\mathrm{HRD}}$} & \multicolumn{2}{l}{$M_{\mathrm{Kiel}}$}\\
(short) &  \multicolumn{2}{l}{[K]} & \multicolumn{2}{l}{[cm\,/\,s$^2$]} & \multicolumn{2}{l}{} & \multicolumn{2}{l}{[\Rsol/100]} &\multicolumn{2}{l}{[\Lsol]} & \multicolumn{2}{l}{[\Msol]} & \multicolumn{2}{l}{[\Msol]} & \multicolumn{2}{l}{[\Msol]} \\
\noalign{\smallskip}
\hline
\noalign{\smallskip}
GD\,1323 & 112744 & 934 & 5.89 & 0.01 & -1.00 & 0.01    &$ 8.80 $&$^{+0.01}_{-0.0100}$&$ 1120$&$^{+360}_{-250}$&$ 0.22 $&$\pm0.07         $&     0.52 & 0.05 & 0.55 & 0.02  \\ \noalign{\smallskip}
J0703$-$3403 & 105839 & 734 & 6.77 & 0.02 & -1.14 & 0.02 & $3.02$&$\pm0.11$ & $103$&$^{+9}_{-8}$ & $0.20$&$\pm0.02$ & 0.58 & 0.06 & 0.53 & 0.05 \\ \noalign{\smallskip}
J0803$-$0347 & 86893 & 4625 & 6.86 & 0.12 & -1.63 & 0.05        &$ 2.69 $&$^{+0.13}_{-0.12}$&$ 37 $&$^{+10}_{-\phantom{0}8}$&$ 0.19 $&$^{+0.07}_{-0.05}$&      0.56 & 0.06 & 0.50 & 0.05   \\ \noalign{\smallskip}
J0809$-$3647 & 103607 & 754 & 6.68 & 0.02 & -1.18 & 0.02 & $3.22$&$^{+0.12}_{-0.11}$ & $108$&$^{+9}_{-8}$ & $0.18$&$\pm0.02$ & 0.57 & 0.06 & 0.52 & 0.06 \\ \noalign{\smallskip}
J1207$+$5401 & 84918 & 11184 & 6.54 & 0.23 & -0.99 & 0.08       &$ 4.50 $&$^{+0.50}_{-0.40}$&$ 90 $&$^{+70}_{-50}$&$ 0.26 $&$^{+0.20}_{-0.12}$&        0.49 & 0.09 & 0.44 & 0.07  \\ \noalign{\smallskip}
J1322$-$3838 & 80000 & 10000 & 6.00 & 0.50 & -2.00 & 0.50       &$ 6.90 $&$^{+0.60}_{-0.50}$&$ 170 $&$^{+120}_{-\phantom{0}80}$&$ 0.17 $&$^{+0.38}_{-0.12}$&   0.42 & 0.07 & 0.42 & 0.07   \\ \noalign{\smallskip}
J1323$-$4805 & 97563 & 808 & 6.87 & 0.02 & -1.36 & 0.02 &$ 2.78 $&$^{+0.11}_{-0.10}$&$ 63 $&$^{+6}_{-5}$&$ 0.21 $&$\pm0.02      $&     0.58 & 0.06 & 0.52 & 0.06       \\ \noalign{\smallskip}
J1811$+$1936 & 83354 & 7075 & 6.71 & 0.17 & -2.33 & 0.26        &$ 2.13 $&$^{+0.13}_{-0.12}$&$ 20 $&$^{+9}_{-7}$&$ 0.09 $&$^{+0.04}_{-0.03}$&  0.59 & 0.09 & 0.46 & 0.05       \\ \noalign{\smallskip}
J1904$-$1701 & 81784 & 8778 & 6.76 & 0.26 & -2.07 & 0.24        &$ 2.17 $&$^{+0.14}_{-0.12}$&$ 19 $&$^{+11}_{-8}$&$ 0.10 $&$^{+0.09}_{-0.05}$& 0.59 & 0.08 & 0.46 & 0.06       \\ \noalign{\smallskip}
J1907$+$3640 & 80924 & 10831 & 7.06 & 0.28 & -1.18 & 0.11       &$ 2.46 $&$^{+0.19}_{-0.15}$&$ 23 $&$^{+17}_{-11}$&$ 0.25 $&$^{+0.24}_{-0.13}$&        0.56 & 0.11 & 0.51 & 0.06  \\ \noalign{\smallskip}
J1945$-$4459 & 87559 & 1978 & 7.10 & 0.06 & -1.81 & 0.07        &$ 2.37 $&$\pm0.05 $&$ 29.8 $&$^{+3}_{-2.8}$&$ 0.26 $&$\pm0.04      $&     0.58 & 0.04 & 0.53 & 0.03  \\ \noalign{\smallskip}
J2101$-$0527 & 48904 & 265 & 8.14 & 0.06 & -0.58 & 0.04 &$ 1.28 $&$\pm0.01 $&$ 0.838 $&$^{+0.023}_{-0.022}$&$ 0.82 $&$^{+0.13}_{-0.11}$&   0.72 & 0.03 & 0.75 & 0.03   \\ \noalign{\smallskip}
J2108$+$2750 & 80116 & 9176 & 7.00 & 0.26 & -2.01 & 0.16        &$ 2.62 $&$^{+0.18}_{-0.15}$&$ 26 $&$^{+15}_{-11}$&$ 0.25 $&$^{+0.22}_{-0.12}$&        0.42 & 0.07 & 0.50 & 0.06  \\ \noalign{\smallskip}
J2235$-$4943 & 75000 & 2161 & 6.06 & 0.04 & -0.75 & 0.02        &$ 7.10 $&$^{+0.80}_{-0.70}$&$ 143 $&$^{+39}_{-29}$&$ 0.21 $&$^{+0.06}_{-0.04}$&       0.41 & 0.04 & 0.40 & 0.03  \\ \noalign{\smallskip}
\hline\hline
\end{tabular}
\end{table}
\end{center}

\begin{center}
\begin{table*}
\caption{\label{table:DO} Effective temperatures, surface gravities, logarithmic C/He abundance ratios (by number), radii, luminosities, gravity, HRD, and Kiel masses of the H-deficient WDs.}
\renewcommand{\tabcolsep}{1.8mm}
\begin{tabular}{l r@{$\pm$}l r@{$\pm$}l l r@{}l r@{}l r@{}l l r@{$\pm$}l} 
\hline\hline
\noalign{\smallskip}
Name &  \multicolumn{2}{l}{\Teff} & \multicolumn{2}{l}{\logg} & $\log$\,(C/He) & \multicolumn{2}{l}{$R$} & \multicolumn{2}{l}{$L$} & \multicolumn{2}{l}{$M_{\mathrm{grav}}$} & $M_{\mathrm{HRD}}$ & \multicolumn{2}{l}{$M_{\mathrm{Kiel}}$}\\
(short) & \multicolumn{2}{l}{[K]} & \multicolumn{2}{l}{[cm\,/\,s$^2$]} & & \multicolumn{2}{l}{[\Rsol/100]} &\multicolumn{2}{l}{[\Lsol]} & \multicolumn{2}{l}{[\Msol]} & [\Msol] & \multicolumn{2}{l}{[\Msol]} \\
\noalign{\smallskip}
\hline
\noalign{\smallskip}
J0143$+$5841 & 116903 & \phantom{00}631 & 7.87 & 0.04  &        &$ 1.08 $&$\pm0.04 $&$ 19.6 $&$^{+1.4}_{-1.3}$&$ 0.31 $&$\pm0.04 $&        0.88$\pm$0.02 & 0.69 & 0.03  \\ \noalign{\smallskip}
J0334$+$1646 & 98383 & \phantom{0}1085 & 7.87 & 0.03 & $-1.70\pm0.30$   &$ 1.64 $&$^{+0.10}_{-0.09}$&$ 22.8 $&$^{+3.1}_{-2.6}$&$ 0.73 $&$^{+0.11}_{-0.09}$&        0.65$\pm$0.08 & 0.67 & 0.05   \\ \noalign{\smallskip}
J0436$-$0654 & 54451 & \phantom{000}77 & 8.31 & 0.03 &  &$ 1.50 $&$\pm0.01 $&$ 1.79 $&$\pm0.04      $&$ 1.68 $&$^{+0.13}_{-0.12}$& 0.59$\pm$0.10 & 0.82 & 0.06  \\ \noalign{\smallskip}
J0519$-$0443 & 72686 & \phantom{0}3993  & 7.63  & 0.12 & & $1.59$ &$\pm 0.04$ & $6.3$&$^{+1.6}_{-1.4}$ & $0.39$&$^{+0.13}_{-0.10}$& 0.61$\pm$0.10 & 0.55&0.08 \\ \noalign{\smallskip}
J0602$-$1351 & 100000 & 10000 & 7.50 & 0.50 & $-0.50\pm0.20$    &$ 1.99 $&$\pm0.11 $&$ 36 $&$^{+18}_{-13}$&$ 0.46 $&$^{+0.47}_{-0.23}$&    0.58$\pm$0.12 & 0.56 & 0.13  \\ \noalign{\smallskip}
J0736$+$1444 & 81319 & \phantom{0}1599 & 7.92 & 0.03 & $-1.50\pm0.30$   &$ 1.48 $&$^{+0.02}_{-0.02}$&$ 8.7 $&$^{+0.8}_{-0.7}$&$ 0.67 $&$^{+0.06}_{-0.05}$& 0.66$\pm$0.07 & 0.66 & 0.05   \\ \noalign{\smallskip}
J0751$-$0158 & 57013 & \phantom{00}395 & 7.90 & 0.05 &  &$ 2.97 $&$^{+0.12}_{-0.11}$&$ 8.4 $&$\pm0.7    $&$ 2.60 $&$^{+5.60}_{-1.80}$& $<0.51$    & 0.61 & 0.09        \\ \noalign{\smallskip}
J1824$+$2931 & 54730 & \phantom{000}66 & 8.27 & 0.03 &  &$ 1.34 $&$^{+0.02}_{-0.02}$&$ 1.44 $&$^{+0.05}_{-0.04}$&$ 1.21 $&$^{+0.10}_{-0.09}$&  0.66$\pm$0.02 & 0.80 & 0.02  \\ \noalign{\smallskip}
J1917$-$2014 & 112649 & \phantom{0}2839 & 8.00 & 0.10 & &$ 3.05 $&$^{+0.18}_{-0.16}$&$ 135 $&$^{+22}_{-19}$&$ 3.40 $&$^{+1.00}_{-0.80}$&       0.53$\pm$0.04 & 0.74 & 0.05  \\ \noalign{\smallskip}
J1943$+$5217 & 70601 & \phantom{00}679 & 7.56 & 0.03 & $-2.20\pm0.30$ &$ 1.50 $&$\pm0.01 $&$ 5.01 $&$\pm0.21      $&$ 0.30 $&$^{+0.02}_{-0.02}$& 0.63$\pm$0.08 & 0.53 & 0.09   \\ \noalign{\smallskip}
J2115$-$6158 & 70279 & \phantom{00}914 & 7.56 & 0.03 &  &$ 1.56 $&$^{+0.03}_{-0.03}$&$ 5.3 $&$\pm0.4    $&$ 0.32 $&$^{+0.03}_{-0.02}$& 0.61$\pm$0.08 & 0.52 & 0.10     \\ \noalign{\smallskip}
\hline\hline
\end{tabular}
\end{table*}
\end{center}

\end{appendix}

\end{onecolumn}

\end{document}